\documentclass[sn-mathphys-num]{sn-jnl}

\usepackage{graphicx}%
\usepackage{multirow}%
\usepackage{amsmath,amssymb,amsfonts}%
\usepackage{amsthm}%
\usepackage{mathrsfs}%
\usepackage[title]{appendix}%
\usepackage{xcolor}%
\usepackage{textcomp}%
\usepackage{manyfoot}%
\usepackage{booktabs}%
\usepackage{algorithm}%
\usepackage{algorithmicx}%
\usepackage{algpseudocode}%
\usepackage{listings}%
\usepackage{subfigure}
\usepackage{dsfont}
\usepackage{braket}
\usepackage{tikz}
\usepackage{enumitem}
\usepackage{caption}
\usepackage[normalem]{ulem}

\def\ket#1{| #1 \rangle}

\begin{document}

\pagecolor{white}

\title[Grokking as an entanglement transition in tensor network machine learning]{Grokking as an entanglement transition in tensor network machine learning}


\author[1,2]{\fnm{Domenico} \sur{Pomarico}}\email{domenico.pomarico@ba.infn.it}

\author*[1,2]{\fnm{Alfonso} \sur{Monaco}}\email{alfonso.monaco@ba.infn.it}

\author[1,2]{\fnm{Giuseppe} \sur{Magnifico}}\email{giuseppe.magnifico@uniba.it}

\author[1,2]{\fnm{Antonio} \sur{Lacalamita}}\email{antonio.lacalamita@ba.infn.it}

\author[1,2]{\fnm{Ester} \sur{Pantaleo}}\email{ester.pantaleo@uniba.it}

\author[2,3]{\fnm{Loredana} \sur{Bellantuono}}\email{loredana.bellantuono@ba.infn.it}

\author[2,4]{\fnm{Sabina} \sur{Tangaro}}\email{sonia.tangaro@ba.infn.it}

\author[1,2]{\fnm{Tommaso} \sur{Maggipinto}}\email{tommaso.maggipinto@ba.infn.it}

\author[1,2]{\fnm{Marianna} \sur{La Rocca}}\email{marianna.larocca@uniba.it}

\author[5,6]{\fnm{Ernesto} \sur{Picardi}}\email{ernesto.picardi@uniba.it}

\author[2,7]{\fnm{Nicola} \sur{Amoroso}}\email{nicola.amoroso@uniba.it}

\author[5,6]{\fnm{Graziano} \sur{Pesole}}\email{graziano.pesole@uniba.it}

\author[1,2]{\fnm{Sebastiano} \sur{Stramaglia}}\email{sebastiano.stramaglia@ba.infn.it}
\equalcont{These authors contributed equally to this work.}


\author[1,2]{\fnm{Roberto} \sur{Bellotti}}\email{roberto.bellotti@ba.infn.it}
\equalcont{These authors contributed equally to this work.}

\affil[1]{\orgdiv{Dipartimento Interateneo di Fisica}, \orgname{Università degli Studi di Bari}, \orgaddress{\city{Bari}, \postcode{I-70125}, \state{Italy}}}

\affil[2]{\orgdiv{Istituto Nazionale di Fisica Nucleare}, \orgname{Sezione di Bari}, \orgaddress{\city{Bari}, \postcode{I-70125}, \state{Italy}}}

\affil[3]{\orgdiv{Dipartimento di Biomedicina Traslazionale e Neuroscienze (DiBraiN)}, \orgname{Università degli Studi di Bari}, \orgaddress{\city{Bari}, \postcode{I-70124}, \state{Italy}}}

\affil[4]{\orgdiv{Dipartimento Di Scienze Del Suolo, Della Pianta e Degli Alimenti}, \orgname{Università degli Studi di Bari}, \orgaddress{\city{Bari}, \postcode{I-70125}, \state{Italy}}}

\affil[5]{\orgdiv{Dipartimento di Bioscienze, Biotecnologie e Biofarmaceutica}, \orgname{Università degli Studi di Bari}, \orgaddress{\city{Bari}, \postcode{I-70125}, \state{Italy}}}

\affil[6]{\orgdiv{Istituto di Biomembrane, Bioenergetica e Biotecnologie Molecolari}, \orgname{Consiglio Nazionale delle Ricerche}, \orgaddress{\city{Bari}, \postcode{I-70126}, \state{Italy}}}

\affil[7]{\orgdiv{Dipartimento di Farmacia-Scienze del Farmaco}, \orgname{Università degli Studi di Bari}, \orgaddress{\city{Bari}, \postcode{I-70125}, \state{Italy}}}


\abstract{
Grokking is a intriguing phenomenon in machine learning where a neural network, after many training iterations with negligible improvement in generalization, suddenly achieves high accuracy on unseen data. By working in the quantum-inspired machine learning framework based on tensor networks, we numerically prove that grokking phenomenon can be related to an entanglement dynamical transition in the underlying quantum many-body systems, consisting in a one-dimensional lattice with each site hosting a qubit. Two datasets are considered as use case scenarios, namely fashion MNIST and gene expression communities of hepatocellular carcinoma. In both cases, we train Matrix Product State (MPS) to perform binary classification tasks, and we analyse the learning dynamics. We exploit measurement of qubits magnetization and correlation functions in the MPS network as a tool to identify meaningful and relevant gene subcommunities, verified by means of enrichment procedures.}




\maketitle

\section{Introduction}\label{sec1}


Quantum computing has been emerging as a transformative paradigm weaving together diverse scientific disciplines ranging from computer science to quantum physics. At the intersection of quantum computing and artificial intelligence, quantum machine learning promises computational advantages for specific learning tasks through controlled manipulations of quantum states \cite{arrazola, banchiscience, chinaphotonics, vakili2024, benedetti, hibat, Gili_2023, qgeneralize, holmes2024, Peters2023generalization, bowles2023contextuality, eisert2024}. However, current quantum devices face significant challenges, particularly due to quantum noise and decoherence that affect preparation and manipulation of quantum states and limits their sizes and capabilities \cite{ibm_dqpt, PRXQuantum.2.010324}. These limitations have led to growing interest in hybrid quantum-classical strategies and quantum-inspired approaches \cite{MPScircuit, Rudolph_2024, Miller_2024, schuhmacher2024hybridtreetensornetworks, Khosrojerdi_2025}. Tensor networks, mathematical structures originally developed for studying the properties of quantum many-body systems, have emerged as a powerful framework in this context. They serve dual purposes: as classical tools for designing and testing quantum circuits, and as standalone quantum-inspired algorithms that capture quantum-like correlations while being efficiently implementable on standard classical computers. This latter application has led to significant developments in machine learning, where tensor networks provide an intriguing perspective on the relationship between expressivity, entanglement, and generalization in both quantum and classical models \cite{stoudenmire2017supervisedlearningquantuminspiredtensor, Huggins_2019, Felser_2021, Dborin_2022, Ballarin2023entanglemententropy, Collura2021, PhysRevX.8.011006, glasser2020, Gallego2022, PhysRevB.99.155131, chen2023machinelearningtreetensor}.

Within the framework of conventional machine learning the sudden gain in classification performances for unseen data is known as grokking. It is intended as a benign generalization beyond a finite training dataset memorization in an over-parameterized regime \cite{power2022grokkinggeneralizationoverfittingsmall, liu2022understandinggrokkingeffectivetheory, liu2023grokkingcompressionnonlinearcomplexity, miller2024grokkingneuralnetworksempirical, varma2023explaininggrokkingcircuitefficiency, huang2024unifiedviewgrokkingdouble}, effectively giving rise to a competition between slow and fast degrees of freedom \cite{varma2023explaininggrokkingcircuitefficiency, huang2024unifiedviewgrokkingdouble, doi:10.1073/pnas.1806579115, seroussi}, which leads to a transition characterized by a synergistic behavior among extracted features \cite{rubin2024grokkingorderphasetransition, clauw2024informationtheoreticprogressmeasuresreveal}.


The description of training dynamics is well captured by neural tangent kernel \cite{NEURIPS2018}, a theoretical framework that has been extended to quantum neural networks \cite{PRXQuantum.3.030323, PhysRevLett.130.150601, zhang2023dynamicalphasetransitionquantum, zhang2024curserandomquantumdata, zhang2024quantumdatadrivendynamicaltransitionquantum}. In the quantum context, grokking manifests as a dynamical phase transition between frozen-kernel and frozen-error phases, characterized by a transcritical bifurcation \cite{PRXQuantum.3.030323, zhang2023dynamicalphasetransitionquantum}. This description has been further enriched to account for varying dataset cardinalities \cite{zhang2024quantumdatadrivendynamicaltransitionquantum}.


Biological data and drug discovery applications in pharmacology inherently involve complex, large-scale datasets. These challenges can be addressed through combined approaches that exploits both quantum models and big data machine learning tools \cite{banchiscience, chinaphotonics, vakili2024, pharmaQML}, with particularly promising applications in cancer classification and risk assessment \cite{Pomarico2025, repetto2024quantumenhancedstratificationbreast, breastQML, recurrence, survival}. In general, implementing the large quantum circuits required for these applications remains challenging on current quantum hardware \cite{zhang2024quantumdatadrivendynamicaltransitionquantum, repetto2024quantumenhancedstratificationbreast}, with only initial demonstrations achieved through quantum annealing approaches \cite{dwave2, dwave1, multiomicsQML}.


The management of of a much higher computational complexity through Matrix Product State (MPS) \cite{stoudenmire2017supervisedlearningquantuminspiredtensor, guo2021neuraltangentkernelmatrix, Ballarin2023entanglemententropy} is essential with respect to machine learning applications involving datasets endowed by a high number of features \cite{PhysRevX.8.011006, glasser2020, Gallego2022, PhysRevB.99.155131, chen2023machinelearningtreetensor}. An increasing complexity of the considered dataset, quantified in terms of the required number of qubits and parameters, may lead towards a drop in generalizability \cite{zhang2024curserandomquantumdata}. The increasing dimension of the space for solution inherently includes barren plateaus \cite{nevenbarren, Larocca2022diagnosingbarren, larocca2023}, identified with the entanglement volume law and managed by means of projective measurement able to decode nonlocal correlations in the variational circuit characterizing the area law \cite{wiersema2023, PhysRevX.7.031016, PhysRevB.100.134203, PhysRevB.99.174205, PhysRevLett.125.030505, PhysRevB.101.104301, PRXQuantum.2.010352, BAO2021168618, fazio2022, ge2024identifyingentanglementphasesbipartite}. More generally the transition is referred to a volume to sub-volume law, as determined for the experimental framework of ion traps \cite{Sierant2022dissipativefloquet, PhysRevLett.125.070606, PhysRevX.10.041020}. This case is properly suited for the targeted quantum machine learning framework, since an area law typifies systems endowed with local interactions, while the exploited data encoding in a one-dimansional lattice of qubits cannot ensure the absence of long range entanglement before features extraction. 



In this paper we investigate the grokking phenomenon in MPS machine learning \cite{stoudenmire2017supervisedlearningquantuminspiredtensor}. We numerically show that grokking manifests through two concurrent signatures: an improvement in test data classification performance and a distinct change in the entanglement entropy scaling along the one-dimensional qubit chain. Starting from randomly initialized MPS, we numerically identify a transition in the entanglement spectrum pointing out a connection with eigenvalue evaporation \cite{PhysRevA.81.052324, Torlai_2014, Facchi_2019}.

\begin{figure}
    \centering
    \subfigure[]{\includegraphics[width=0.9\linewidth]{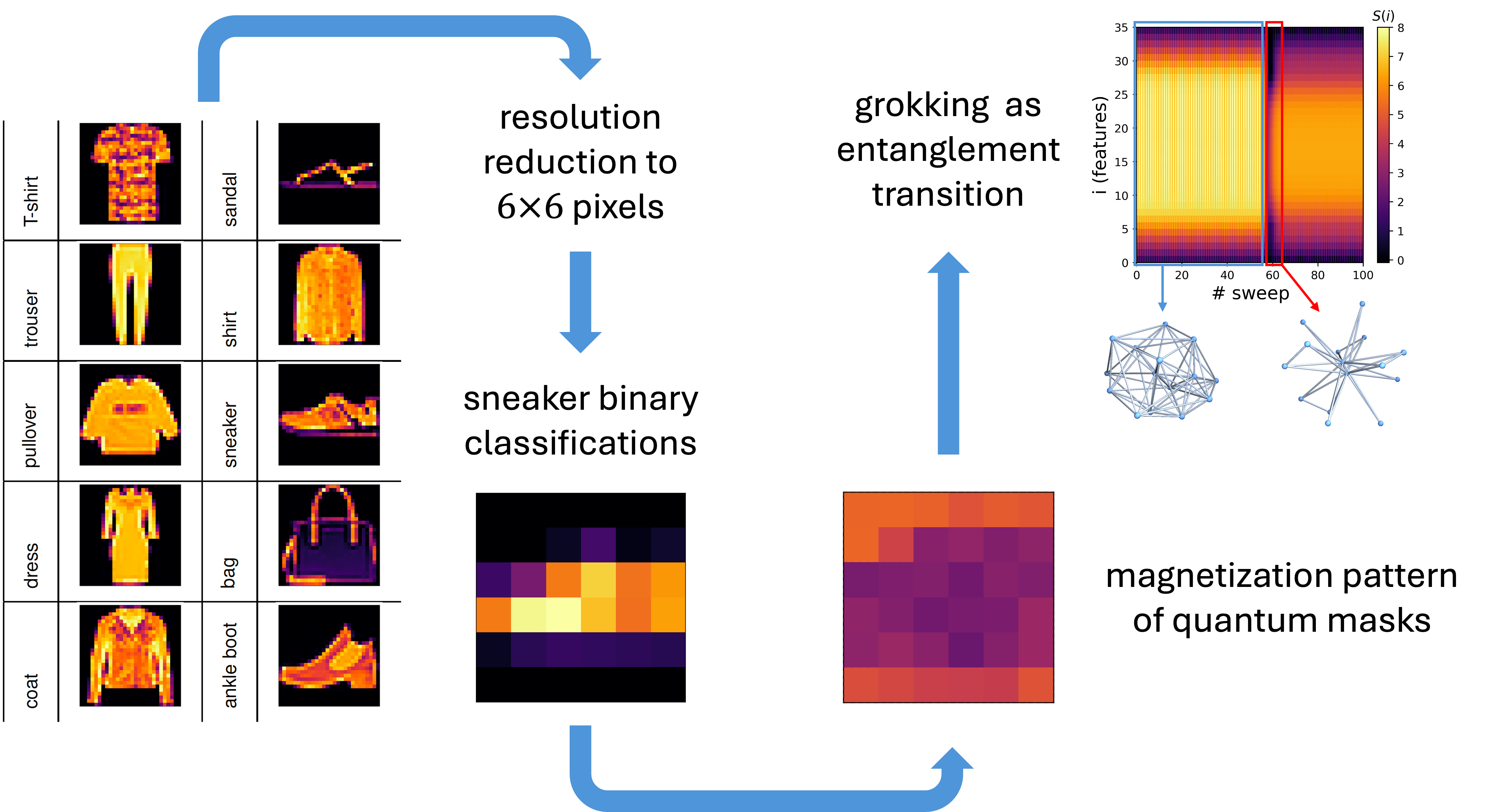}}
    \subfigure[]{\includegraphics[width=0.9\linewidth]{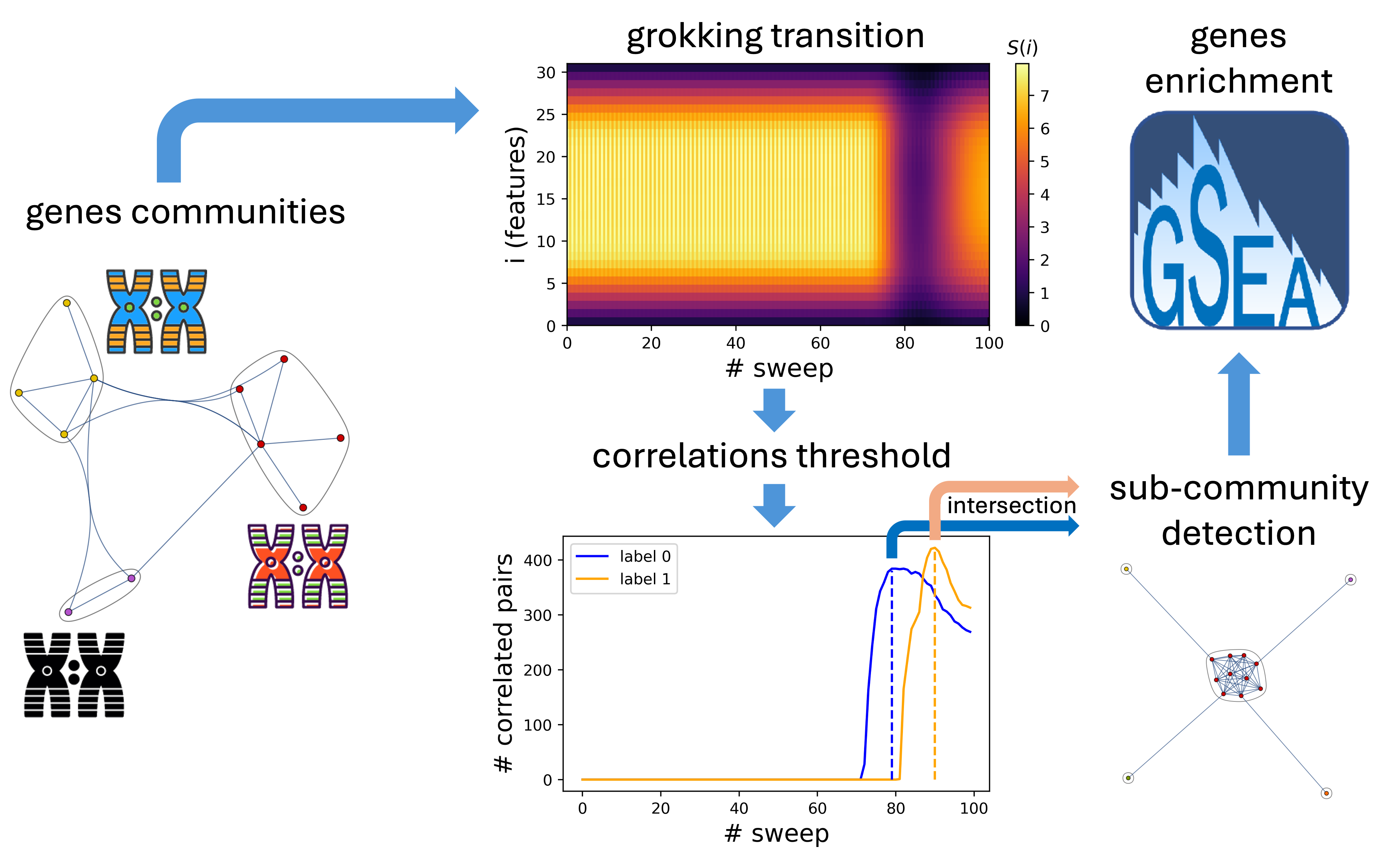}}
    \caption{A flowchart representing the adopted workflow, based on fashion MNIST for panel (a) and gene communities identified in \cite{genescomm} for panel (b). The grokking behavior is characterized for each binary classification involving sneaker with a reduced resolution version of fashion MNIST, as resumed in panel (a). Features extraction emerges with magnetization patterns corresponding to generalization and to the entanglement transition from volume-law of the initial random MPS. In panel (b) for each community features subset we ascertain whether or not the entanglement transition takes place, such that a correlation threshold can be imposed to select gene expressions pairs during the features extraction stage associated with grokking. Once the maximal number of pairs is identified for each class, their intersection defines the sub-community, which are further evaluated in terms of gene set enrichment analysis.}
    \label{fig:flowchart}
\end{figure}

We consider two datasets, as use case scenarios, namely fashion MNIST \cite{glasser2020} and gene expressions from gene expression omnibus (GEO) of hepatocellular carcinoma (HCC), partitioned in communities as defined in \cite{genescomm}. The workflow adopted for the two datasets is shown in Fig. \ref{fig:flowchart}. Fashion MNIST is referred to Fig. \ref{fig:flowchart}(a), which show features extractions as magnetization patterns in the MPS masks. Corresponding to this gain in generalization performances that we are going to discuss in the following, an entanglement entropy transition, from volum- to area-law scaling, takes place: in the volume-law phase we pictorially represent the three-dimensional graph edges as a description of how entanglement connects each element in the one-dimensional lattice in a all-to-all fashion beyond the spatial proximity, while at grokking the detected sub-volume law partially unravels a more localized entanglement structure.



The workflow regarding the gene expressions communities classifications is shown in Fig. \ref{fig:flowchart}(b). We focus on the two biologically meaningful communities described in \cite{genescomm}, highlighting the presence of the associated grokking transition. This critical behavior is characterized in terms of entanglement entropy scaling. Furthermore, after the training phase, we exploit pairwise quantum-like correlations among gene expressions to determine relevant gene sub-communities. These sub-communities are evaluated through gene set enrichment analysis (GSEA), aiming at the discovery of new candidate biomarkers for diagnosis and prognosis \cite{hepatology, genescomm}, e.g. the immune microenvironment \cite{immune_cancer, immune1, immune2, immune3} or metabolic abnormalities and alcohol consumption \cite{SHIN2023152134} with related serum biomarkers \cite{YI201813}. Another example used in clinical practice is referred to alpha-fetoprotein biomarker, correlated with the observed enrichment of vascular endothelial growth factor \cite{AFP_endothelial, AFP}. Hormones and estrogen enrichments are included as well, even if the latter is endowed with a controversial role as both a carcinogen and protective effect in liver \cite{BALDISSERA201667}. These characterizations are supported by the proposed sub-communities detection, which underlines the enrichment of Wnt pathways.


The manuscript is structured as follows. In Section \ref{sec2} we introduce MPS as a mask for machine learning recognitions. In Section \ref{sec3} we highlight for the fashion MNIST dataset the emerging features extraction as magnetization patterns. In Section \ref{sec4} relevant gene sub-communities are detected through quantum correlations along the qubits lattice corresponding to the entanglement entropy transition.


\section{Methods}\label{sec2}

\begin{figure}
    \centering
    \includegraphics[width=0.7\linewidth]{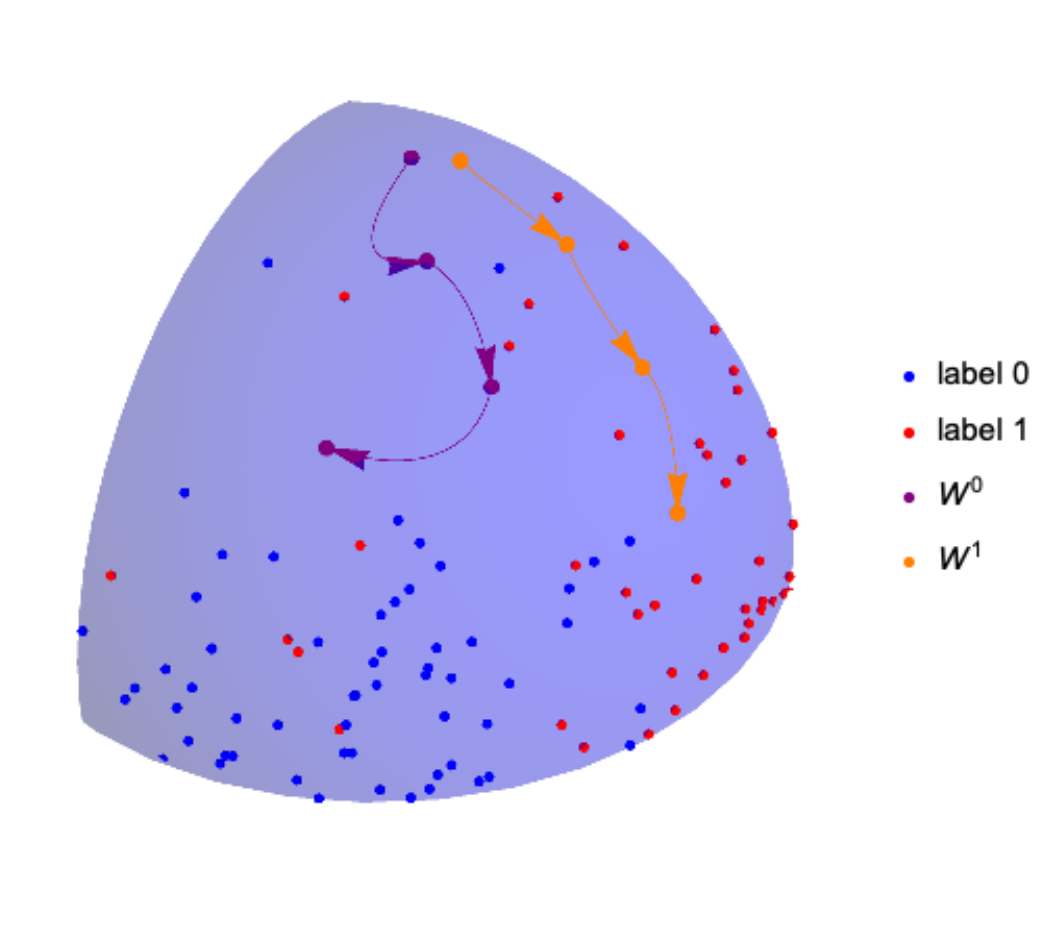}
    \caption{Representation of training dynamics with gradient flow for quantum masks $W^\ell$, introduced in Eq. \eqref{eq::pred}. Blue and red points are referred to the embedded cases for each class.}
    \label{fig:manifold}
\end{figure}


Tensor network machine learning leverages the compact representation of high-dimensional data using tensor networks, a class of tools and algorithms originally developed in quantum physics and quantum information for studying equilibrium and dynamical properties of quantum many-body systems \cite{PhysRevLett.69.2863, RevModPhys.77.259, SCHOLLWOCK201196}. Among the most common tensor network architectures used for machine learning tasks are MPS, also known as tensor trains in mathematics  \cite{Tucker1966, hackbusch2019, delathauwer2000, oseledets2011}. MPS allow to decompose large multi-dimensional tensors into a chain of smaller, interconnected tensors, having a drastic reduction in the number of variational parameters with negligible loss of expressiveness. A key strength of MPS is their capability to work with a controllable approximation error by adjusting the bond dimension, a parameter that determines the trade-off between computational efficiency and representational power. Their scalability and interpretability make them a compelling choice for bridging quantum-inspired approaches and classical machine learning.

In MPS-based machine learning for classification tasks, each element of the dataset, assuming that $N$ is the number of features, is mapped onto a product state of $N$ qubits arranged on a one-dimensional lattice. Each qubit, representing a two-level quantum-mechanical system, encodes a feature from the data, as described in more details in Appendix \ref{secA1}.

In a classification problem involving $N_{cl}$ classes labeled by the index $\ell$, the MPS defines a predictor by acting on the qubits in the resulting product state as
\begin{equation} \label{eq::pred}
    f^\ell_W (\mathbf{x}) = \sum_{s_1 \dots s_N} W^\ell_{s_1,\dots,s_N} \ket{\phi(x^{(1)})^{s_1}} \otimes \dots \otimes \ket{\phi(x^{(N)})^{s_N}},
\end{equation}
which has to minimize the mean squared error cost function during the training phase
\begin{equation} \label{eq::cost}
    \mathcal{C}(W) = \frac{1}{2} \sum_{\omega=1}^{N_T} \sum_\ell \left( f^\ell_{W}(\mathbf{x}_{\omega}) - y^\ell_{\omega} \right)^2,
\end{equation}
where $N_T$ is the number of training elements $(\mathbf{x}_{\omega}, y^\ell_{\omega})$ with $y^\ell_{\omega}=1$ for an element belonging to the $\ell$-th class and $y^\ell_{\omega}=0$ otherwise. The predicted label is determined by the component showing the highest absolute value $|f^\ell_{W}(\mathbf{x})|$ \cite{chen2023machinelearningtreetensor}. 


During the training phase of the MPS pictorially depicted in Fig. \ref{fig:manifold}, we exploit the gradient descent procedure explained in Appendix \ref{secA1} to perform the cost function optimization. The measured quantities during training dynamics are: reduced density matrix $\varrho^{(\ell)}$ in the label space associated with the tensor index $\ell$, local magnetization for the $i$-th feature $\braket{\sigma_Z^{k,i}}$ projected in the label space for $k=0,1$, as well as correlation functions $\braket{\sigma_Z^{k,i}\sigma_Z^{k,j}}$ between features $(i, j)$. In this way we can introduce
\begin{equation} \label{eq::corr} 
C^k_{i,j} = \frac{\braket{\sigma_Z^{k,i} \sigma_Z^{k,j}} - \braket{\sigma_Z^{k,i}} \braket{\sigma_Z^{k,j}}}{\varrho^{(\ell)}_{k,k}} 
\end{equation}
such that we can select features sub-communities by imposing $|C^k_{i,j}| > t$, where $t$ is a threshold. 

\subsection{Fashion MNIST dataset}\label{subsec2a}

\begin{figure}
    \centering
    \begin{tabular}{cc}
    \begin{tabular}{c|c|c}
         & $28 \times 28$ & $6 \times 6$ \\
        \hline
        \rotatebox{90}{T-shirt} & \includegraphics[width=0.16\linewidth]{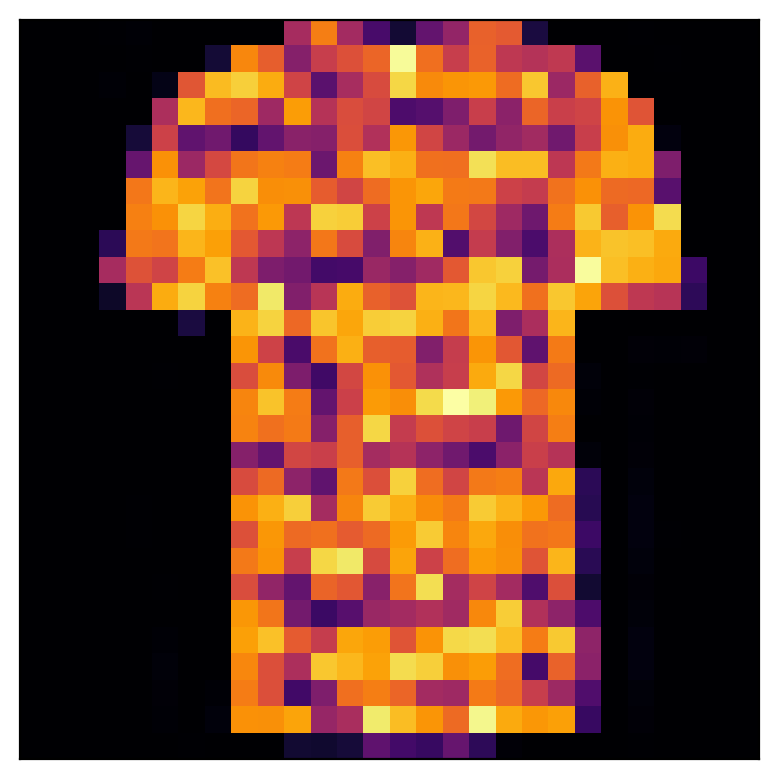} & \includegraphics[width=0.16\linewidth]{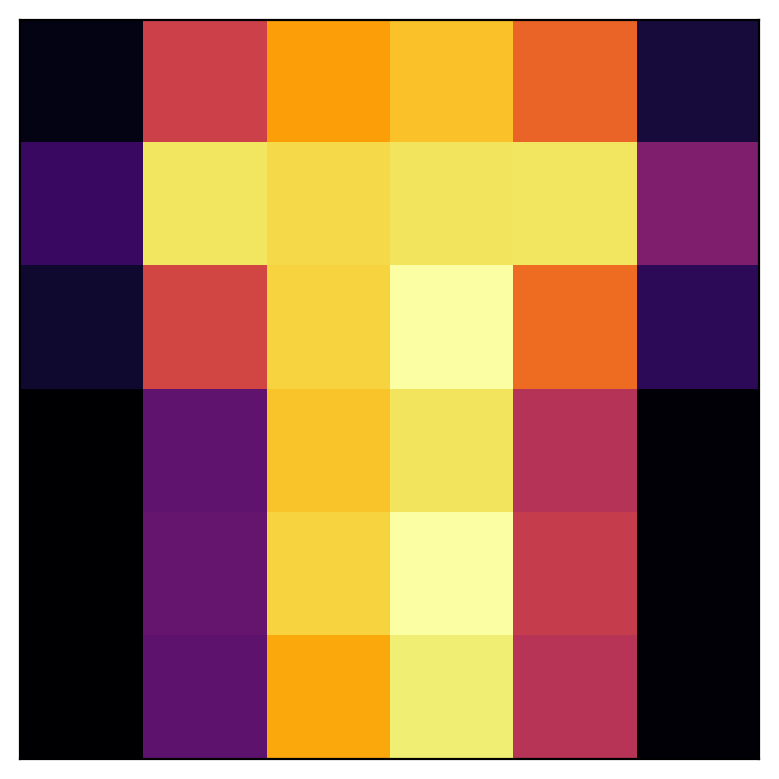} \\
        \hline
        \rotatebox{90}{trouser} &\includegraphics[width=0.16\linewidth]{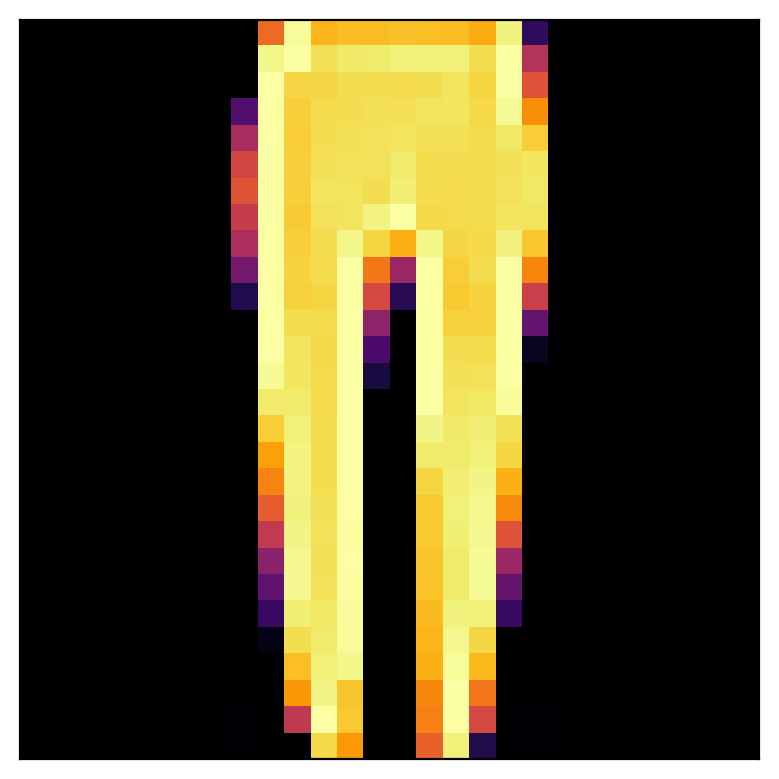} & \includegraphics[width=0.16\linewidth]{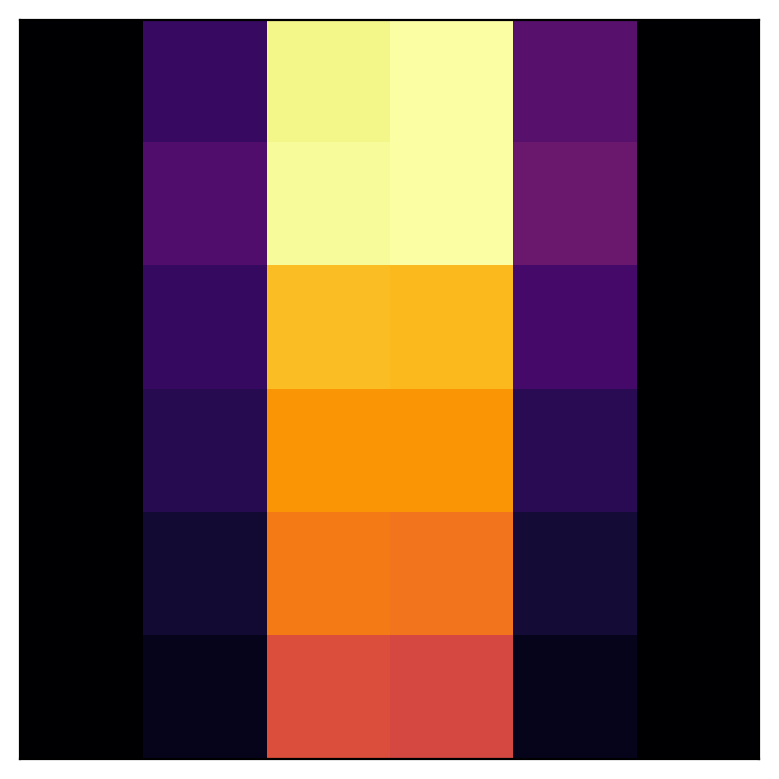} \\
        \hline
        \rotatebox{90}{pullover} &\includegraphics[width=0.16\linewidth]{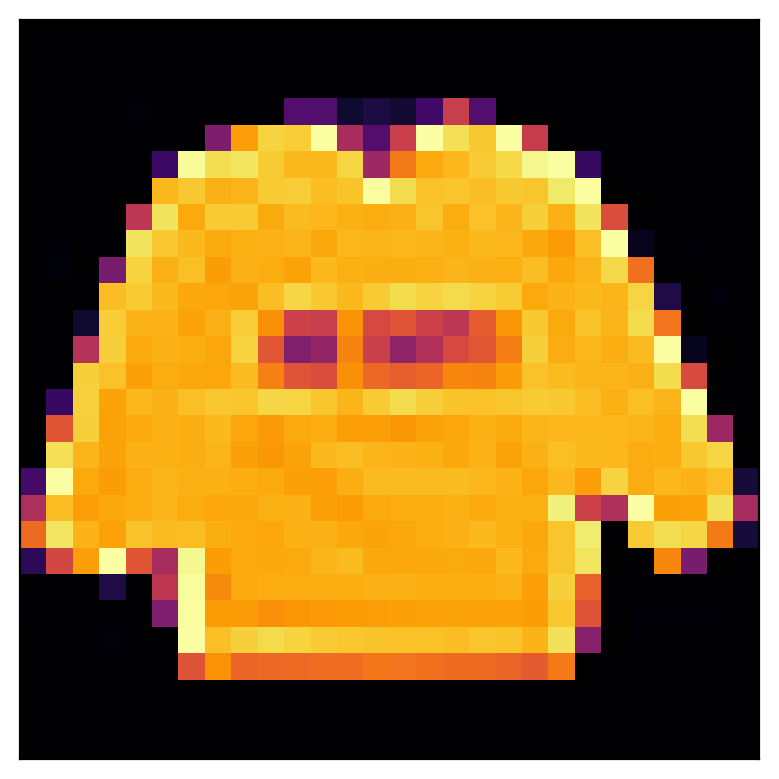} & \includegraphics[width=0.16\linewidth]{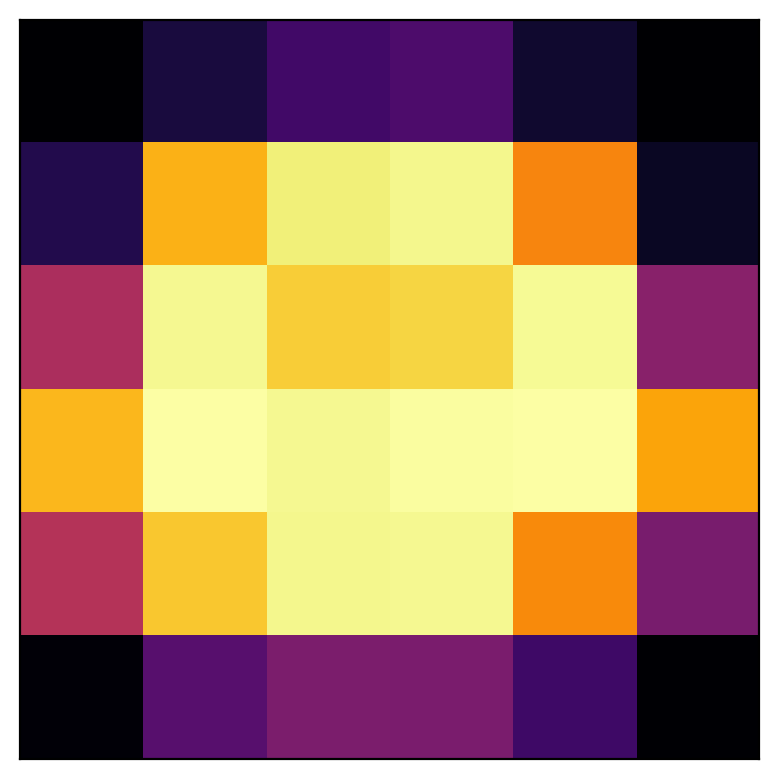} \\
        \hline
        \rotatebox{90}{dress} &\includegraphics[width=0.16\linewidth]{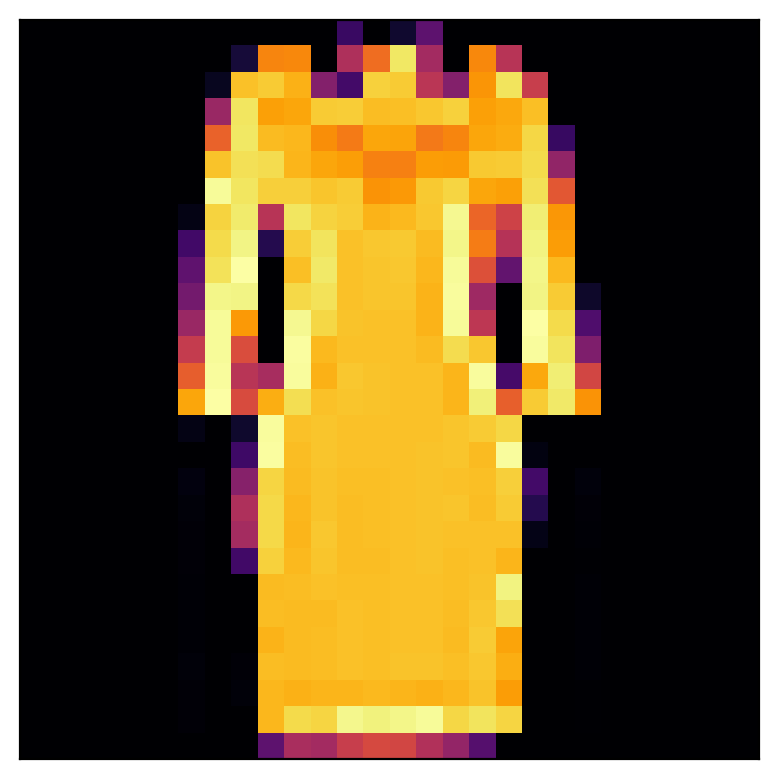} & \includegraphics[width=0.16\linewidth]{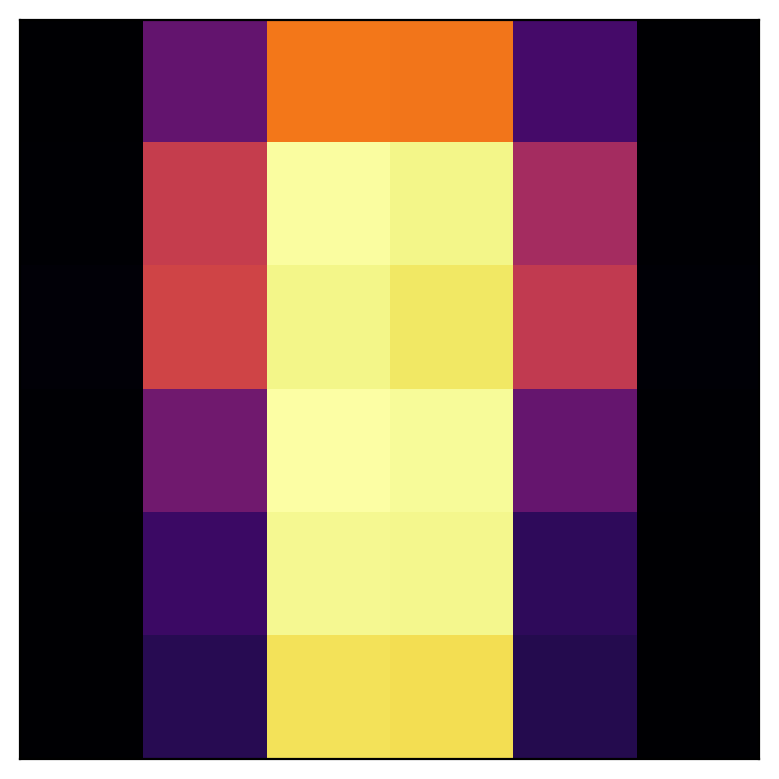} \\
        \hline
        \rotatebox{90}{coat} &\includegraphics[width=0.16\linewidth]{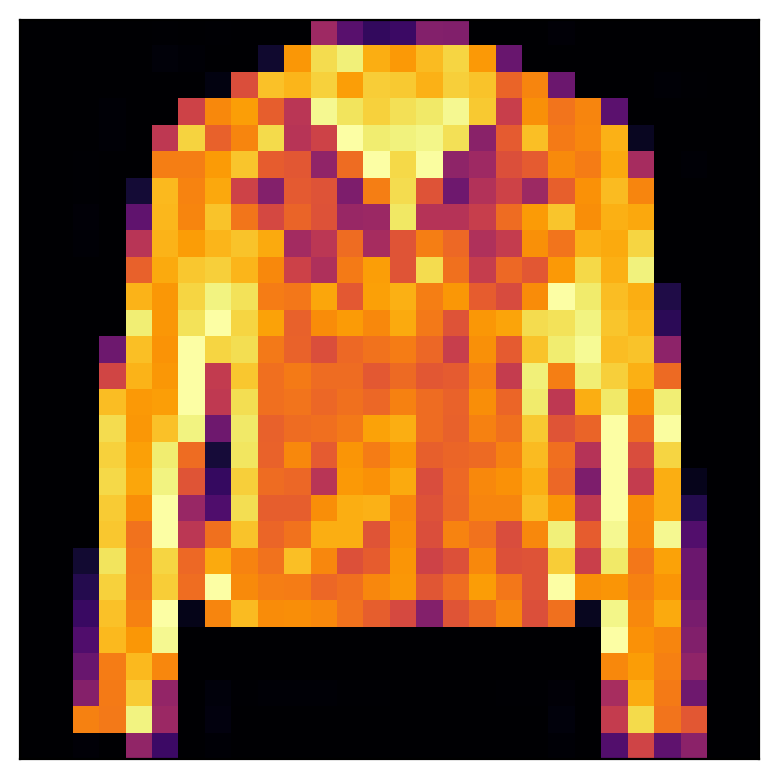} & \includegraphics[width=0.16\linewidth]{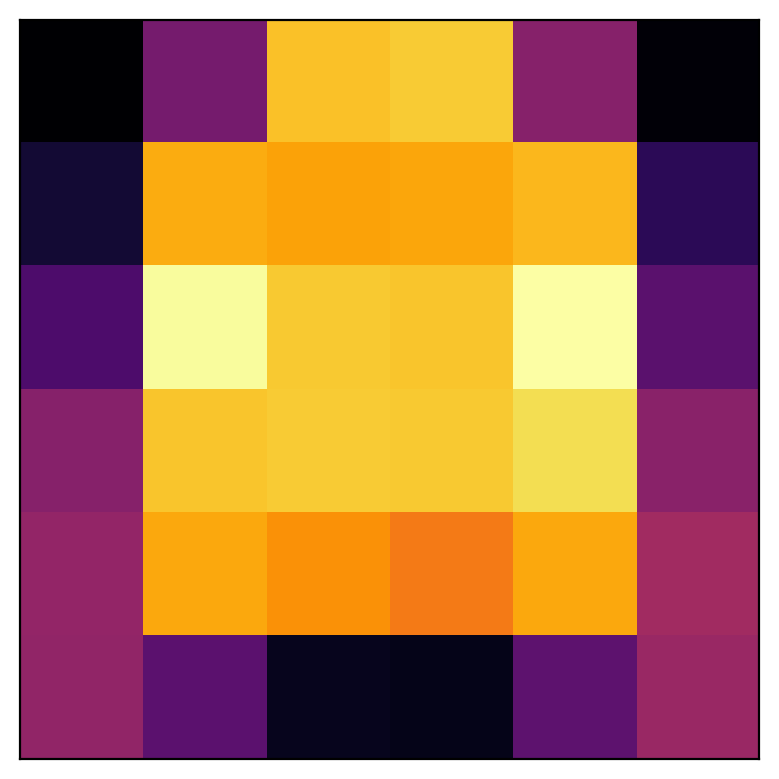}
    \end{tabular}
    &
    \begin{tabular}{c|c|c}
         & $28 \times 28$ & $6 \times 6$ \\
        \hline
        \rotatebox{90}{sandal} & \includegraphics[width=0.16\linewidth]{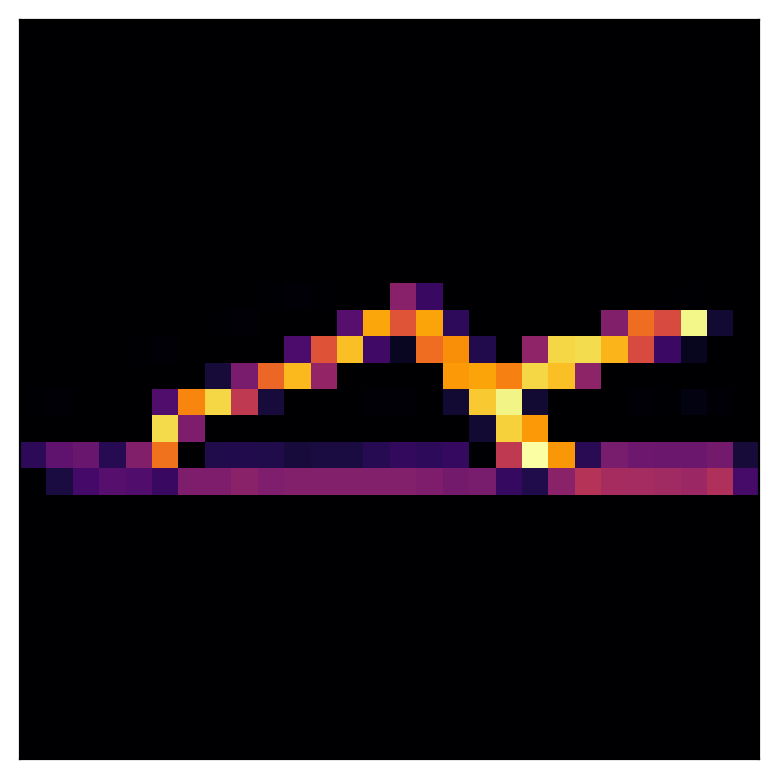} & \includegraphics[width=0.16\linewidth]{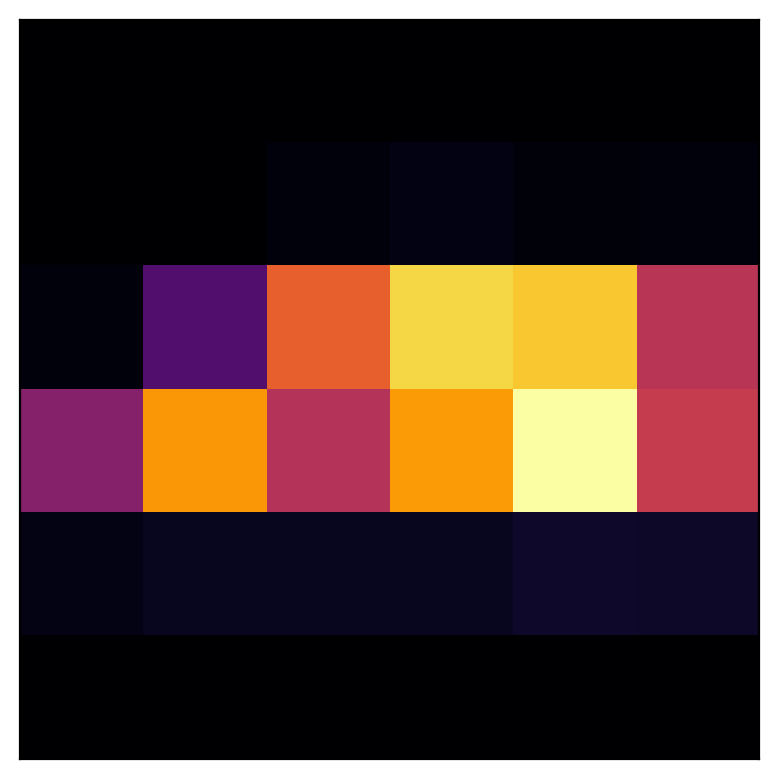} \\
        \hline
        \rotatebox{90}{shirt} &\includegraphics[width=0.16\linewidth]{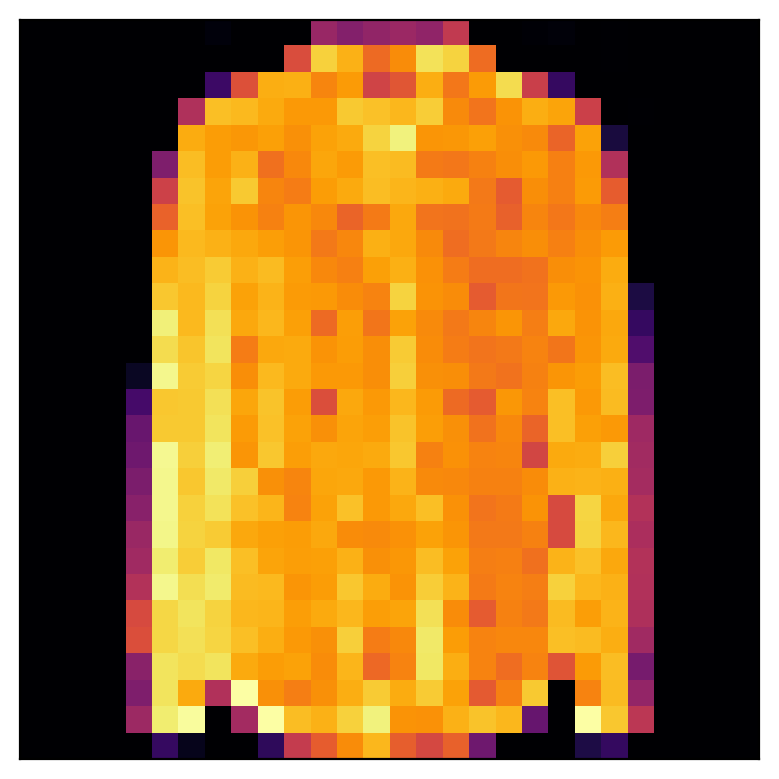} & \includegraphics[width=0.16\linewidth]{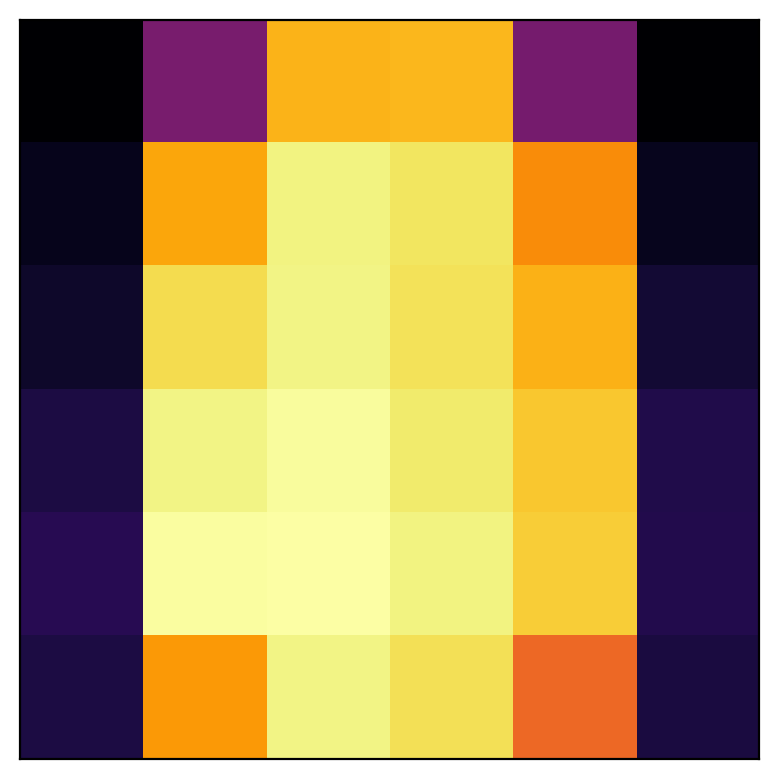} \\
        \hline
        \rotatebox{90}{sneaker} &\includegraphics[width=0.16\linewidth]{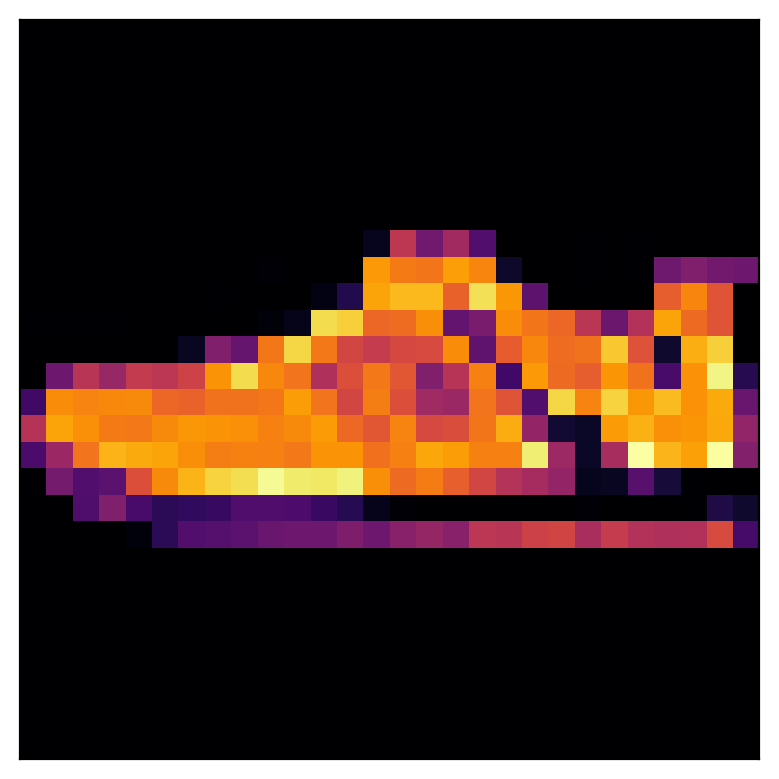} & \includegraphics[width=0.16\linewidth]{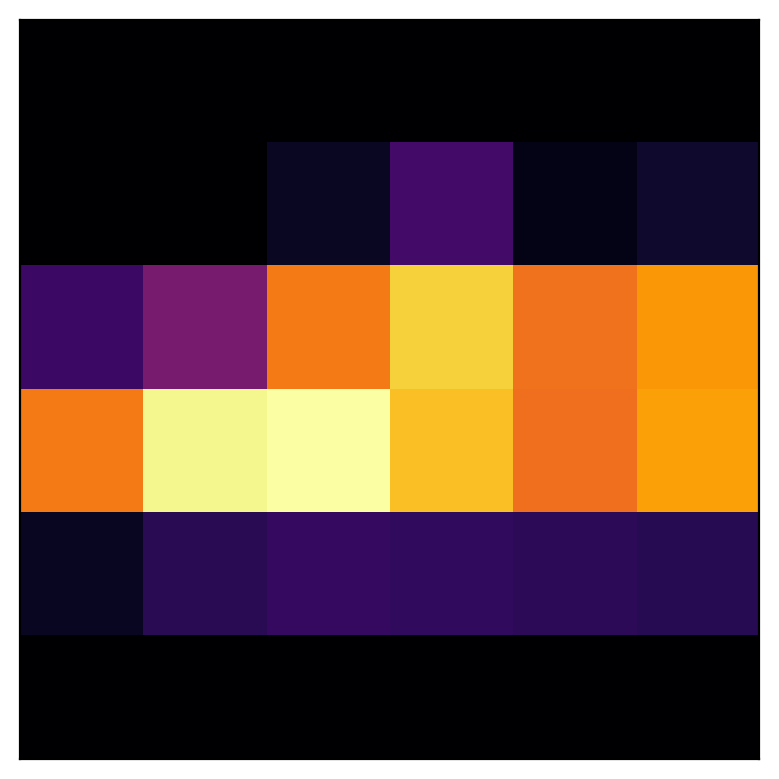} \\
        \hline
        \rotatebox{90}{bag} &\includegraphics[width=0.16\linewidth]{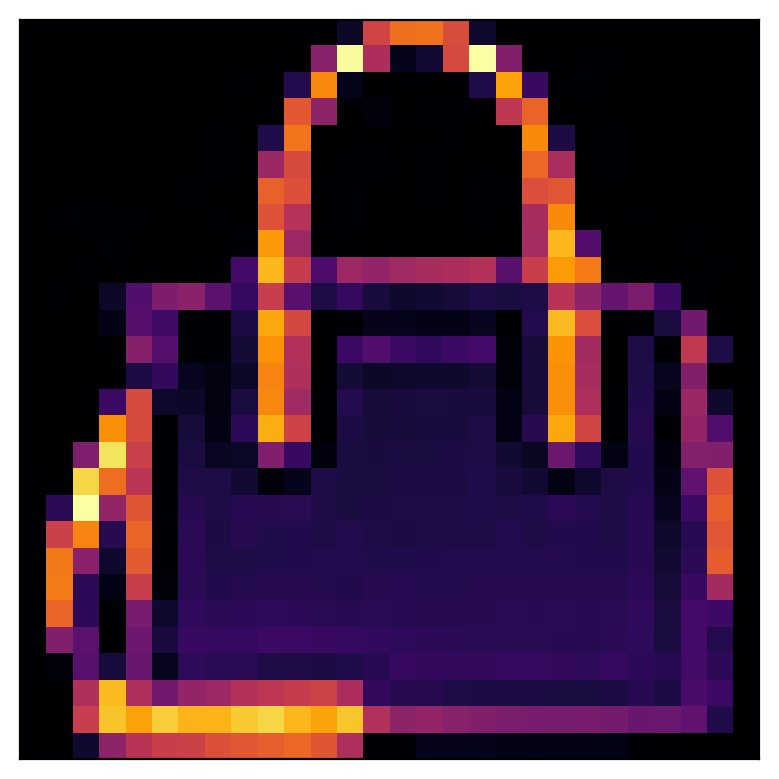} & \includegraphics[width=0.16\linewidth]{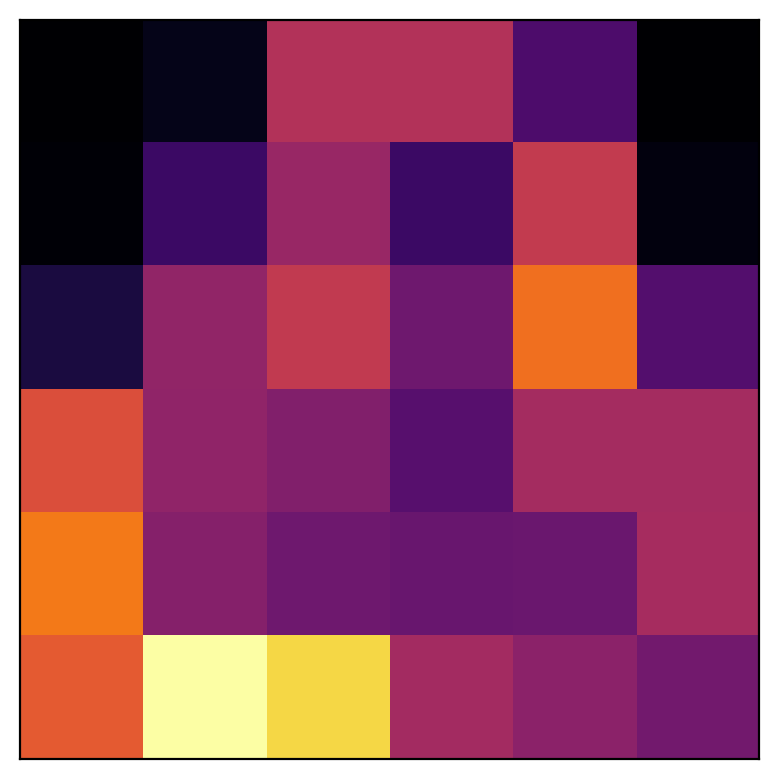} \\
        \hline
        \rotatebox{90}{ankle boot} &\includegraphics[width=0.16\linewidth]{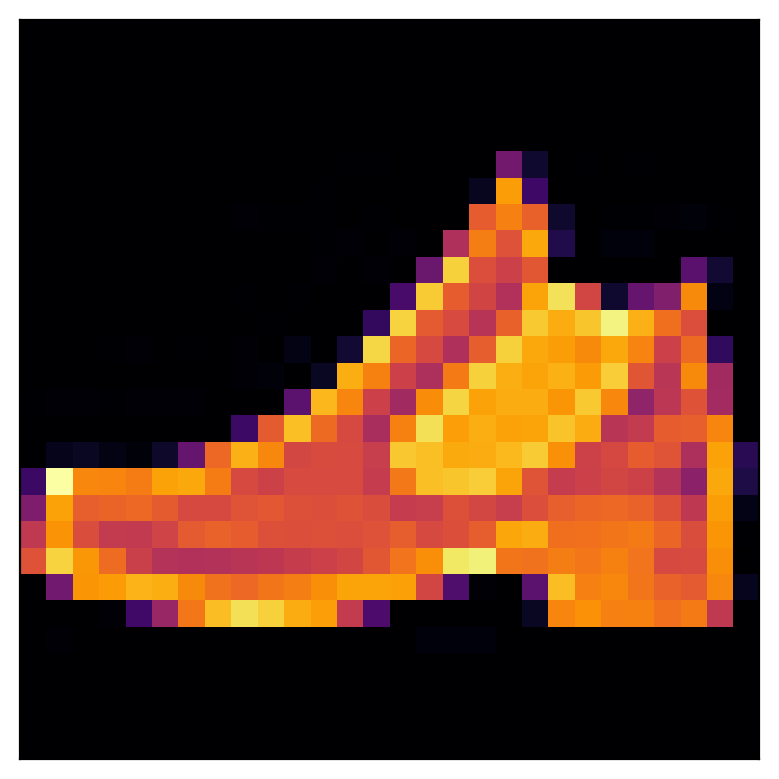} & \includegraphics[width=0.16\linewidth]{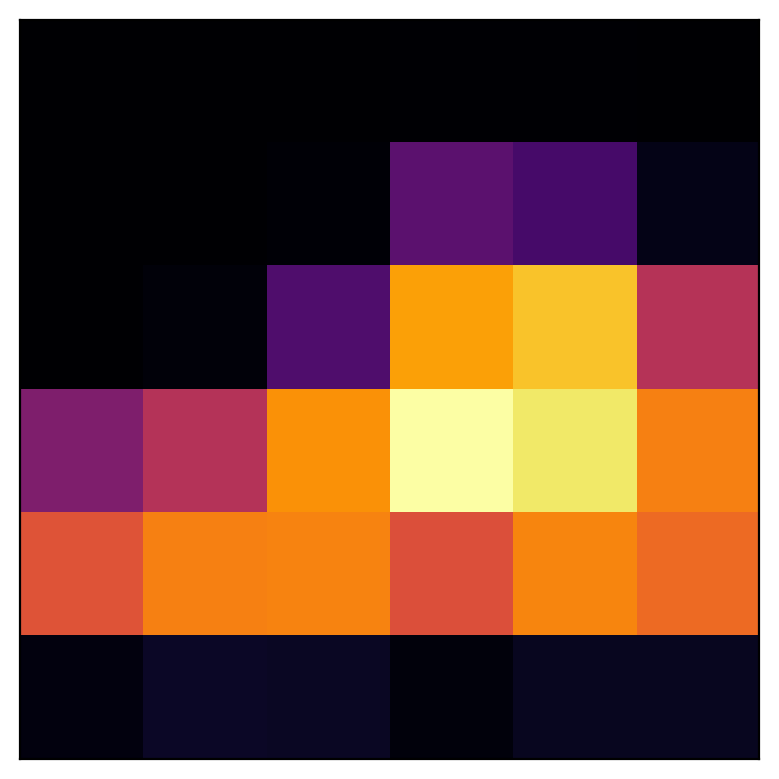}
    \end{tabular}
    \end{tabular}
    \caption{Representative elements of the fashion MNIST dataset, both in the original resolution $28 \times 28$ pixels and compressed version $6 \times 6$.}
    \label{fig:fashion}
\end{figure}

The fashion MNIST dataset is generally considered as a benchmark for machine learning models \cite{glasser2020}. It consists of $60,000$ images, with $28 \times 28$ pixels, referring in a balanced way to 10 labels, namely T-shirt, trouser, pullover, dress, coat, sandal, shirt, sneaker, bag, ankle boot. In order to obtain a dataset with a comparable cardinality of the one presented in Section \ref{subsec2b} and minimize the influence of the latter on the grokking behavior, we use subsamples composed by $1\%$ of the original images, then split in half for train and test. We select a particular class of binary classification problemss by choosing sneakers versus any remaining label, thus namely selecting 60 samples per label. This choice is explained in Fig. \ref{fig:fashion}, since to make the simulation feasible we compress the original images in a $6 \times 6$ pixels format. The characterizing trait of sneaker in the low resolution is based on the first and last black pixel rows, with just one confounding label referring to sandal, making it an easily recognized pattern. In order to map such images in a one-dimansional lattice, we unroll each element as a sequence of rows from top to bottom.

\subsection{Gene expression dataset}\label{subsec2b}

We consider two microarray public datasets, GSE102079 and GSE54236. Gene expression data from liver tissue samples of 152 patients who underwent liver resection compose the GSE102079 dataset. These samples were analyzed using the GPL570 Affymetrix Human Genome U133 Plus 2.0 array. This dataset includes 152 tumor tissues and 91 adjacent liver tissues from patients with HCC and 14 adjacent liver tissues from patients with colorectal cancer metastases who had not undergone chemotherapy. Gene expression data from 156 samples, 78 HCC tumor tissues and 78 adjacent non-tumor tissues are described in the GSE54236 dataset. These data were obtained using the GPL6480 Agilent-014850 Whole Human Genome Microarray. The last dataset is used for independent testing purposes.

Microarray sequencing technologies simultaneously measure the expression level of thousands of genes. Both datasets contain over 10,000 gene expressions, thus in order to reduce the computational complexity in our quantum-inspired approach, we follow the data processing in \cite{genescomm} in two steps: (i) a data normalization step based on multiarray analysis (robust multiarray analysis, RMA) with background correction of the original data, log2 transformation and quantile normalization; (ii) a community detection phase in which we first created the gene co-expression matrix and then applied the Leiden algorithm to find stable gene communities with highly correlated gene expression profiles.

Based on the results reported in \cite{genescomm}, we selected the best performing communities listed in Table \ref{tab1} out of 46 where classical algorithms discriminated HCC from control tissues with an accuracy greater than $90\%$, since the binary classification problem we are interested in are based on recognizing cancer against normal gene expressions measured in the collected tissues.

\begin{table}[t!]
\caption{Summary of the gene communities.}\label{tab1}
\begin{tabular}{c|c|c|c} 
  \hline
  \textbf{Community} & \textbf{$\#$ genes} & \textbf{Community} & \textbf{$\#$ genes} \\ 
  \hline
  C8 & 28 & C29 & 48 \\
  \hline
  C12 & 47 & C30 & 32 \\
  \hline
  C14 & 31 & C31 & 25 \\
  \hline
  C15 & 25 & C32 & 35 \\
  \hline
  C16 & 34 & C33 & 35 \\
  \hline
  C17 & 26 & C35 & 31 \\
  \hline
  C23 & 31 & C40 & 29 \\
  \hline
  C24 & 23 & C41 & 48 \\
  \hline
  C27 & 36 & C42 & 33 \\
  \hline
  C28 & 35 & C43 & 32 \\
\end{tabular}
\end{table}

\section{Features extraction in magnetization patterns}\label{sec3}

The training of the MPS corresponds to the training of a mask per label able to distinguish classes in the considered binary classification, as described in more details in Appendix \ref{secA1} and \ref{secA2}. In order to study the grokking phenomenon, we analyze the training dynamics of the mask associated with each label.
 
\begin{figure}
    \centering
    \subfigure[]{\includegraphics[width=\linewidth]{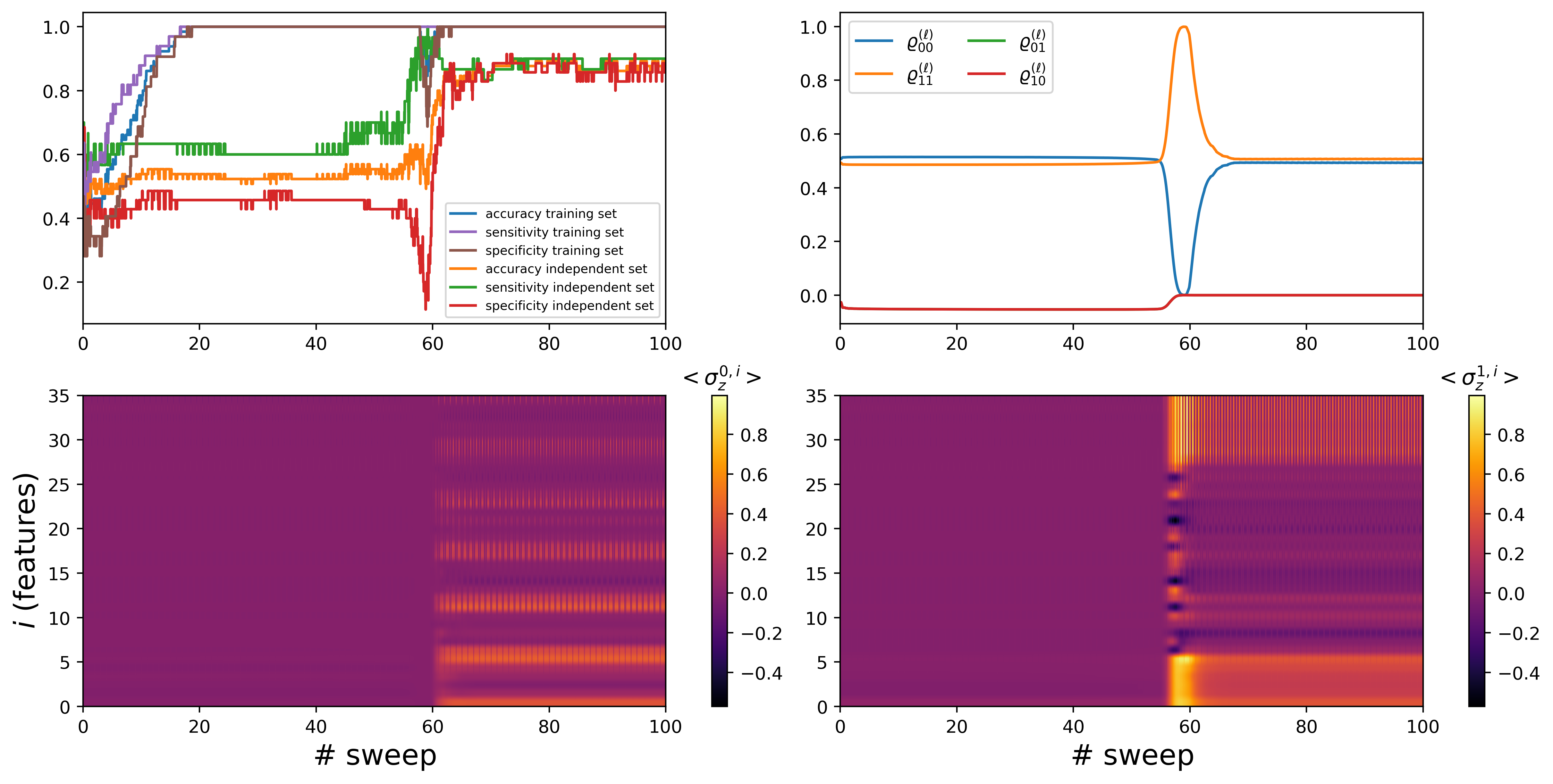}}
    \subfigure[]{\includegraphics[width=0.5\linewidth]{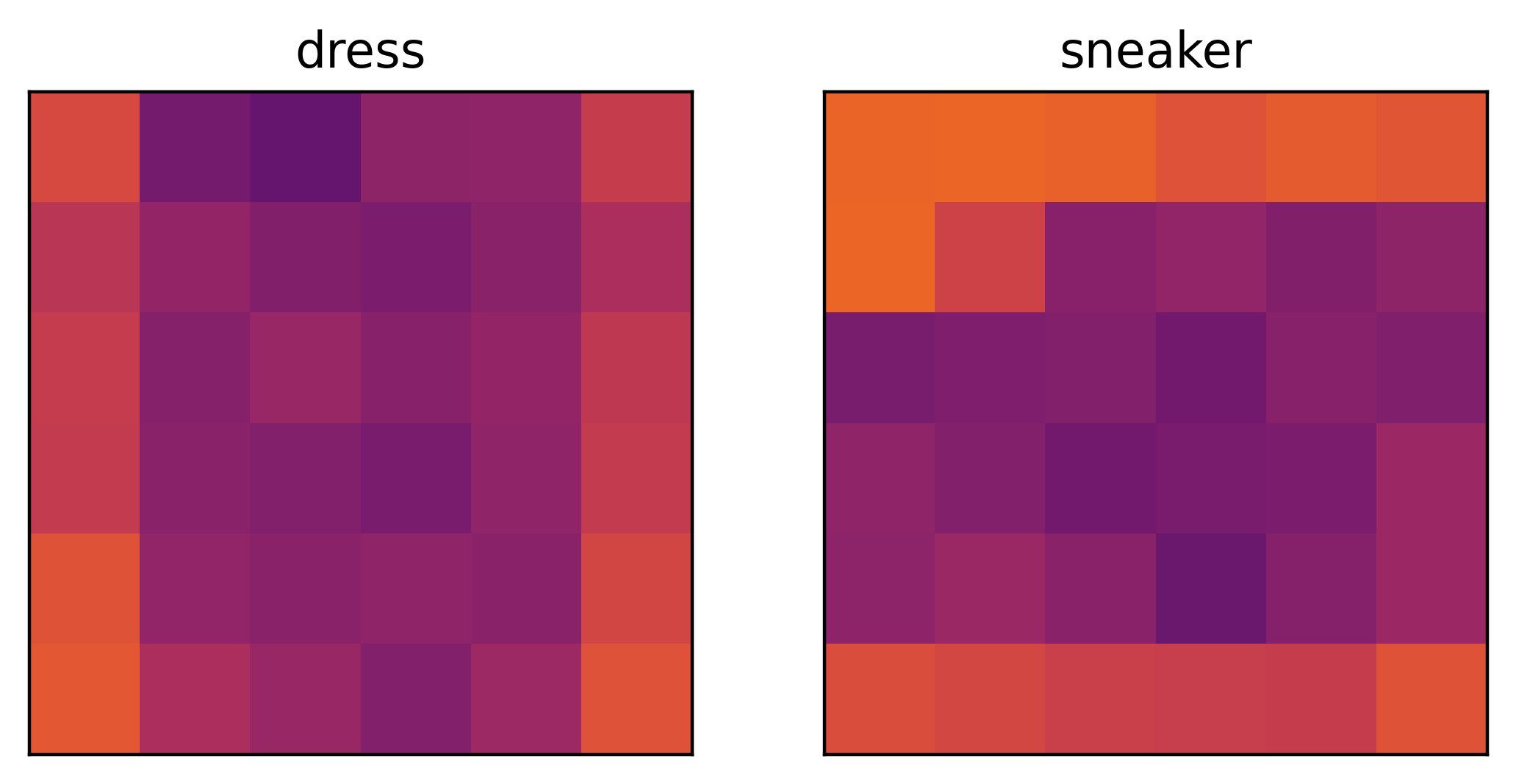}}
    \caption{Grokking for the binary classification dress versus sneaker. In panel (a) every plot is referred to the considered 100 sweeps, with evaluation metrics in the top left, reduced density matrix in the top right and local magnetization per mask at the bottom. In panel (b) the rearranged squared images of the two objects are obtained from the final local magnetizations, with the same color map used in panel (a).}
    \label{fig:dress_sneaker}
\end{figure}

An illustrative example from the first dataset involves the binary classification of dress (label 0) versus sneaker (label 1) in Fig. \ref{fig:dress_sneaker}. In panel (a) every plot is referred to a sweep axis, that can be interpreted as an analogous concept of epoch in conventional machine learning, as briefly explained in Appendix \ref{secA1}. The top left plot shows the behavior of three metrics, namely accuracy, sensitivity (also known as recall) and specificity:
\begin{align}
    \mathrm{accuracy} = & \ \frac{TP + TN}{TP + TN + FP + FN}, \\
    \mathrm{sensitivity} = & \ \frac{TP}{TP+FN} = P(|f_W^1(\mathbf{x})|>|f_W^0(\mathbf{x})| | y=1), \\
    \mathrm{specificity} = & \ \frac{TN}{TN+FP} = P(|f_W^1(\mathbf{x})|<|f_W^0(\mathbf{x})| | y=0),
\end{align}
where $TP$ and $TN$ stand for true positives (number of sneaker correctly classified) and true negatives (number of dress correctly classified), while $FP$ (number of dress identified as sneaker) and $FN$ (number of sneaker identified as dress) are false positives and false negatives, respectively. 

The peculiar grokking behavior emerges after approximately 60 sweeps, since aforementioned metrics increase in an overfitting regime for the training set, while concerning the independent set a sudden jump of sensitivity is followed by the one of specificity, thus reaching a stable trade-off between them yielding an accuracy equal to $88\%$ after 100 sweeps. 

In the top right plot of Fig. \ref{fig:dress_sneaker}(a) the represented elements of the reduced density matrix show that off-diagonal coherence elements vanishing after grokking and the diagonal element $\varrho^{(\ell)}_{11}$ reaching the value 1 corresponding to the maximum of sensitivity, since sneaker is recognized a few sweeps before dress. 


The plots at the bottom of Fig. \ref{fig:dress_sneaker}(a) illustrate the behavior of the two masks in terms of local magnetization of the underlying quantum states. When the sensitivity of the independent set undergoes a significant increase, the subspace associated with the sneaker mask dominates the training dynamics. From there on, the presence of the competing dress mask redistributes the training dynamics between the two components, leading to an increase in specificity. The final masks rearranged in a squared format are shown in panel (b), where we observe positive magnetization corresponding to black pixels in Fig. \ref{fig:fashion}, such that the shape of the corresponding object is recognized. This behavior mimics the output of explainability techniques based on features relevance propagation through vision transformers: the color map representation of this quantity over input images allows to figure out the pixels weight affecting the classifier identification of depicted objects \cite{abnar2020quantifyingattentionflowtransformers, Chefer_2021_CVPR}.

\begin{figure}
    \centering
    \includegraphics[width=\linewidth]{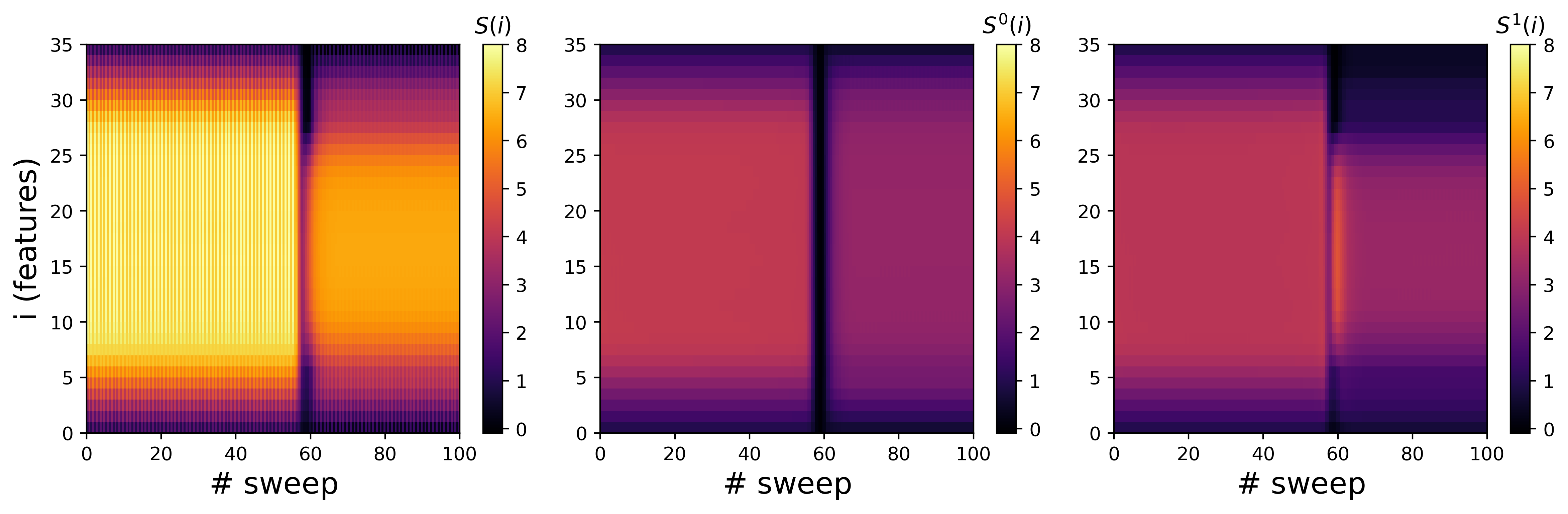}
    \caption{Entanglement entropies $S(i)$, $S^{0}(i)$ and $S^{1}(i)$, from left to right, along the one-dimensional lattice for each bipartition corresponding to the edge separating sites $s_i$ and $s_{i+1}$ in the MPS tensor $W$, $W^{0}$ and $W^{1}$, respectively.}
    \label{fig:entropy}
\end{figure}

The dynamics of the local magnetization leading to a mask matching with the pattern typifying the targeted label is determined also by analyzing the modification of entanglement along the one-dimensional lattice. We quantify this dynamical behavior through Von Neumann entropy \cite{Bengtsson_Zyczkowski_2006}, considered for the global and projected state into a specific label subspace, $S(i)$ and $S^\ell (i)$ respectively, evaluated corresponding to each bipartition obtained at a specific edge in the MPS between consecutive features $(i, i+1)$. 

\begin{figure}
    \centering
    \includegraphics[width=\linewidth]{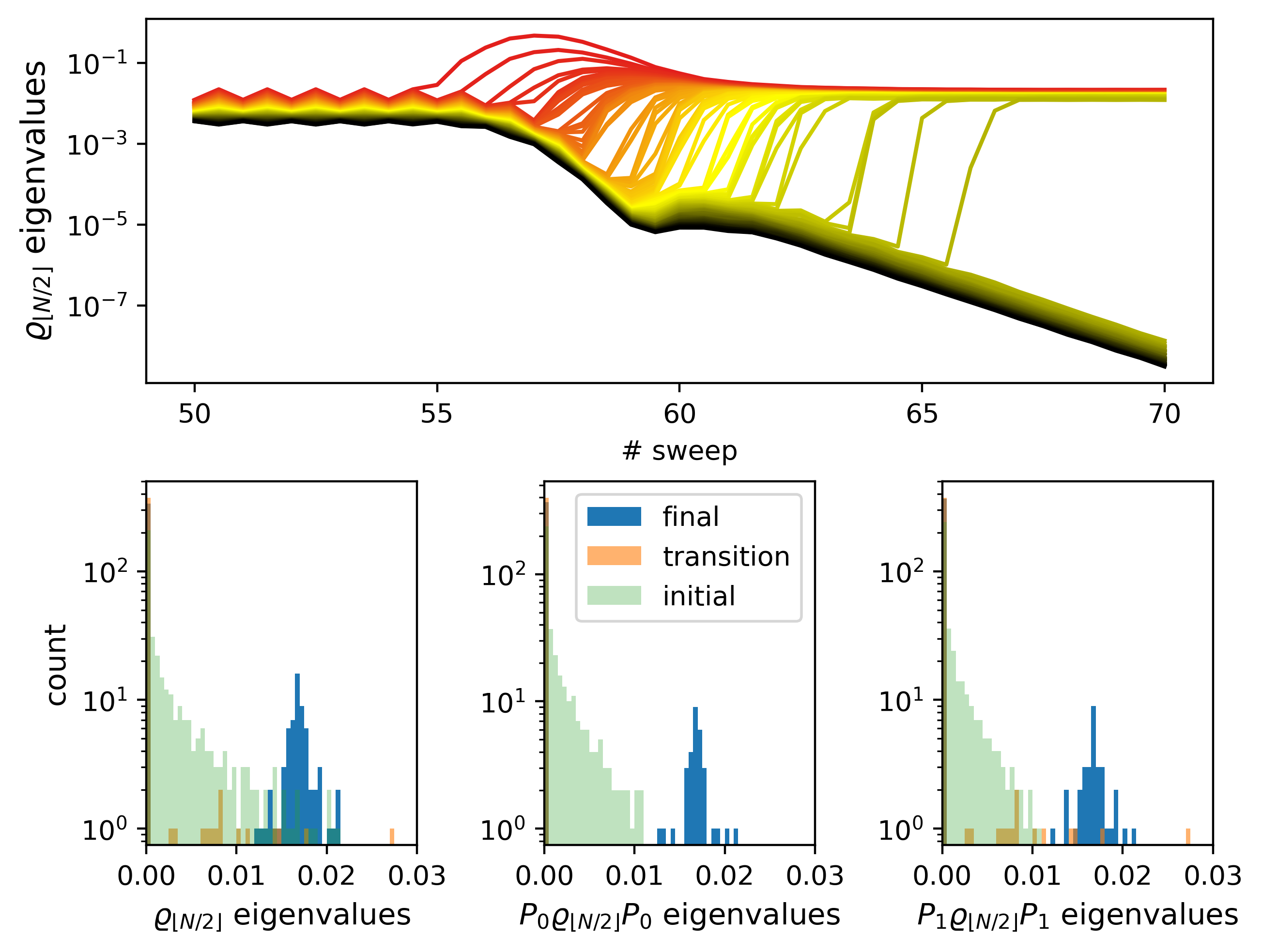}
    \caption{Spectrum dynamics of the density matrix $\varrho_{\lfloor N/2 \rfloor}$ referring to the bipartition along the edge joining the sites pair $(\lfloor N/2 \rfloor, \lfloor N/2 \rfloor+1)$. In the top panel the highest 100 eigenvalues with respect to the sweep number are denoted with graded colors from red to black. Lower panels represent the histogram distribution of such eigenvalues in the plot at the left, while histograms at the center and at the right describe eigenvalues distribution of the aforementioned density matrix projected in each label subspace.}
    \label{fig:evl}
\end{figure}

In Fig. \ref{fig:entropy} we show collected results about entanglement entropies concerning the binary classification problem for dress versus sneaker. We start with each tensor of the MPS randomly sampled according to a Gaussian distribution: this initial condition causes an all-to-all entanglement of qubits beyond their spatial proximity with a volume-law behavior for the entropy, which grows linearly with the distance from the boundary, reaching a saturation in the middle of the chain, highlighted in the first plot of Fig. \ref{fig:entropy}. Corresponding to the first mask transition during training, represented for the sneaker label in Fig. \ref{fig:dress_sneaker}(a), the tensor components $W^{0}$ for the dress label are not involved in the training dynamics for a few sweeps before the 60-th sweep, as we can verify with a vanishing entropy $S^{0}(i)$ along the lattice, in this transition condition expressing the drained dynamics in the opposite label subspace, as shown in Fig. \ref{fig:evl}. After this sweep the off-diagonal coherence elements of $\varrho^{(\ell)}$ vanish, thus implying independent masks. The tensor components $W^{1}$ also shows an almost vanishing entropy corresponding to the spins “switched-on” by the pattern recognition associated with a shape of sneaker, namely the first and last row of resized and unrolled image in Fig \ref{fig:fashion}, which become almost dynamically decoupled during the following sweeps. This decoupling is easily understood if we figure out that once some characterizing features of a pattern are extracted, the loss function will forbid to forget the acquired memory because of gradient descent. This phenomenon concentrates the norm of the reduced state in the specific label subspace along those features directions spanning the extracted pattern, thus causing some eigenvalues to be significantly higher than other meaningless directions with respect to the considered classification, as represented in Fig. \ref{fig:evl}.

In order to monitor variations at the level of the density matrix spectrum, we consider the bipartition associated with sites pair $(\lfloor N/2 \rfloor, \lfloor N/2 \rfloor+1)$. We denote the density matrix referred to this bipartition by $\varrho_{\lfloor N/2 \rfloor}$, whose spectrum dynamics corresponding to the sweeps involved in the transition is shown in Fig. \ref{fig:evl}. In the top panel we detect a sequence of eigenvalues evaporation \cite{Facchi_2019}, i.e. a split off from the continuous distribution of eigenvalues,  that activate the grokking transition. The first eigenvalues that become detach from the bulk reach high values (up to 0.5 approximately), but then converge towards a separated group centered at 0.018. In the lower panels we can verify that the spectrum at the end of training sweeps is given by the combination of the two spectra referring to the two labels (with projections $P_0$ and $P_1$), because of the vanishing off-diagonal elements of the reduced density matrix $\varrho^{(\ell)}$ in the label space. This statement holds true at the transition sweep. Instead the initial random state is endowed with coherence between label subspaces, thus it is characterized by a more complex behavior. 

\begin{figure}
    \centering
    \includegraphics[width=\linewidth]{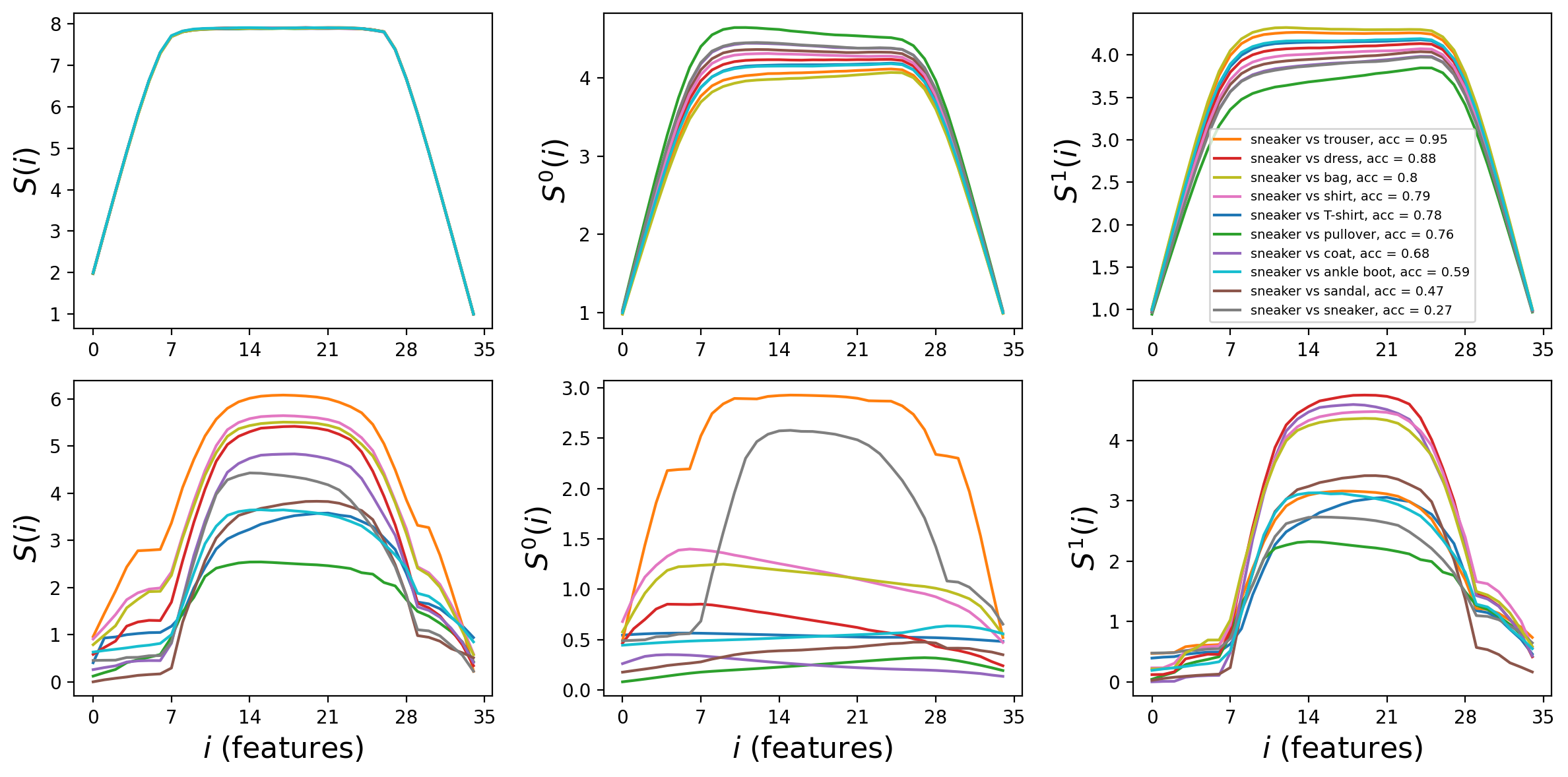}
    \caption{Entanglement entropy along the lattice evaluated corresponding each bipartition defined with respect to the link joining the features pair $(i, i+1)$. The top panels are referred to the initial random tensor $W$, yielding a volume law, while the bottom panels describe the grokking condition, highlighting a sharp entropy reduction, typifying a subvolume law entanglement transition. The legend reports the ranking of each binary classification problem for the fashion MNIST dataset with respect to accuracy.}
    \label{fig:volumearea}
\end{figure}

The characterization of sneaker images is investigated in the remaining fashion MNIST binary classifications, as represented in Fig. \ref{fig:volumearea}. In the top panels the initial randomized condition is depicted, with a volume law consisting in a linearly increasing entropy from both boundaries up to an upper bound imposed by the bond dimension, as previously mentioned in Fig. \ref{fig:entropy}. In the bottom panels the entropy profiles at the grokking condition underline the entanglement transition, with a sharp entropy reduction, typifying a subvolume law entanglement behavior. 


\section{Correlations reveal enriched gene sub-communities}\label{sec4}

The emerging grokking transition is further investigated in gene expression datasets about HCC presented in Section \ref{subsec2b}. The feature extraction involved at this training stage promotes the identification of gene sub-communities encapsulating the information required for generalization. The flowchart scheme in Fig. \ref{fig:flowchart}(b) exemplifies the workflow adopted in order to achieve this goal. As presented in Section \ref{subsec2b}, we consider the best performing communities determined during training in \cite{genescomm}. The generalization test requires the reduction of the exploited gene expression in GSE102079 to the intersection with common ones in GSE54236 independent sets. During training dynamics we monitor the emergence of an entanglement transition, such that we can take into account correlations, introduced in Eq. \eqref{eq::corr}, corresponding to a critical grokking condition. For increasing thresholds $t \in [0,1]$ with step size equal to 0.1, we consider absolute values of correlations $|C^k_{i,j}| > t$ in order to detect both correlated and anti-correlated feature pairs. The number of these pairs is maximal corresponding to the grokking transition, even if for each label $k=0,1$ the sweep may vary. The sub-community is identified by the intersection of these two maximal list of features pairs. We evaluate the effective information content of sub-communities obtained for each threshold through the GSEA gene enrichment procedure (\url{https://www.gsea-msigdb.org/gsea/msigdb/human/annotate.jsp}).

\subsection{Entanglement transition for meaningful gene communities}\label{subsec4a}

\begin{figure}
    \centering
    \subfigure[]{\includegraphics[width=\linewidth]{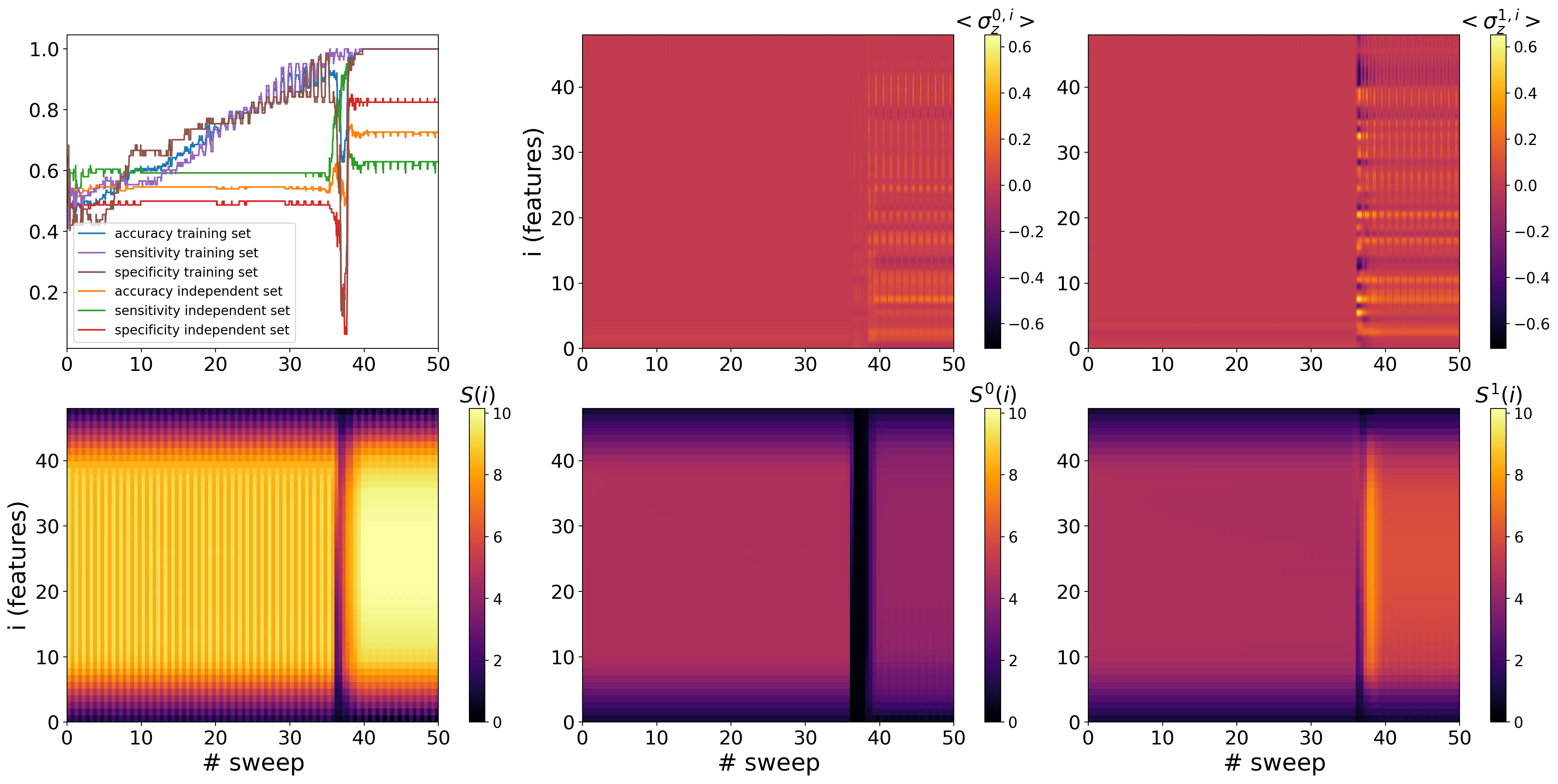}}
    \subfigure[]{\includegraphics[width=\linewidth]{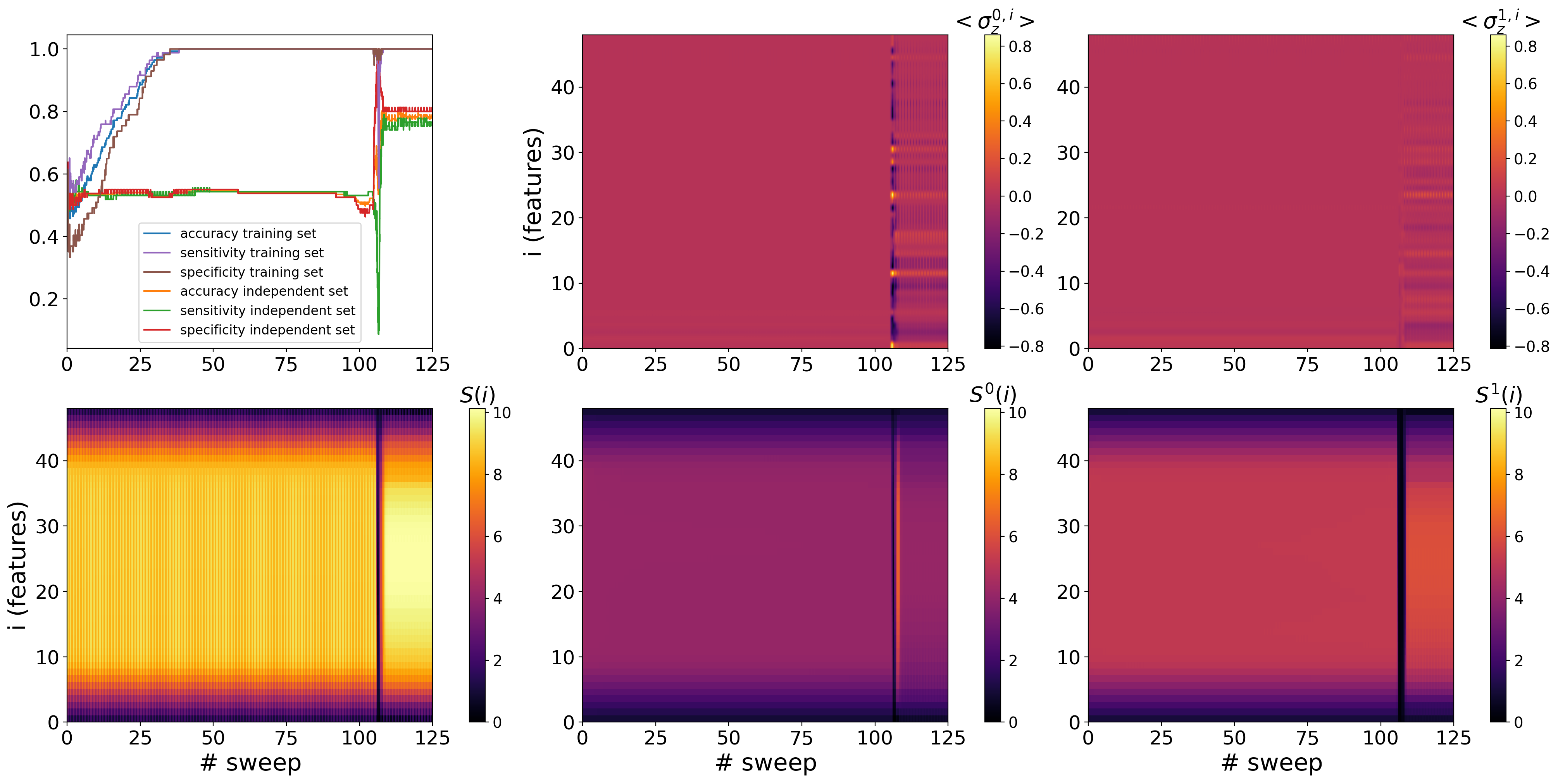}}
    \caption{Grokking in the HCC classification for the two biologically meaningful communities C29 and C41 determined in \cite{genescomm}, referred to panel (a) and (b), respectively. In both cases the top left plot describes evaluation metrics for the training and independent set. The two density plots at the top show the local magnetization dynamics projected in the two label subspaces. At the bottom of each panel entanglement entropies $S(i), S^0(i), S^1(i)$ are shown.}
    \label{fig:C29C41}
\end{figure}

\begin{figure}
    \centering
    \subfigure[]{\includegraphics[width=0.8\linewidth]{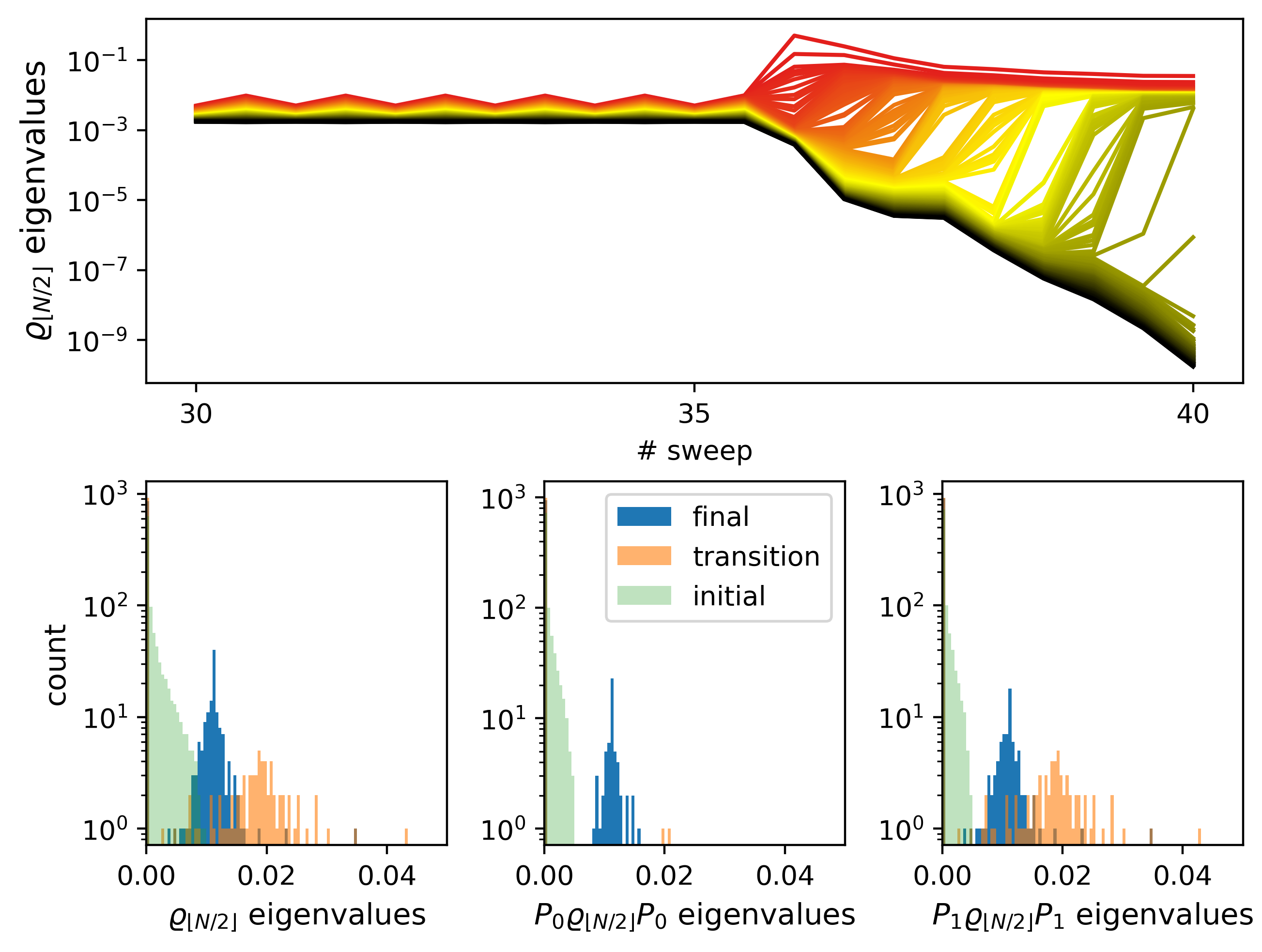}}
    \subfigure[]{\includegraphics[width=0.8\linewidth]{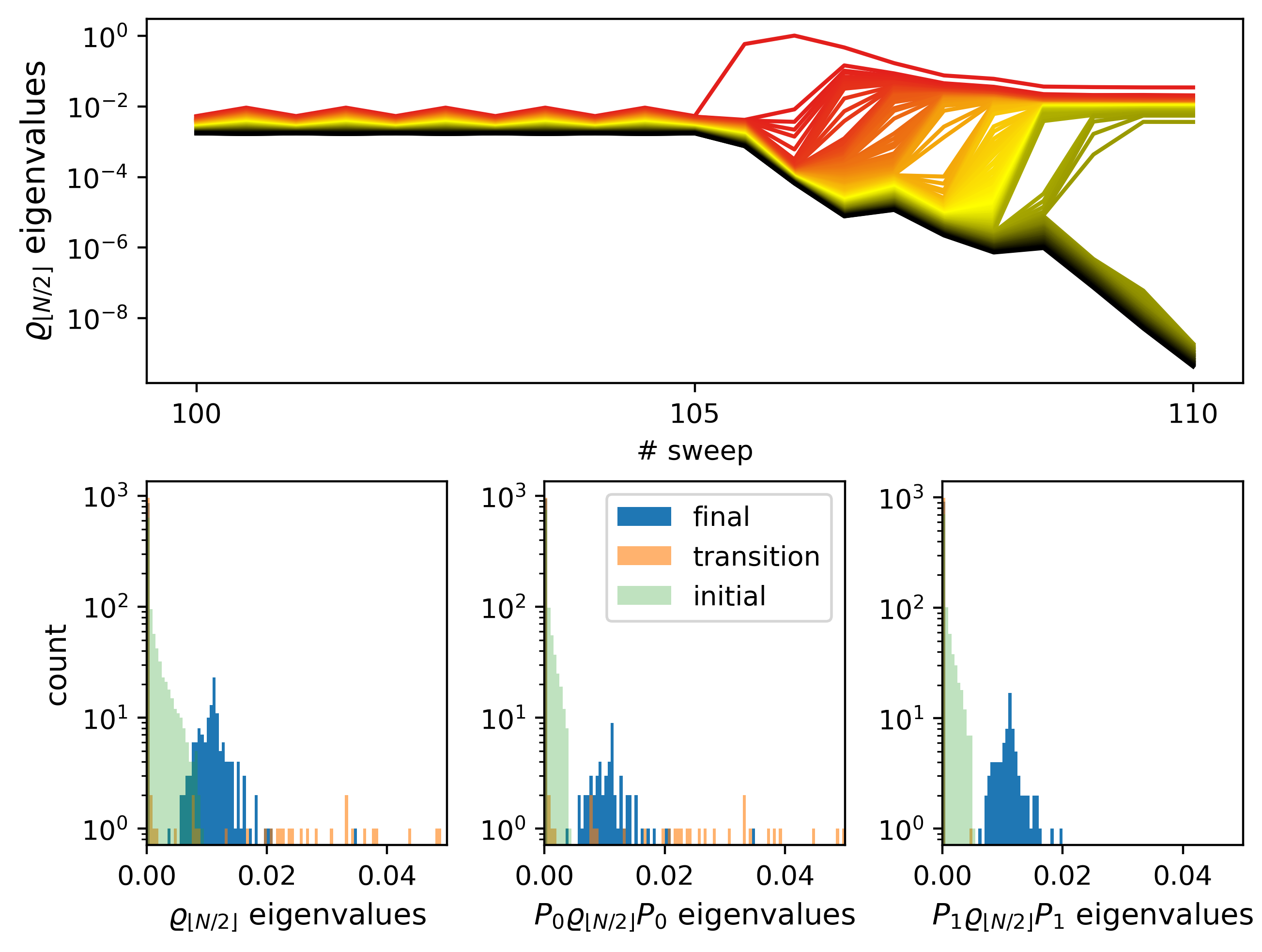}}
    \caption{Spectrum dynamics of the density matrix $\varrho_{\lfloor N/2 \rfloor}$ for the half lattice bipartition of communities C29 and C41 in panel (a) and (b), respectively. The highest 200 eigenvalues with respect to the sweep number are denoted with graded colors from red to black in the top panel. Histogram of eigenvalues distribution in the bottom left panel, while the one of $\varrho_{\lfloor N/2 \rfloor}$ projected in each label subspace at the center and right panels.}
    \label{fig:C29C41_evaporation}
\end{figure}

The training dynamics of each gene community in Table \ref{tab1} undergoes an entanglement transition, with an effective grokking in independent sets classification observed in almost $60\%$ of cases in terms of a sufficient gain in performances with respect to a random classifier. This means that grokking implies an entanglement transition, but the viceversa does not hold true.

We firstly focus on the two most meaningful community according to a biological perspective, namely C29 and C41 as determined in \cite{genescomm}. In both cases the lattice is composed of $N=48$ qubits, a size requiring an increased bond dimension $\chi=1000$, since there is no more notion of proximity, like for pixels in Section \ref{sec3}. In Fig. \ref{fig:C29C41} we resume the training dynamics observed for both communities, with a grokking transition after a different number of sweeps. In panel (a) the training dynamics for C29 shows the gain in generalization performance approximately corresponding to the same sweep of the maximal performance in accuracy for the training set. The sensitivity of independent set classification experiences a sudden growth because the magnetization patterns firstly emerge for the MPS mask $W^1$. Immediately after the competing specificity rises because of the switched feature extraction in $W^0$, thus leading to the establishment of a steady classification performance. A reversed behavior is observed in panel (b) for C41, with magnetization patterns formation affecting the control mask before the HCC one, so involving a specificity growth followed by an increased sensitivity. In this case their stable value after grokking is almost equal, while in the previous community C29 their difference is much more pronounced. 

Entanglement entropy for these communities are characterized by a different dynamics with respect to the remaining communities, since the maximum value at half lattice overcomes the saturation value of the initial random MPS. In both case this behavior is driven by the entropy $S^1(i)$ after grokking, while the entropy $S^0(i)$ of the control mask keeps the reduction even after the peak for C41.

\begin{figure}
    \centering
    \subfigure[]{\includegraphics[width=0.8\linewidth]{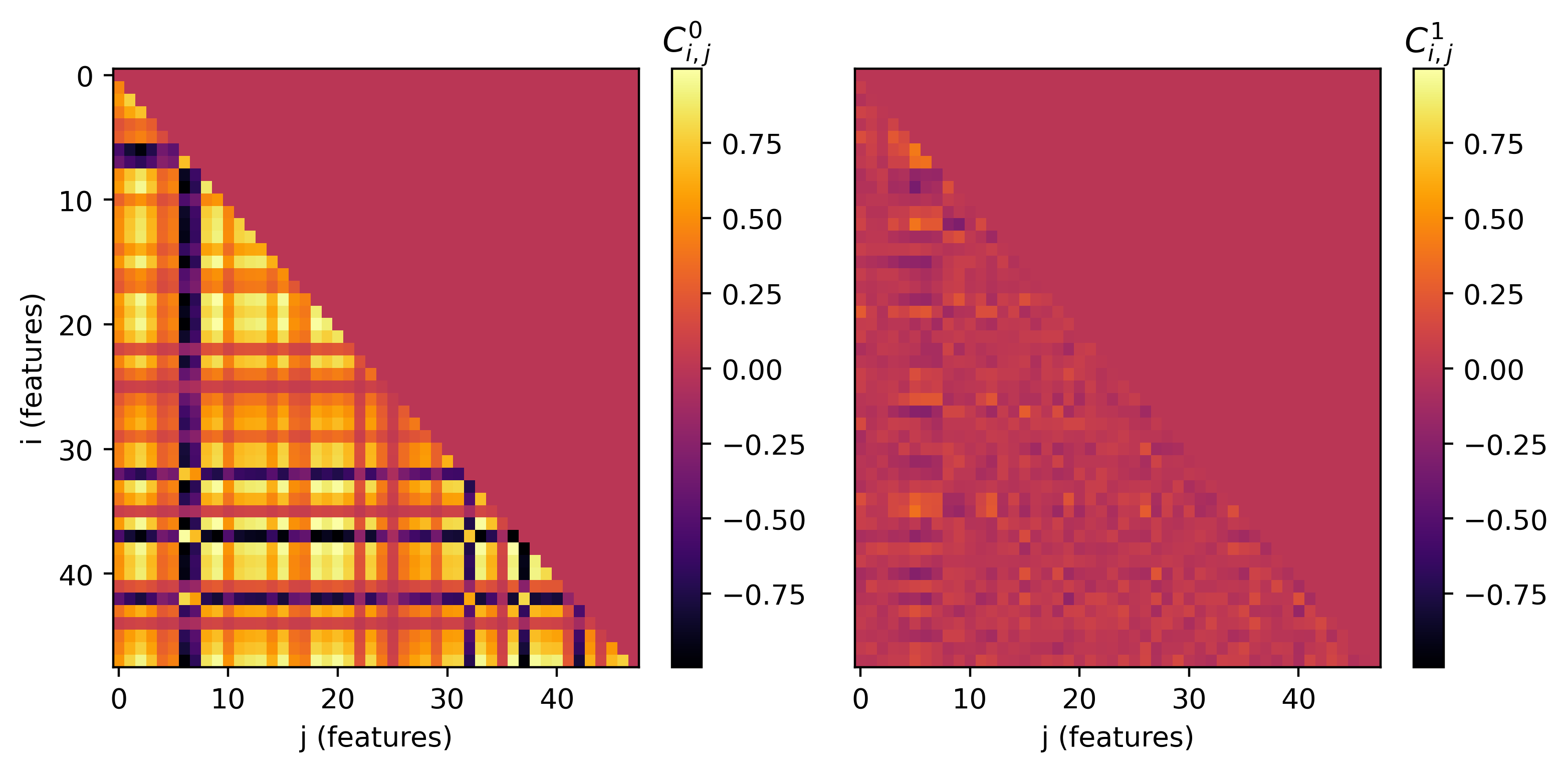}}
    \subfigure[]{\includegraphics[width=0.8\linewidth]{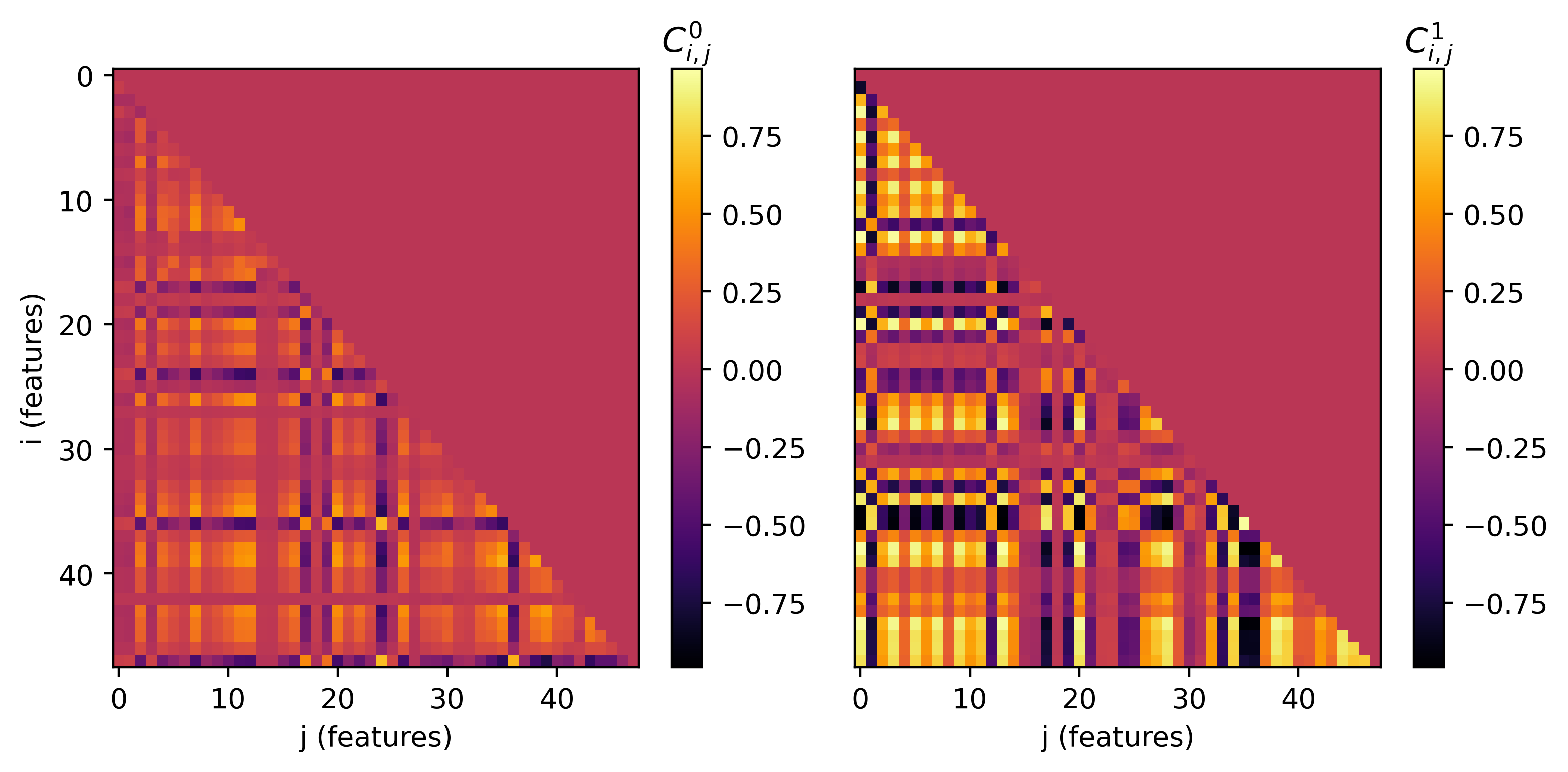}}
    \caption{Correlations pattern at grokking for C29 and C41, in panel (a) and (b), respectively. Left density plots are referred to the projection in the control label subspace $C^0_{i,j}$, while right ones to the HCC subspace $C^1_{i,j}$.}
    \label{fig:C29C41_correlations}
\end{figure}

The increased entanglement entropy at half lattice after the grokking transition is further investigated by considering the spectrum dynamics of the density matrix $\varrho_{\lfloor N/2 \rfloor}$, as represented in Fig. \ref{fig:C29C41_evaporation}. A similar behavior with respect to the one presented in Section \ref{sec3} is observed, even if C29 shows in panel (a) a quite different dynamics, since there is directly the evaporation of a group of eigenvalues, while in panel (b) there is firstly a single eigenvalue getting detached from the bulk for C41, more similarly to what represented in Fig. \ref{fig:evl}. In both cases the number of sweeps required to complete the transition is reduced to 5, namely one half of those required for the cases analyzed in Section \ref{sec3}. 

The lower histograms in both panels depict initial, transition and final eigenvalues distributions, by setting the transition sweep to 37 for C29 and to 107 for C41. We choose the transition sweep by selecting the minimum entropy along the lattice, namely when the HCC mask for C29 and the control one for C41 drain the dynamics.


In the histograms of Fig.  \ref{fig:C29C41_evaporation}, we observe that, at the transition, the eigenvalue distribution in the complementary subspace becomes entirely compressed towards near-zero values. The stable final configuration corresponds to a cluster of eigenvalues confined within a compact support, separated from the bulk, as previously noted in Fig.  \ref{fig:evl}.

An important characterization of the grokking transition for our purposes relies on correlations, evaluated according to prescriptions presented in Appendix \ref{secA1}. Absolute values of correlations reach the associated maxima corresponding to the grokking features extraction, depicted for C29 and C41 in Fig. \ref{fig:C29C41_correlations}. We properly select the sweep for each mask to capture maximum absolute values of correlations and we can verify that the dynamics draining at the transition influences correlations because of $\varrho_{kk}^{(\ell)}$, namely the norm of the reduced state projected in the $k$-th label subspace. Thus the mask firstly experiencing the magnetization patterns formation shows less pronounced correlations peaks. Nevertheless we are able to recognize some features pairs that remain almost uncorrelated, since the proposed flowchart in Fig. \ref{fig:flowchart}(b) will be able to detect sub-communities by varying the threshold for absolute values of correlations, thus highlighting the persistency of pairs in an independent or correlated behavior.


\subsection{General behavior of entropies and gene enrichments}\label{subsec4b}

\begin{figure}
\centering 
\captionof{table}[foo]{Linear fit for each community of entanglement entropy dependence on the distance from the left boundary of the one-dimensional lattice, $\log S=q + c \log i$.\label{tab:left}}
\includegraphics[width=\linewidth]{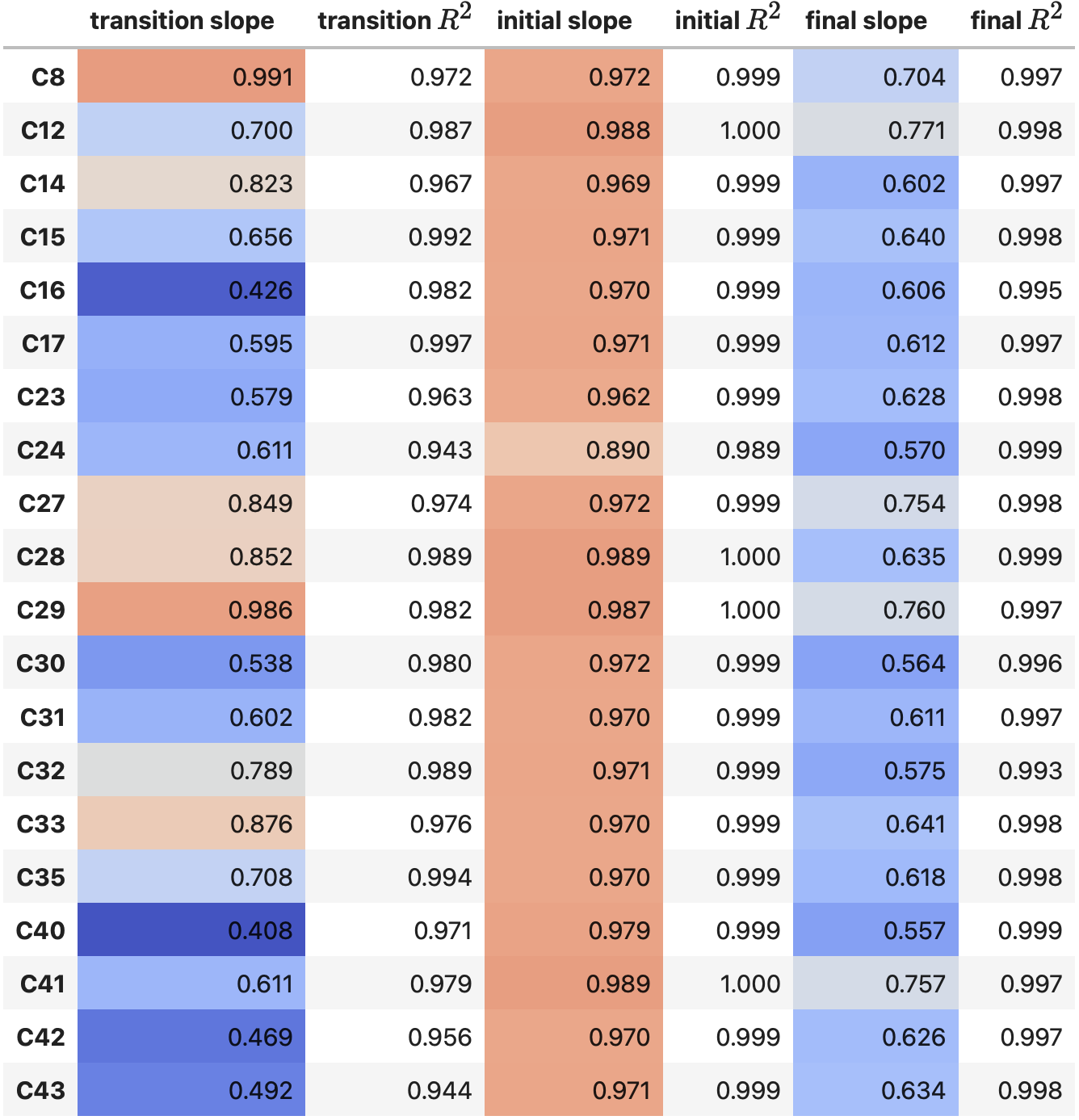}
\end{figure}

\begin{figure}
\centering
\captionof{table}[foo]{Linear fit for each community of entanglement entropy dependence on the distance from the right boundary of the one-dimensional lattice, $\log S=q + c \log(N-i)$. \label{tab:right}}
\includegraphics[width=\linewidth]{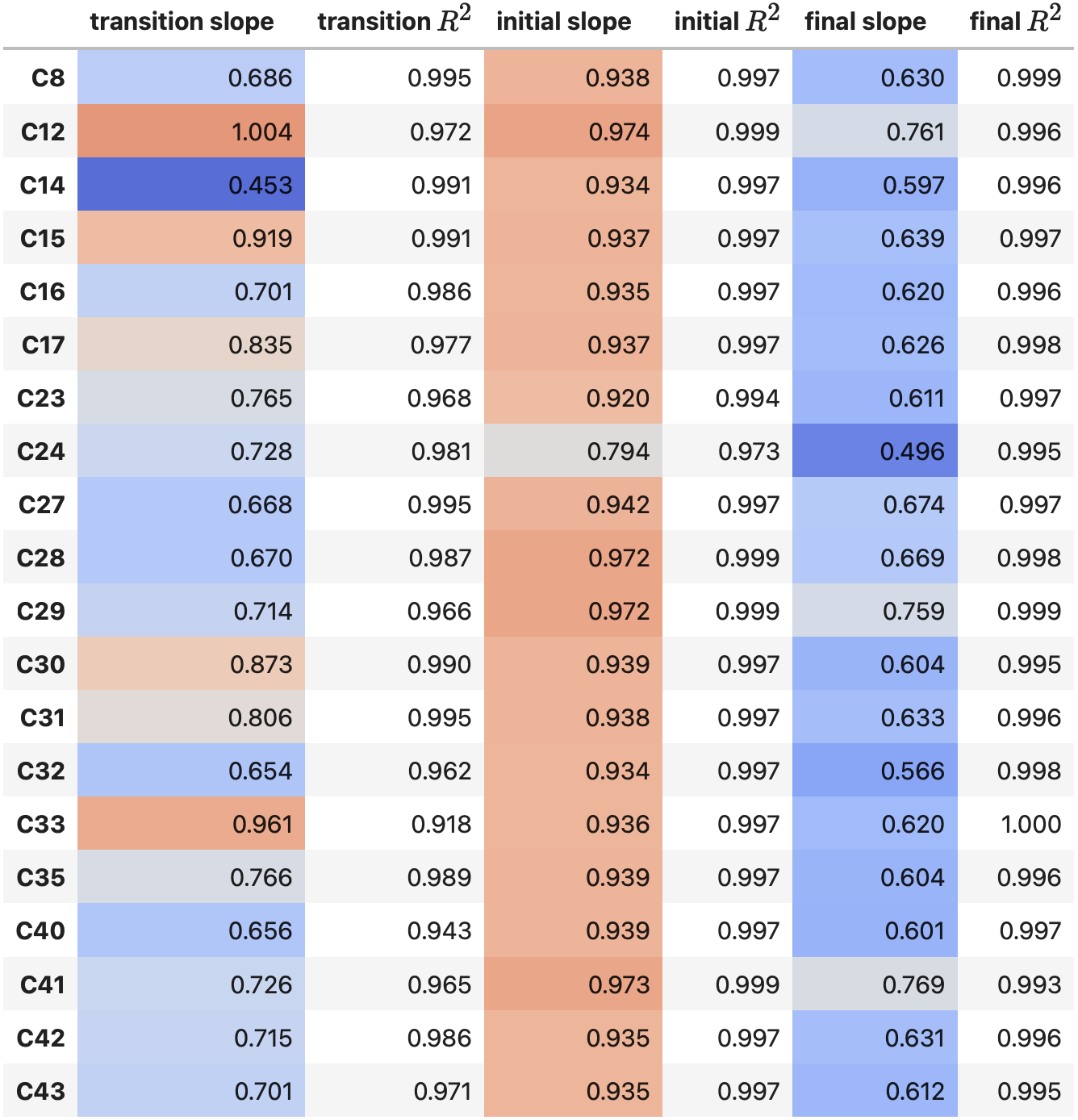}
\end{figure}

Entanglement transitions are observed in every gene community, even if an effective grokking is not ensured. In order to estimate the dependence on the distance from each boundary of the one-dimensional lattice and to determine a transition from a volume to a sub-volume law of entanglement entropy \cite{Sierant2022dissipativefloquet}, we apply a linear fit of the corresponding logarithms, such that the slope is able to quantify this information. 

We collect fitting results in Tables \ref{tab:left} and \ref{tab:right}, referring to Appendix \ref{secA3} for more details. The volume law of the initial random MPS is easily recognized, since a slope close to unity is verified in all cases, with a unique exception for C24 caused by the bond dimension $\chi=100$, sufficient for the smallest community (see Table \ref{tab1}). The comparison of such linear fits considers the same number of lattice sites from both boundary for every gene community and the bond dimension imposes an entropy saturation value. For this reason the entropy linear growth for C24 experiences a curvature before remaining cases. Moreover reported coefficients of determination establish the quality of this linear dependence for initial random MPS. This means that entanglement spreads along the lattice beyond the spatial proximity of the qubits.

At the grokking transition a more heterogenous condition emerges, because the feature extraction is highly dependent on the particular ordering of gene expressions along the lattice: e.g. a group of independent features, endowed with correlations among themselves, can cause a not concave behavior of entanglement entropy. Slope values collected in this particular condition vary in a considerable way, but in general at least one boundary shows a significant reduction with respect to linear dependence characterizing a volume law. This less regular trend is captured by coefficients of determination as well, since they quantify in general a slight reduction with respect to the perfect fitting of the aforementioned volume law.

At the end of the considered training dynamics final states entanglement entropies generally show a much more regular behavior. Sloped on both sides are more close with respect to their values for each community, up to slight deviations of C12, C29 and C41 driven by the lattice size, as well as in general a sub-volume law emerges with a renewed high quality for each fit testified by coefficients of determination close to unity. We define such behaviors as sub-volume laws because an area law requires a logarithmic dependence, thus we cannot establish a local interaction regime after grokking, which could be unreasonable given the absence of any proximity notion among gene expressions. We partially unravel through features extraction the intricate entanglement typifying the initial random MPS, actually enhancing our gene sub-community detection purposes.


\begin{figure}
\centering
\captionof{table}[foo]{Evaluation metrics after magnetization transition in independent sets. \label{tab:metrics}}
\includegraphics[width=\linewidth]{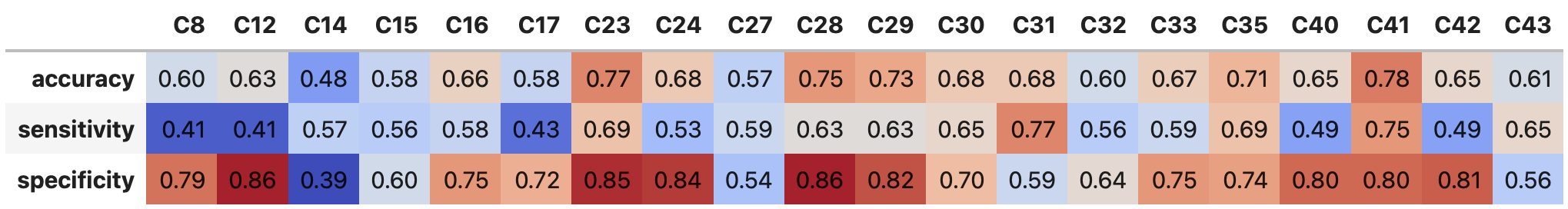}
\end{figure}

\begin{figure}[b!]
\centering
\captionof{table}[foo]{New enriched gene subsets within sub-communities detected through the sequence of correlation thresholds. \label{tab:enrich}}
\includegraphics[width=\linewidth]{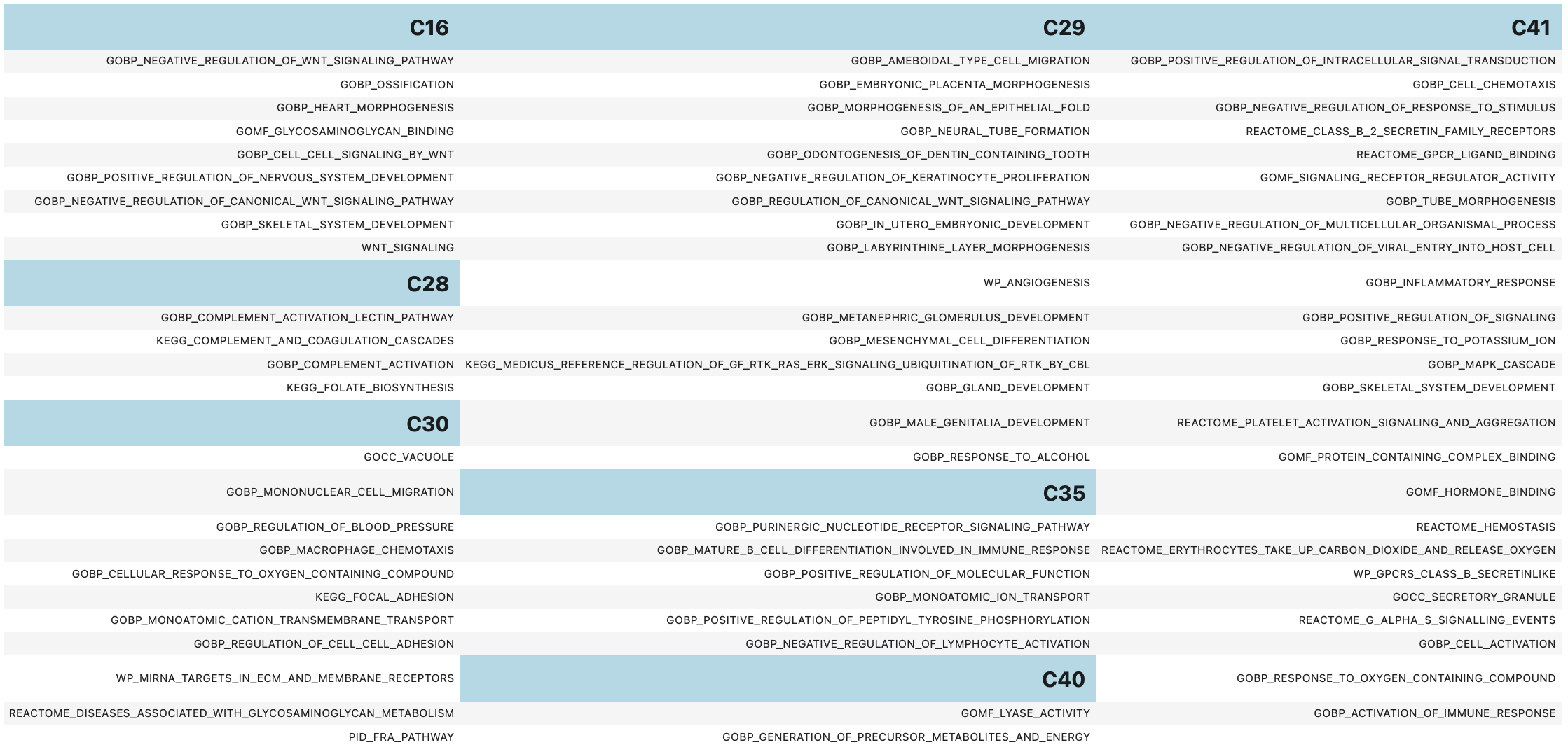}
\end{figure}

A check for the identified sub-communities is based on GSEA. We apply such tests to communities characterized by a proper generalizability, quantified as an accuracy higher than 0.65 in independent sets classifications listed in Table \ref{tab:metrics}, namely C16, C23, C24, C28, C29, C30, C31, C33, C35, C40, C41 and C42.

The variation of the correlations thresholds $t$ from 0.1 to 1, with step size equal to 0.1, allows us to collect a sequence of sub-communities for each case. Following the definition of these sequences of gene expressions, we input them for each fixed threshold one by one into the GSEA database by setting a false discovery rate equal to $5\times 10^{-2}$ for the computation of overlaps with hallmark gene sets of the molecular signature database (MSigDB), canonical pathways and gene ontology \cite{genescomm}. The preset maximal number of significant overlaps is equal to 100, so the definition for correlation thresholded sub-communities of new enriched gene subsets is referred to this upper bound, reached by C29, C35 and C41.

We resume the overall collection of aforementioned new enriched gene subsets in Table \ref{tab:enrich}, while the communities baseline for enrichments without thresholds is discussed in Appendix \ref{secA4}. The community C41 shows the highest number of new enrichments and it is endowed with the best accuracy and more balanced trade-off between sensitivity and specificity in Table \ref{tab:metrics}.

The main focus of our analysis resides in the discovery of new candidate biomarkers for diagnosis and prognosis \cite{hepatology, genescomm}. The immune microenvironment \cite{immune_cancer, immune1, immune2, immune3} plays a significant role also at the level of sub-communities listed in Table \ref{tab:enrich}. Metabolic abnormalities (obesity or diabetes) and alcohol consumption \cite{SHIN2023152134} represent a resilient characterization with respect to the baseline, e.g. in C29. Often this condition intertwines with the possible use of serum biomarkers to diagnose the presence of cancer \cite{YI201813}. An enforcement of enriched Wnt pathways \cite{wnt} is observed in the selected sub-communities, which is confirmed in both C16 and C29.

\section{Conclusion}\label{sec5}

By exploiting the quantum-inspired machine learning framework based on tensor networks, we numerically prove that grokking phenomenon, with its emerging generalization gain in classification performances, can be related to an entanglement dynamical transition in the underlying quantum many-body systems.

We consider two datasets as case studies: Fashion MNIST and gene expression data related to hepatocellular carcinoma communities \cite{genescomm}. In both cases, we employ Matrix Product States (MPS) as tensor networks to perform binary classification tasks and analyze the learning dynamics. For the second dataset, we leverage measurements of qubit magnetization and correlation functions within the MPS network to identify significant and relevant gene subcommunities, which are further validated through enrichment analysis.

\backmatter



\bmhead{Acknowledgements} We are grateful to Cosmo Lupo and Andrea De Girolamo for useful discussions and comments. Computational resources were provided by ReCaS Bari \cite{recas}. Some icons in Fig. \ref{fig:flowchart} by \url{https://freeicons.io/profile/433683}.

\section*{Declarations}

\bmhead{Funding} D.P. acknowledges the support by PNRR MUR project CN00000041-“National Center for Gene Therapy and Drugs based on RNA Technology”. S.S. was supported by the project “Higher-order complex systems modeling for personalized medicine,” Italian Ministry of University and Research (funded by MUR, PRIN 2022-PNRR, code P2022JAYMH, CUP: H53D23009130001). Authors were supported by the Italian funding within the “Budget MIUR - Dipartimenti di Eccellenza 2023 - 2027” (Law 232, 11 December 2016) - Quantum Sensing and Modelling for One-Health (QuaSiModO), CUP:H97G23000100001. Authors want to thank the project “Genoma mEdiciNa pERsonalizzatA –GENERA”, local project code T3-AN-04 – CUP H93C22000500001, financed under the Health Development and Cohesion Plan 2014-2020, Trajectory 3 “Regenerative, predictive and personalized medicine” - Action line 3.1 “Creation of a precision medicine program for the mapping of the human genome on a national scale”, referred to in the Notice of the Ministry of Health published in the Official Journal no. 46 of 24 February 2021. Authors want to thank the Funder: Project funded under the National Recovery and Resilience Pan (NRRP), Mission 4 Component 2 Investment 1.4 - Call for tender No. 3138 of 16 December 2021 of Italian Ministry of University and Research funded by the European Union - NextGenerationEU (award number/project code: CN00000013), and Concession Decree No. 1031 of 17 February 2022 adopted by the Italian Ministry of University and Research (CUP: D93C22000430001), Project title: “National Centre for HPC, Big Data and Quantum Computing”.

\bmhead{Conflict of interest/Competing interests} The authors have no competing interests to declare that are relevant to the content of this article.

\bmhead{Data availability} The data that support the findings of this study are openly available in Kaggle at \url{https://www.kaggle.com/datasets/zalando-research/fashionmnist} and Gene Expression Omnibus (GEO) at \url{https://www.ncbi.nlm.nih.gov/geo/}.

\bmhead{Code availability} The code that supports the findings of this study is available from the authors, upon reasonable request.

\bmhead{Author contribution} Conceptualization: DP, AM, GM; Software: DP; Data Curation: DP, AL; Methodology: DP, AM, GM, NA; Formal analysis and investigation: DP; Visualization: DP; Funding acquisition: RB, SS, GP, EPi; Resources: DP, AM, GM, AL; Supervision: DP, AM, GM; Writing - original draft preparation: DP, AM, GM; Writing - review and editing: DP, AM, GM, AL, EPa, LB, ST, TM, MLR, EPi, NA, GP, SS, RB.


\begin{appendices}

\section{Matrix product states}\label{secA1}

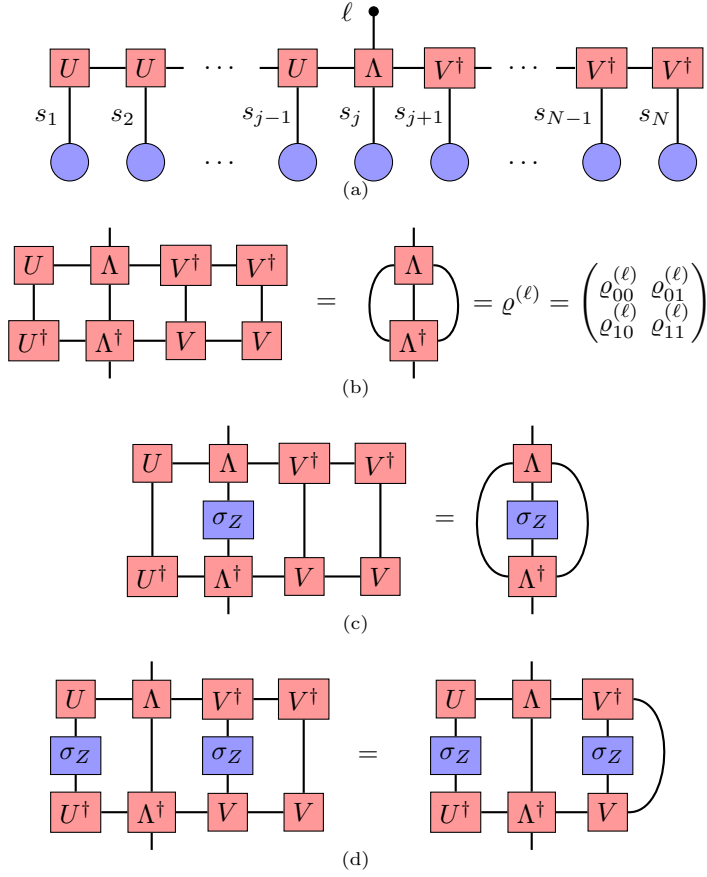
\begin{figure}[t!]
\begin{center}
\begin{tabular}{c}
    \subfigure[]{
    \begin{tikzpicture}
    \node[draw, fill=red!40, shape=rectangle, minimum width=0.5cm, minimum height = 0.5cm] (v0z) at (-1,0.5) {$U$};
    \node[draw, fill=red!40, shape=rectangle, minimum width=0.5cm, minimum height = 0.5cm] (v0b) at (0,0.5) {$U$};
    \node[draw, fill=red!40, shape=rectangle, minimum width=0.5cm, minimum height = 0.5cm] (v0a) at (2,0.5) {$U$};
    \node[draw, fill=red!40, shape=rectangle, minimum width=0.5cm, minimum height = 0.5cm] (v0c) at (3,0.5) {$\Lambda$};
    \node[draw, fill=red!40, shape=rectangle, minimum width=0.5cm, minimum height = 0.5cm] (v2a) at (4,0.5) {$V^\dagger$};
    \node[draw, fill=red!40, shape=rectangle, minimum width=0.5cm, minimum height = 0.5cm] (v2b) at (6,0.5) {$V^\dagger$};
    \node[draw, fill=red!40, shape=rectangle, minimum width=0.5cm, minimum height = 0.5cm] (v2z) at (7,0.5) {$V^\dagger$};
    \node[draw, fill=blue!40, shape=circle, inner sep=5 pt] (v00z) at (-1,-0.75) {};
    \node[draw, fill=blue!40, shape=circle, inner sep=5 pt] (v00) at (0,-0.75) {};
    \node[draw, fill=blue!40, shape=circle, inner sep=5 pt] (v00a) at (2,-0.75) {};
    \node[draw, fill=blue!40, shape=circle, inner sep=5 pt] (v00b) at (3,-0.75) {};
    \node[draw, fill=blue!40, shape=circle, inner sep=5 pt] (v22a) at (4,-0.75) {};
    \node[draw, fill=blue!40, shape=circle, inner sep=5 pt] (v22c) at (6,-0.75) {};
    \node[draw, fill=blue!40, shape=circle, inner sep=5 pt] (v22z) at (7,-0.75) {};
    \draw [thick] (v0b) -- (v00)
    (v0c) -- (v00b)
    (v0c) -- (v2a)
    (v0c) -- (v0a)
    (v2a) -- (v22a)
    (v0a) -- (v00a)
    (v0z) -- (v00z)
    (v0b) -- (v0z)
    (v2b) -- (v2z)
    (v2z) -- (v22z)
    (v2b) -- (v22c);
    \draw [thick] (3,0.75) -- (3,1.25);
    \draw [thick] (0.27,0.5) -- (0.5,0.5);
    \draw [thick] (1.5,0.5) -- (1.75,0.5);
    \draw [thick] (4.32,0.5) -- (4.55,0.5);
    \draw [thick] (5.4,0.5) -- (5.65,0.5);
    \node [] (t) at (5,0.5) {$\dots$};
    \node [] (td) at (5,-0.75) {$\dots$};
    \node [] (tb) at (1,0.5) {$\dots$};
    \node [] (tbd) at (1,-0.75) {$\dots$};
    \draw[black,fill=black] (3,1.25) circle (.4ex);
    \node [] (t3) at (2.65,1.25) {$\ell$};
    \node [] (t2) at (-1.3,-0.15) {$s_1$};
    \node [] (t2) at (-0.3,-0.15) {$s_2$};
    \node [] (t2) at (1.6,-0.15) {$s_{j-1}$};
    \node [] (t2) at (2.7,-0.15) {$s_j$};
    \node [] (t2) at (3.6,-0.15) {$s_{j+1}$};
    \node [] (t6) at (5.5,-0.15) {$s_{N-1}$};
    \node [] (t6) at (6.7,-0.15) {$s_N$};
    \end{tikzpicture}} \\
    \subfigure[]{
    \begin{tikzpicture}
    \node[draw, fill=red!40, shape=rectangle, minimum width=0.5cm, minimum height = 0.5cm] (v0b) at (2,0.5) {$U$};
    \node[draw, fill=red!40, shape=rectangle, minimum width=0.5cm, minimum height = 0.5cm] (v0c) at (3,0.5) {$\Lambda$};
    \node[draw, fill=red!40, shape=rectangle, minimum width=0.5cm, minimum height = 0.5cm] (v1) at (4,0.5) {$V^\dagger$};
     \node[draw, fill=red!40, shape=rectangle, minimum width=0.5cm, minimum height = 0.5cm] (v2b) at (5,0.5) {$V^\dagger$};
    \node[draw, fill=red!40, shape=rectangle, minimum width=0.5cm, minimum height = 0.5cm] (v00b) at (2,-0.5) {$U^\dagger$};
    \node[draw, fill=red!40, shape=rectangle, minimum width=0.5cm, minimum height = 0.5cm] (v00c) at (3,-0.5) {$\Lambda^\dagger$};
    \node[draw, fill=red!40, shape=rectangle, minimum width=0.5cm, minimum height = 0.5cm] (v11) at (4,-0.5) {$V$};
     \node[draw, fill=red!40, shape=rectangle, minimum width=0.5cm, minimum height = 0.5cm] (v22b) at (5,-0.5) {$V$};
    \draw [thick] (v0b) -- (v0c)
    (v0c) -- (v1)
    (v1) -- (v2b)
    (v0b) -- (v00b)
    (v0c) -- (v00c)
    (v1) -- (v11)
    (v2b) -- (v22b)
    (v00b) -- (v00c)
    (v00c) -- (v11)
    (v11) -- (v22b);
    \draw [thick] (3,0.75) -- (3,1);
    \draw [thick] (3,-0.775) -- (3,-1);
    \node [] (t2) at (5.85,0) {$=$};
     \node[draw, fill=red!40, shape=rectangle, minimum width=0.5cm, minimum height = 0.5cm] (v0d) at (7,0.5) {$\Lambda$};
    \node[draw, fill=red!40, shape=rectangle, minimum width=0.5cm, minimum height = 0.5cm] (v00d) at (7,-0.5) {$\Lambda^\dagger$};
    \draw [thick] (v0d) -- (v00d);
    \draw [thick] (v0d) to [bend left=90] (v00d);
    \draw [thick] (v0d) to [bend right=90] (v00d);
    \draw [thick] (7,0.75) -- (7,1);
    \draw [thick] (7,-0.775) -- (7,-1);
    \node [] (t3) at (9.35,0) {\( \displaystyle = \varrho^{(\ell)} = \begin{pmatrix}
        \varrho^{(\ell)}_{00} & \varrho^{(\ell)}_{01} \\
        \varrho^{(\ell)}_{10} & \varrho^{(\ell)}_{11}
    \end{pmatrix} \)};
    \end{tikzpicture}} \\
    \subfigure[]{
    \begin{tikzpicture}
    \node[draw, fill=red!40, shape=rectangle, minimum width=0.5cm, minimum height = 0.5cm] (v0b) at (2,0.5) {$U$};
    \node[draw, fill=red!40, shape=rectangle, minimum width=0.5cm, minimum height = 0.5cm] (v0c) at (3,0.5) {$\Lambda$};
    \node[draw, fill=red!40, shape=rectangle, minimum width=0.5cm, minimum height = 0.5cm] (v1) at (4,0.5) {$V^\dagger$};
     \node[draw, fill=red!40, shape=rectangle, minimum width=0.5cm, minimum height = 0.5cm] (v2b) at (5,0.5) {$V^\dagger$};
    \node[draw, fill=red!40, shape=rectangle, minimum width=0.5cm, minimum height = 0.5cm] (v00b) at (2,-1) {$U^\dagger$};
    \node[draw, fill=red!40, shape=rectangle, minimum width=0.5cm, minimum height = 0.5cm] (v00c) at (3,-1) {$\Lambda^\dagger$};
    \node[draw, fill=red!40, shape=rectangle, minimum width=0.5cm, minimum height = 0.5cm] (v11) at (4,-1) {$V$};
    \node[draw, fill=red!40, shape=rectangle, minimum width=0.5cm, minimum height = 0.5cm] (v22b) at (5,-1) {$V$};
    \node[draw, fill=blue!40, shape=rectangle, minimum width=0.5cm, minimum height = 0.5cm] (vint) at (3,-0.25) {$\sigma_Z$};
    \draw [thick] (v0b) -- (v0c)
    (v0c) -- (v1)
    (v1) -- (v2b)
    (v0b) -- (v00b)
    (v1) -- (v11)
    (v0c) -- (vint)
    (vint) -- (v00c)
    (v2b) -- (v22b)
    (v00b) -- (v00c)
    (v00c) -- (v11)
    (v11) -- (v22b);
    \draw [thick] (3,0.75) -- (3,1);
    \draw [thick] (3,-1.275) -- (3,-1.5);
    \node [] (t2) at (5.85,-0.25) {$=$};
    \node[draw, fill=red!40, shape=rectangle, minimum width=0.5cm, minimum height = 0.5cm] (v0d) at (7,0.5) {$\Lambda$};
    \node[draw, fill=red!40, shape=rectangle, minimum width=0.5cm, minimum height = 0.5cm] (v00d) at (7,-1) {$\Lambda^\dagger$};
    \node[draw, fill=blue!40, shape=rectangle, minimum width=0.5cm, minimum height = 0.5cm] (vintb) at (7,-0.25) {$\sigma_Z$};
    \draw [thick] (v0d) to [bend right=90] (v00d)
    (v0d) -- (vintb)
    (vintb) -- (v00d)
    (v0d) to [bend left=90] (v00d);
    \draw [thick] (7,0.75) -- (7,1);
    \draw [thick] (7,-1.275) -- (7,-1.5);
    \end{tikzpicture}} \\
    \subfigure[]{
    \begin{tikzpicture}
    \node[draw, fill=red!40, shape=rectangle, minimum width=0.5cm, minimum height = 0.5cm] (v0b) at (2,0.5) {$U$};
    \node[draw, fill=red!40, shape=rectangle, minimum width=0.5cm, minimum height = 0.5cm] (v0c) at (3,0.5) {$\Lambda$};
    \node[draw, fill=red!40, shape=rectangle, minimum width=0.5cm, minimum height = 0.5cm] (v1) at (4,0.5) {$V^\dagger$};
     \node[draw, fill=red!40, shape=rectangle, minimum width=0.5cm, minimum height = 0.5cm] (v2b) at (5,0.5) {$V^\dagger$};
    \node[draw, fill=red!40, shape=rectangle, minimum width=0.5cm, minimum height = 0.5cm] (v00b) at (2,-1) {$U^\dagger$};
    \node[draw, fill=red!40, shape=rectangle, minimum width=0.5cm, minimum height = 0.5cm] (v00c) at (3,-1) {$\Lambda^\dagger$};
    \node[draw, fill=red!40, shape=rectangle, minimum width=0.5cm, minimum height = 0.5cm] (v11) at (4,-1) {$V$};
    \node[draw, fill=red!40, shape=rectangle, minimum width=0.5cm, minimum height = 0.5cm] (v22b) at (5,-1) {$V$};
    \node[draw, fill=blue!40, shape=rectangle, minimum width=0.5cm, minimum height = 0.5cm] (vint) at (4,-0.25) {$\sigma_Z$};
    \node[draw, fill=blue!40, shape=rectangle, minimum width=0.5cm, minimum height = 0.5cm] (vint2) at (2,-0.25) {$\sigma_Z$};
    \draw [thick] (v0b) -- (v0c)
    (v0c) -- (v1)
    (v1) -- (v2b)
    (v0b) -- (vint2)
    (vint2) -- (v00b)
    (v0c) -- (v00c)
    (v1) -- (vint)
    (vint) -- (v11)
    (v2b) -- (v22b)
    (v00b) -- (v00c)
    (v00c) -- (v11)
    (v11) -- (v22b);
    \draw [thick] (3,0.75) -- (3,1);
    \draw [thick] (3,-1.275) -- (3,-1.5);
    \node [] (t2) at (5.85,-0.25) {$=$};
    \node[draw, fill=red!40, shape=rectangle, minimum width=0.5cm, minimum height = 0.5cm] (v0e) at (7,0.5) {$U$};
    \node[draw, fill=red!40, shape=rectangle, minimum width=0.5cm, minimum height = 0.5cm] (v0d) at (8,0.5) {$\Lambda$};
    \node[draw, fill=red!40, shape=rectangle, minimum width=0.5cm, minimum height = 0.5cm] (v1b) at (9,0.5) {$V^\dagger$};
     \node[draw, fill=red!40, shape=rectangle, minimum width=0.5cm, minimum height = 0.5cm] (v00e) at (7,-1) {$U^\dagger$};
    \node[draw, fill=red!40, shape=rectangle, minimum width=0.5cm, minimum height = 0.5cm] (v00d) at (8,-1) {$\Lambda^\dagger$};
    \node[draw, fill=red!40, shape=rectangle, minimum width=0.5cm, minimum height = 0.5cm] (v11b) at (9,-1) {$V$};
    \node[draw, fill=blue!40, shape=rectangle, minimum width=0.5cm, minimum height = 0.5cm] (vintb) at (9,-0.25) {$\sigma_Z$};
    \node[draw, fill=blue!40, shape=rectangle, minimum width=0.5cm, minimum height = 0.5cm] (vint2b) at (7,-0.25) {$\sigma_Z$};
    \draw [thick] (v0e) -- (v0d)
    (v0d) -- (v1b)
    (v0e) -- (vint2b)
    (vint2b) -- (v00e)
    (v0d) -- (v00d)
    (v1b) -- (vintb)
    (vintb) -- (v11b)
    (v00e) -- (v00d)
    (v00d) -- (v11b)
    (v1b) to [bend left=90] (v11b);
    \draw [thick] (8,0.75) -- (8,1);
    \draw [thick] (8,-1.275) -- (8,-1.5);
    \end{tikzpicture}}
\end{tabular}
\end{center}
\caption{Diagrammatic representation in panel (a) of the predictor in Eq. \eqref{eq::pred} with the decomposition of the tensor $W$ as the MPS introduced in Eq. \eqref{eq::Wmps} in mixed-canonical form \cite{SCHOLLWOCK201196}. The sum over repeated indices $s_1, s_2, \dots, s_N$ refers to tensors leg contraction. The use of an orthogonality center $\Lambda$ allows us to optimized tensor contraction in the evaluation of physical quantities: reduced density matrix in the label space in panel (b), local magnetization in panel (c) and correlation functions in panel (d). \label{fig::mps}}
\end{figure}

We exploit the training of a quantum mask $W$ able to recognize two categorical classes for each encoded array $\mathbf{x}$ of classical data
\begin{equation}
\mathbf{x} \in \mathbb{R}^N \mapsto \ket{\Phi(\mathbf{x})} = \bigotimes_{j=1}^{N} \ket{\phi(x^{(j)})} \in \mathbb{R}^{2^N},
\end{equation} 
where each qubit state reads $\ket{\phi(x^{(j)})}=\left( \cos(x^{(j)}), \ \sin(x^{(j)}) \right)^\intercal$ \cite{stoudenmire2017supervisedlearningquantuminspiredtensor}. We will implement the recognition of the two categories through the MPS decomposition of the mask $W$.

MPS format is based on truncated singular value decomposition (SVD), $M=U \Sigma V^\dagger$, keeping the $\chi$ (bond dimension) highest singular values and iterating for each link of the one-dimensional lattice \cite{SCHOLLWOCK201196}:
\begin{equation} \label{eq::Wmps}
    W^\ell_{s_1,\dots,s_N} = \sum_{a_1,\dots,a_N} U_{s_1,a_1} U_{s_2,a_2}^{a_1} \dots U_{s_{j-1},a_{j-1}}^{a_{j-2}}  \Lambda_{s_j}^{a_{j-1},\ell,a_j} V_{s_{j+1},a_j}^{\dagger a_{j+1}} \dots V_{s_{N-1},a_{N-2}}^{\dagger a_{N-1}} V_{s_N,a_{N-1}}^{\dagger}
\end{equation}
with position $j$ for the orthogonality center $\Lambda$ \cite{PhysRevLett.124.037201, evenbly2022} and represented in Fig. \ref{fig::mps}(a). The gradient descent of the cost function introduced in Eq. \eqref{eq::cost} is based on a two-sites update
\begin{center}
\begin{tikzpicture}
    \node[draw, fill=red!40, shape=rectangle, minimum width=0.5cm, minimum height = 0.5cm] (v0b) at (2,0.5) {$U$};
    \node[draw, fill=red!40, shape=rectangle, minimum width=0.5cm, minimum height = 0.5cm] (v0c) at (3,0.5) {$\Lambda$};
    \node[draw, fill=red!40, shape=rectangle, minimum width=0.5cm, minimum height = 0.5cm] (v1) at (4,0.5) {$V^\dagger$};
     \node[draw, fill=red!40, shape=rectangle, minimum width=0.5cm, minimum height = 0.5cm] (v2b) at (5,0.5) {$V^\dagger$};
    \draw [thick] (v0b) -- (v0c)
    (v0c) -- (v1)
    (v1) -- (v2b);
    \draw [thick] (3,0.75) -- (3,1.25);
    \draw [thick] (5,0.225) -- (5,-0.25);
    \draw [thick] (2,0.25) -- (2,-0.25);
    \draw [thick] (3,0.25) -- (3,-0.25);
    \draw [thick] (4,0.225) -- (4,-0.25);
    \node [] (t3) at (2.65,1.25) {$\ell$};
    \node [] (t2) at (2.7,-0.15) {$s_j$};
    \node [] (t5) at (4.5,-0.15) {$s_{j+1}$};
    \node [] (t2) at (5.8,0.5) {$=$};
    \node[draw, fill=red!40, shape=rectangle, minimum width=0.5cm, minimum height = 0.5cm] (v0bB) at (6.5,0.5) {$U$};
    \node[draw, fill=red!40, shape=rectangle, minimum width=1.5cm, minimum height = 0.5cm] (v0cB) at (8,0.5) {$B$};
     \node[draw, fill=red!40, shape=rectangle, minimum width=0.5cm, minimum height = 0.5cm] (v2bB) at (9.5,0.5) {$V^\dagger$};
    \draw [thick] (v0bB) -- (v0cB)
    (v0cB) -- (v2bB);
    \draw [thick] (8,0.75) -- (8,1.25);
    \draw [thick] (9.5,0.225) -- (9.5,-0.25);
    \draw [thick] (6.5,0.25) -- (6.5,-0.25);
    \draw [thick] (7.5,0.25) -- (7.5,-0.25);
    \draw [thick] (8.5,0.25) -- (8.5,-0.25);
    \node [] (t3B) at (7.65,1.25) {$\ell$};
    \node [] (t2B) at (7.2,-0.15) {$s_j$};
    \node [] (t5B) at (9,-0.15) {$s_{j+1}$};
\end{tikzpicture}
\end{center}
referring to a single right step, which leads to a sweep when back at the initial sites pair. The update rule requires the introduction of the following quantity \\
\begin{center}
    \begin{tikzpicture}
    \node[draw, fill=red!40, shape=rectangle, minimum width=0.5cm, minimum height = 0.5cm] (v0b) at (1,0.5) {$U$};
    \node[draw, fill=red!40, shape=rectangle, minimum width=0.5cm, minimum height = 0.5cm] (v2b) at (4,0.5) {$V^\dagger$};
    \node[draw, fill=blue!40, shape=circle, inner sep=5 pt] (v00) at (1,-0.75) {};
    \node[draw, fill=blue!40, shape=circle, inner sep=5 pt] (v00b) at (3,-0.75) {};
    \node[draw, fill=blue!40, shape=circle, inner sep=5 pt] (v22b) at (2,-0.75) {};
    \node[draw, fill=blue!40, shape=circle, inner sep=5 pt] (v22c) at (4,-0.75) {};
    \draw [thick] (v0b) -- (v00)
    (v2b) -- (v22c);
    \draw [thick] (2,0) -- (2,-0.5);
    \draw [thick] (3,0) -- (3,-0.5);
    \draw [thick] (1.27,0.5) -- (1.5,0.5);
    \draw [thick] (3.4,0.5) -- (3.65,0.5);
    \node [] (t2) at (1.7,-0.15) {$s_j$};
    \node [] (t5) at (3.5,-0.15) {$s_{j+1}$};
    \node [] (t2) at (4.8,0) {$=$};
    \node[draw, fill=blue!80, shape=circle, inner sep=5 pt] (v00) at (5.5,-0.75) {};
    \node[draw, fill=blue!40, shape=circle, inner sep=5 pt] (v00b) at (7.5,-0.75) {};
    \node[draw, fill=blue!40, shape=circle, inner sep=5 pt] (v22b) at (6.5,-0.75) {};
    \node[draw, fill=blue!80, shape=circle, inner sep=5 pt] (v22c) at (8.5,-0.75) {};
    \draw [thick] (6.5,0) -- (6.5,-0.5);
    \draw [thick] (7.5,0) -- (7.5,-0.5);
    \draw [thick] (5.5,0) -- (5.5,-0.5);
    \draw [thick] (8.5,0) -- (8.5,-0.5);
    \node [] (t2) at (6.2,-0.15) {$s_j$};
    \node [] (t5) at (8,-0.15) {$s_{j+1}$};
    \node [] (t3) at (10,0) {\( \displaystyle = \quad \ket{\widetilde{\Phi}(\bold{x})} \)};
\end{tikzpicture}
\end{center}
equivalently expressed in the predictor
\begin{equation}
    f^\ell_W (\bold{x}) = \sum_{s_j, s_{j+1}} \sum_{a_{j-1},a_{j+1}} B^{a_{j-1},\ell,a_{j+1}}_{s_j,s_{j+1}} \ket{\widetilde{\Phi}(\bold{x})^{s_j, s_{j+1}}_{a_{j-1},a_{j+1}}},
\end{equation}
such that
\begin{equation}
    \Delta B^\ell = -\frac{\partial \mathcal{C}}{\partial B^\ell} = \sum_{\omega = 1}^{N_T} \ket{\widetilde{\Phi}(\bold{x}_{\omega})} \otimes (y^\ell_{\omega} - f^\ell_{W}(\bold{x}_{\omega})).
\end{equation}

The updated tensor is ruled by the learning rate $\alpha$, $B'^\ell=B^\ell + \alpha \Delta B^\ell$, that we compress according to the bond dimension and rescale in order to keep the state $W$ normalized, since the training dynamics is not unitary. The projection in the subspace spanned by principal components followed by a rescaling effectively implements a projective measurement \cite{wiersema2023}.

We measure the following quantities:
\begin{enumerate}[label=(\roman*)]
	\item reduced density matrix $\varrho^{(\ell)}$ in label space, yielded by the contraction scheme depicted in Fig. \ref{fig::mps}(b);
	\item local magnetization $\braket{\sigma_Z^{k,i}}$ for labels $k=0,1$ and features $i = 1,\dots,N$, whose contraction is shown in Fig. \ref{fig::mps}(c);
	\item correlation function $\braket{\sigma_Z^{k,i} \sigma_Z^{k,j}}$ for labels $k=0,1$ and features pairs $(i, j)$, with contraction scheme represented in Fig. \ref{fig::mps}(d) for features on opposite sides with respect to the orthogonality center, otherwise the same procedure in the previous point (ii) has to be fulfilled.
\end{enumerate}

At each step of the gradient descent procedure, we implement a SVD for the new tensor $B'_{s_i, s_{i+1}}=U_{s_i} \Sigma V^\dagger_{s_{i+1}}$, followed by the contraction of $\Sigma$ with $U$ or $V^\dagger$ to obtain the orthogonality center $\Lambda$ if the step is moving to the left or right, respectively. The truncated matrix $\Sigma$ contains singular values $\{\lambda_j\}_{j=1,\dots,\chi}$, with $\chi$ bond dimension, such that we can easily evaluate the entanglement entropy $S(i)=-\sum_{j=1}^\chi \lambda_j^2 \log \lambda_j^2$. It is possible to project into a specific label subspace by considering the SVD of $B'^{0}_{s_i, s_{i+1}}$ and $B'^{1}_{s_i, s_{i+1}}$, leading to the evaluation of $S^{0}(i)$ and $S^{1}(i)$, respectively.

\section{Fashion MNIST binary classifications}\label{secA2}

\begin{figure}
    \centering
    \begin{tabular}{cc}
    \subfigure[]{\includegraphics[width=0.45\linewidth]{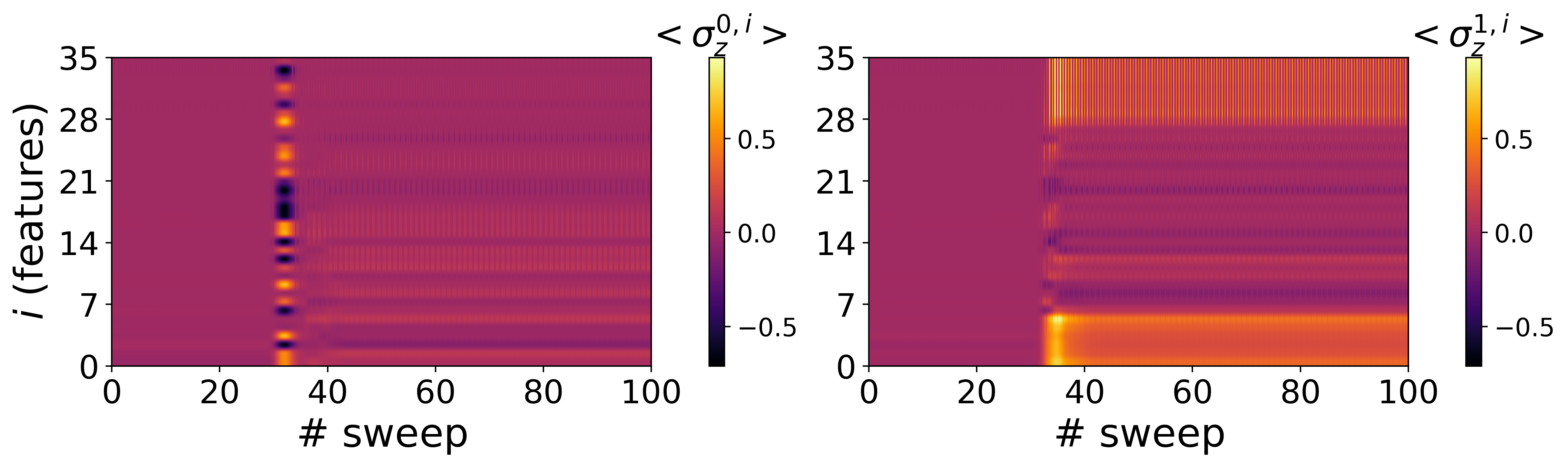}} &
    \subfigure[]{\includegraphics[width=0.45\linewidth]{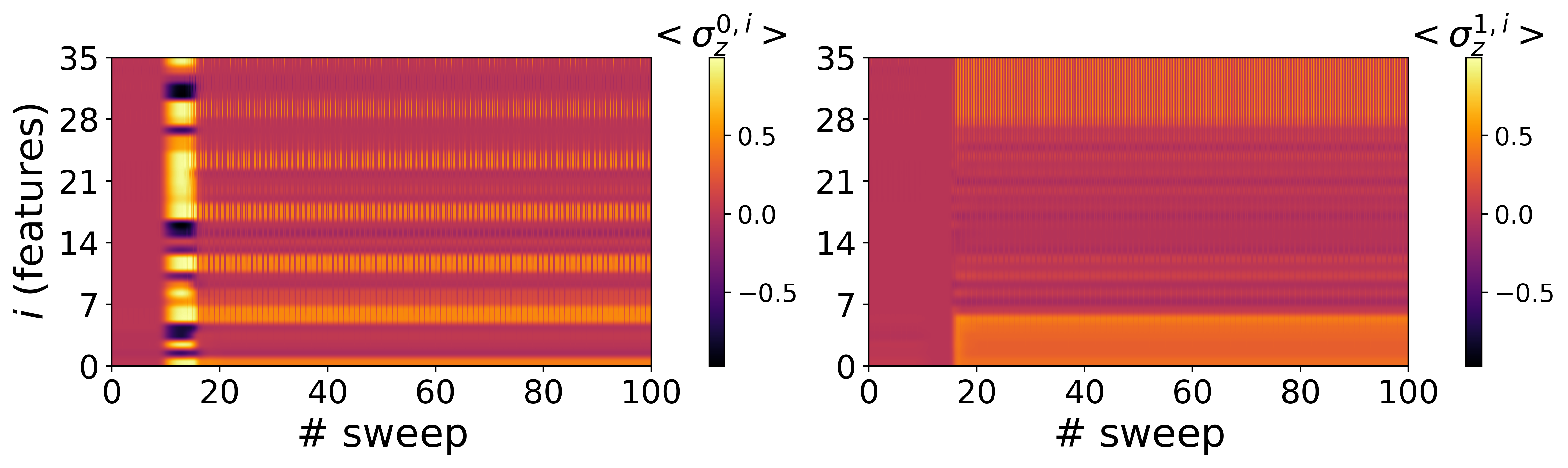}} \\
    \subfigure[]{\includegraphics[width=0.45\linewidth]{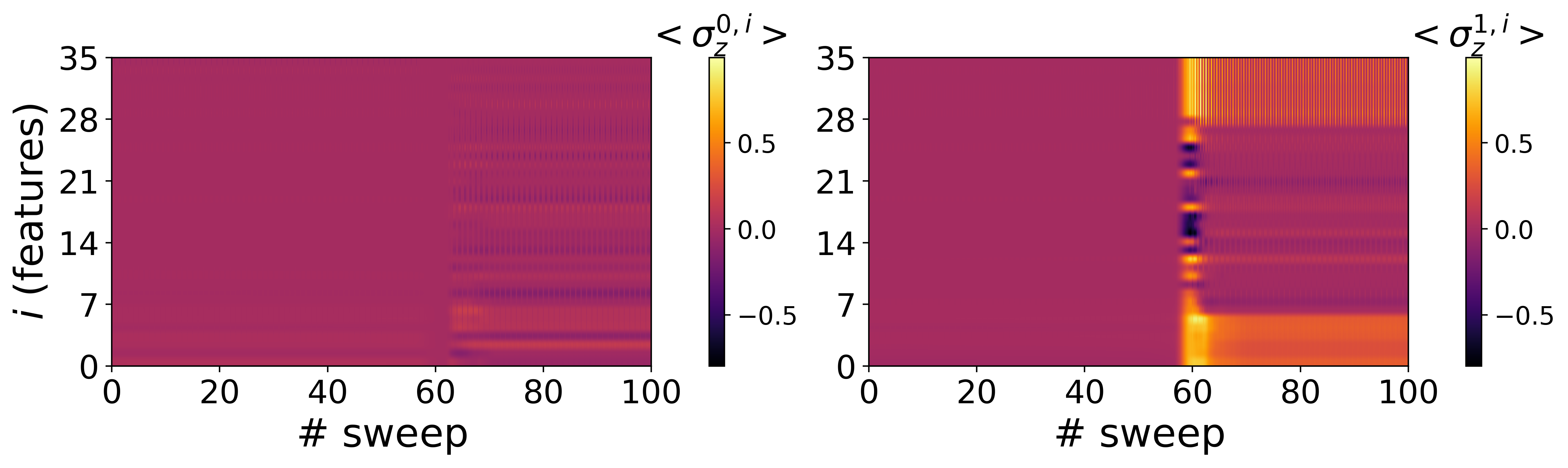}} &
    \subfigure[]{\includegraphics[width=0.45\linewidth]{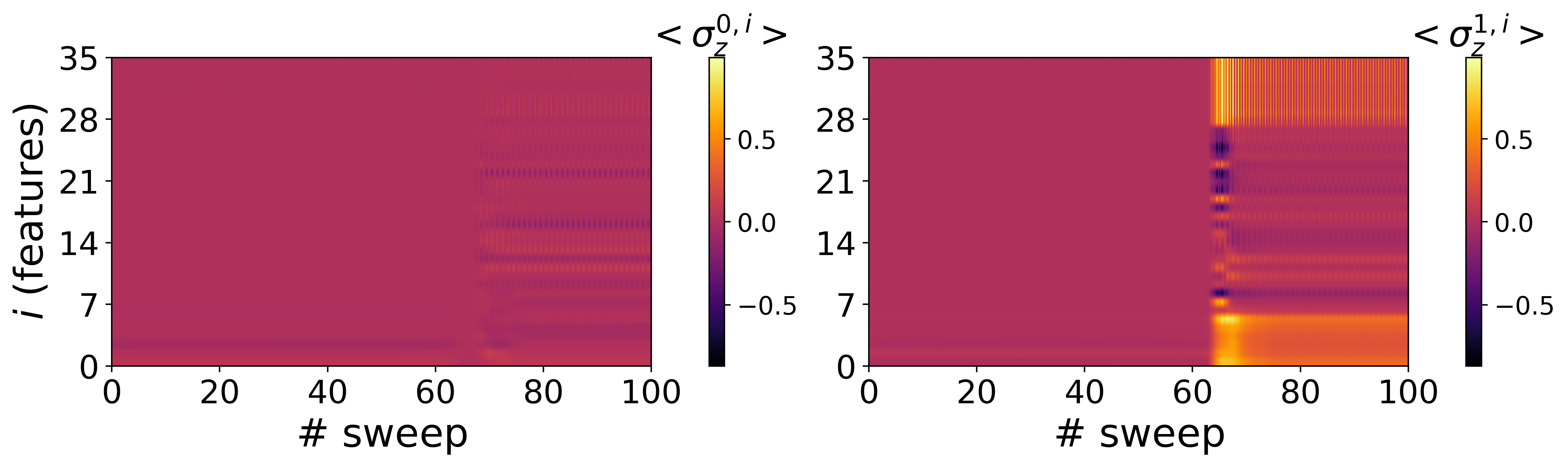}} \\
    \subfigure[]{\includegraphics[width=0.45\linewidth]{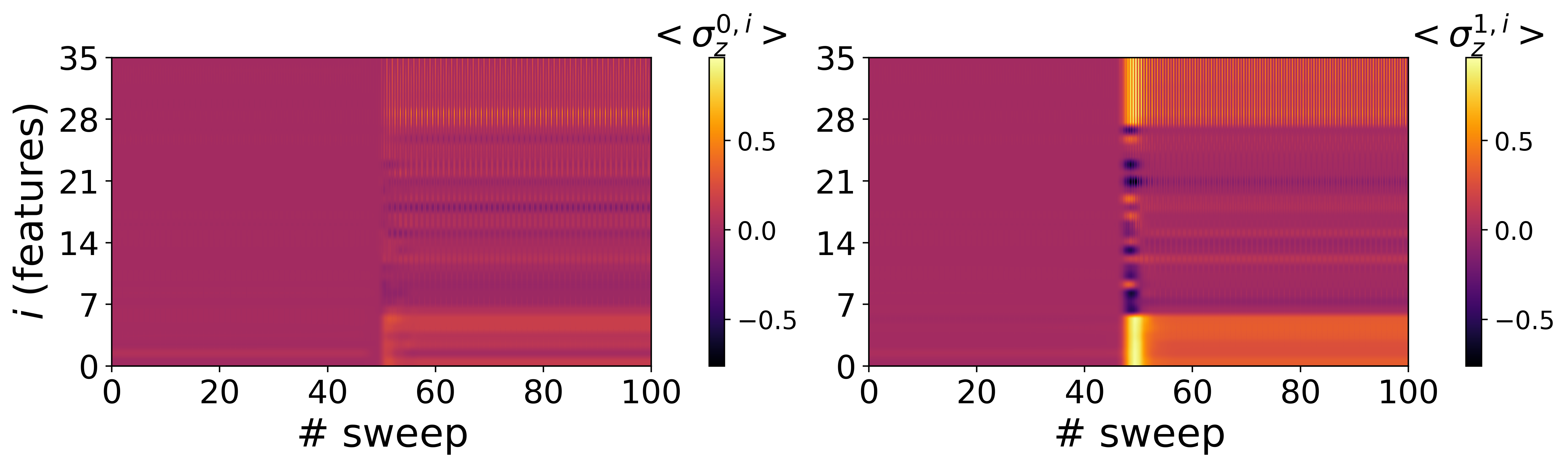}} & \subfigure[]{\includegraphics[width=0.45\linewidth]{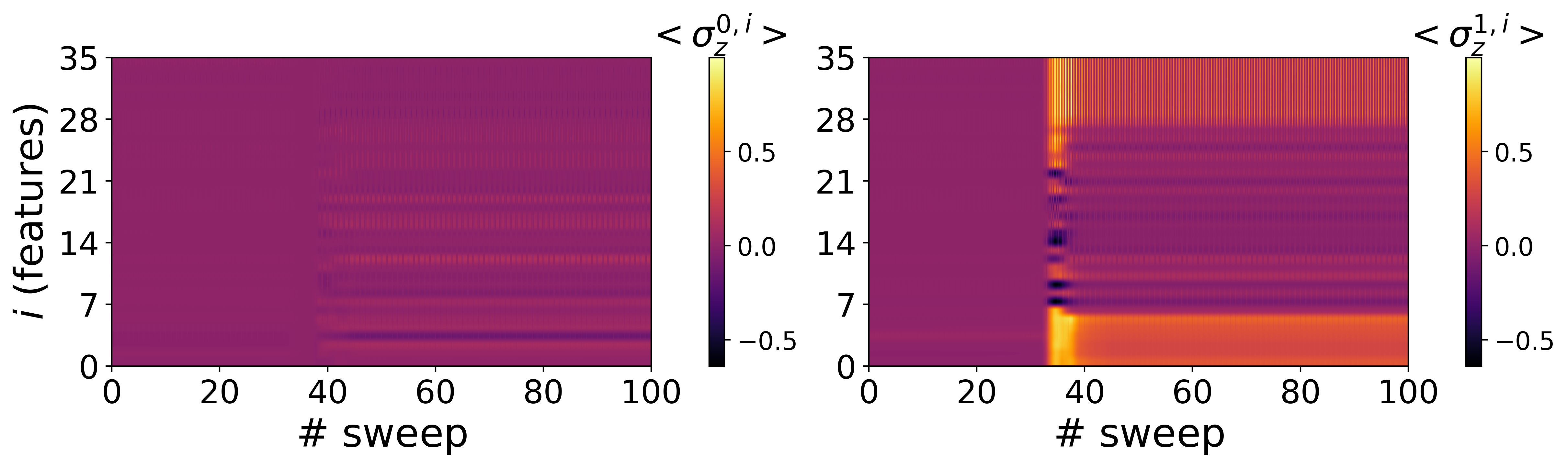}} \\
    \subfigure[]{\includegraphics[width=0.45\linewidth]{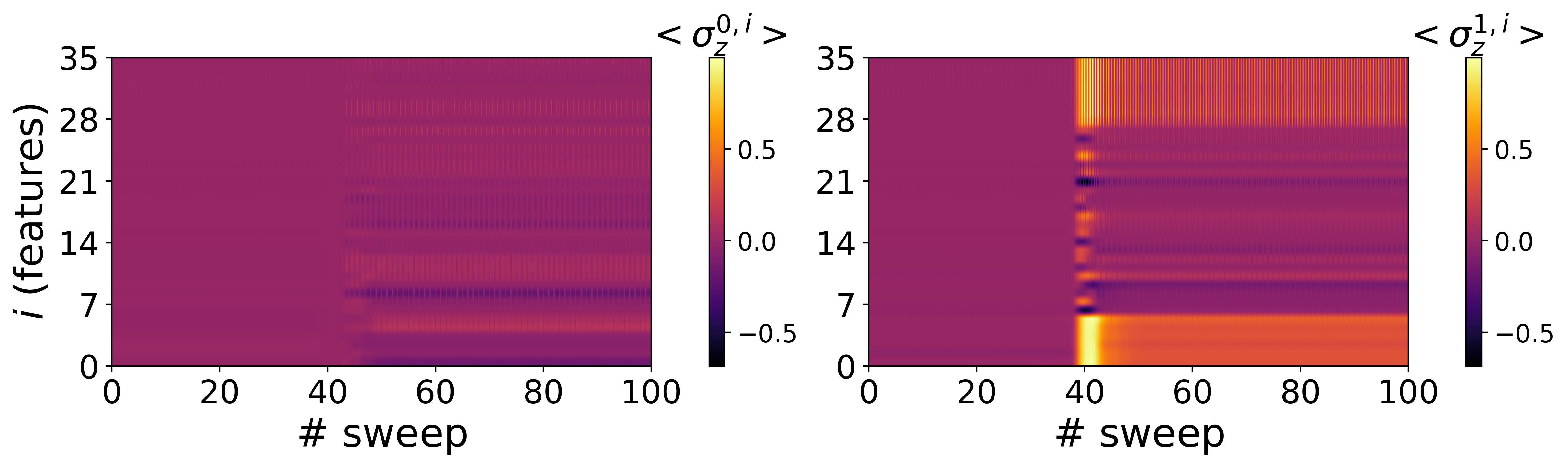}} &
    \subfigure[]{\includegraphics[width=0.45\linewidth]{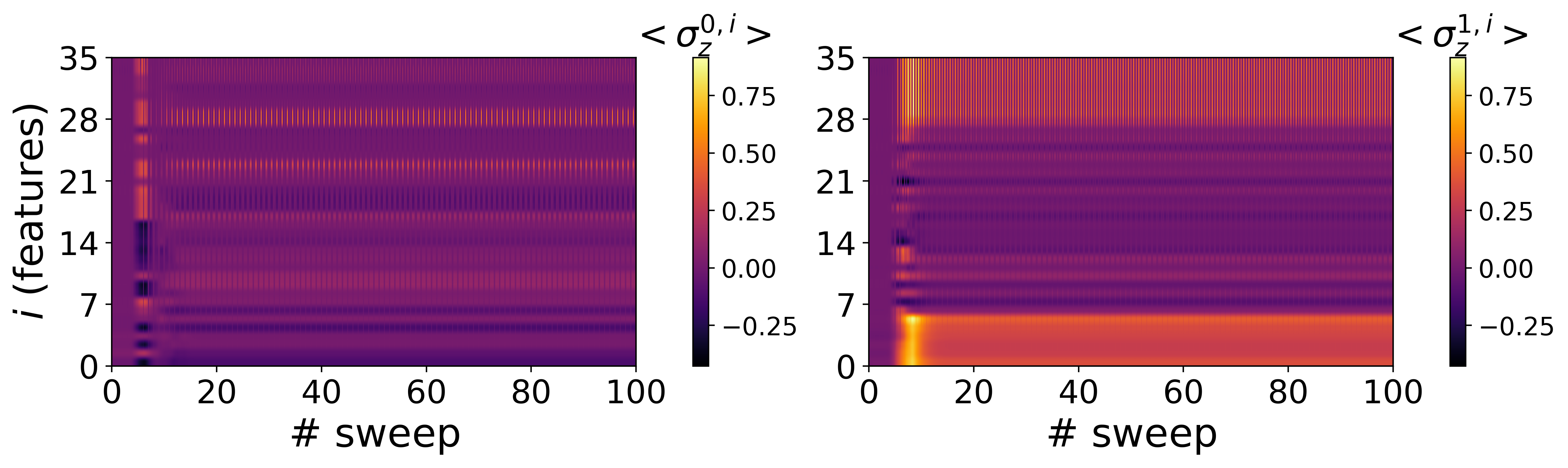}}
    \end{tabular}
    \caption{Magnetization profiles measures at each training step in the remaining binary classification problems involving sneaker available in the fashion MNIST dataset: (a) T-shirt, (b) trouser, (c) pullover, (d) coat, (e) sandal, (f) shirt, (g) bag and (h) ankle boot.}
    \label{fig:binary}
\end{figure}

\begin{figure}
    \centering
    \subfigure[]{\includegraphics[width=\linewidth]{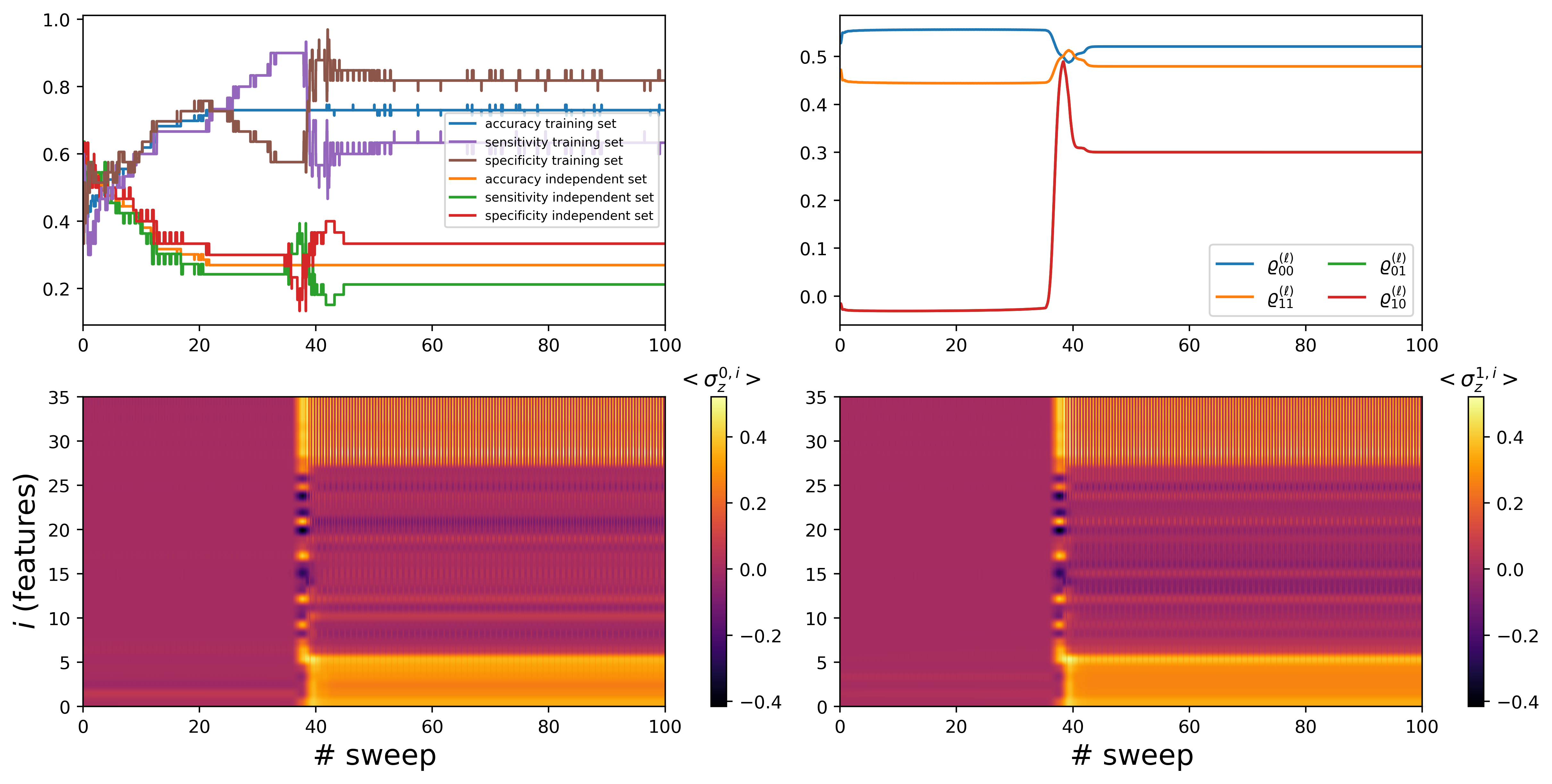}}
    \subfigure[]{\includegraphics[width=\linewidth]{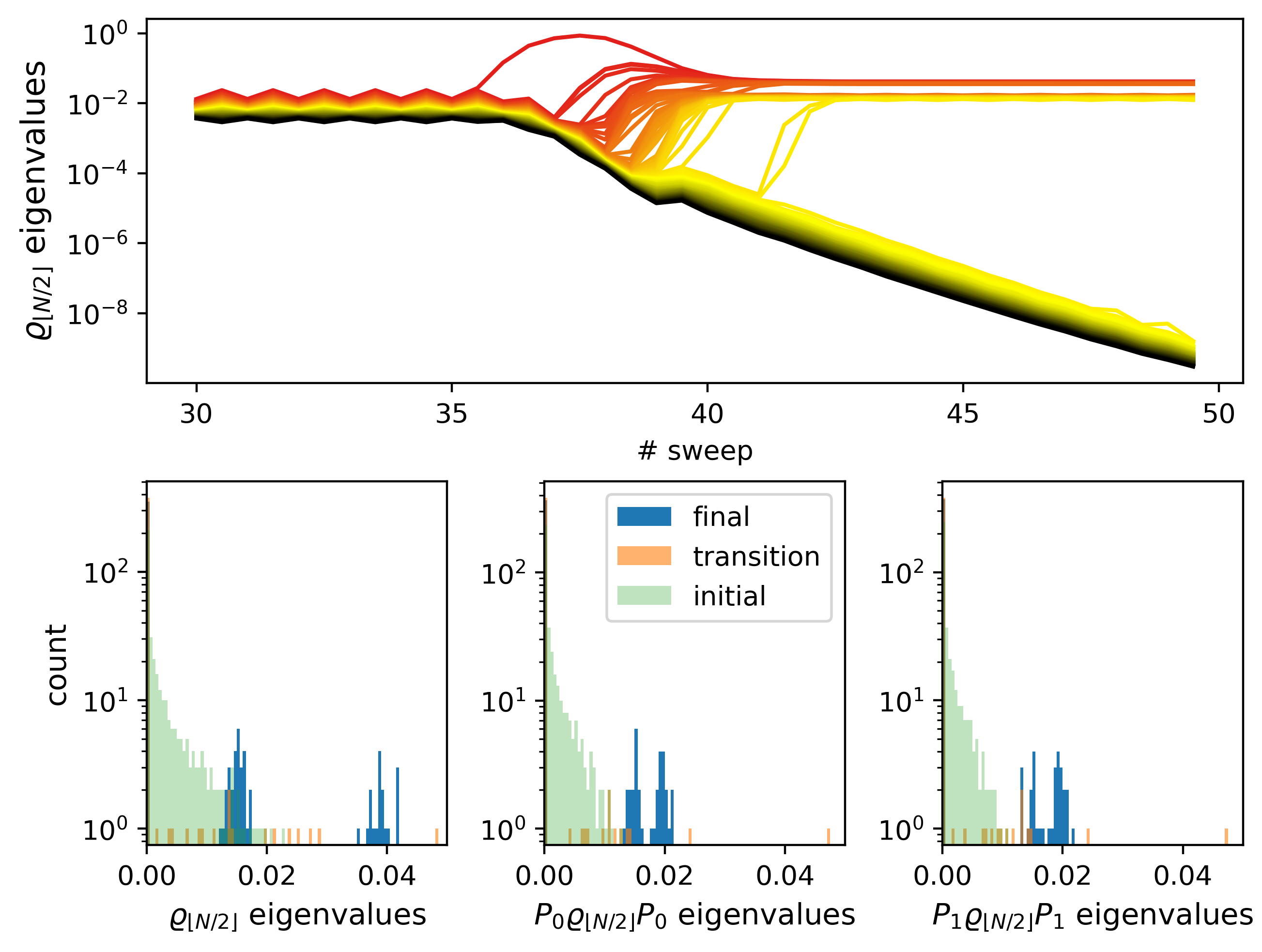}}
    \caption{Artificial fake sneaker versus sneaker classification: panel (a) shows performances, reduced density matrix elements and magnetization profiles, panel (b) represents the entanglement transition at the level of density matrix spectrum concerning the highest 100 eigenvalues.}
    \label{fig:fake}
\end{figure}

The stability of magnetization in the mask training is investigated in each of the possible binary classifications involving sneaker in fashion MNIST dataset. In Fig. \ref{fig:binary} we observe the grokking transition in all remaining cases with respect to the one discussed in Section \ref{sec3}. Since the same lattice size characterizes each classification problem, we choose the same bond dimension $\chi=400$. The wide variation of the sweep number required to observe the transition is affected by the initial random configuration and the specific images subsample. In each case we assign the label 1 to sneaker: in Fig. \ref{fig:binary} each panel confirms the definition of the same mask, with a particular case in panel (e), because sandal represent the most confounding classification, yielding almost the same mask of sneaker. Ankle boot and coat do not show high performances as well, probably because of the shape elongated along the horizontal direction, thus causing the first and last pixel rows behaving in a similar way with sneaker. Instead in panel (b) the mask for trouser is well differentiated, thus leading to the high performance reported in the legend of Fig. \ref{fig:volumearea}, and to a mask converging to the one of dress, discussed in Section \ref{sec3}, because of the similar shape in the $6 \times 6$ pixel format (see Fig. \ref{fig:fashion}). We can underline that the magnetization transition is observed before for sneaker masks in all cases, with the exception of panels (a), (b) and (h) standing for T-shirt, trouser and ankle boot, respectively.

The complete decoherence observed after the grokking transition in the reduced density matrix for the label space for the well performing classification in Section \ref{sec3} is further investigated by considering an artifical ‘‘fake" classification of sneaker versus sneaker. In Fig. \ref{fig:fake} we report a sharply different behavior with respect to Fig. \ref{fig:dress_sneaker} and Fig. \ref{fig:evl}. The coherence in the reduced density matrix reaches a peak corresponding to the magnetization transition shown in Fig. \ref{fig:fake} (a), related with an eigenvalues degeneracy condition with not distinguishable magnetization masks. After the transition the coherence reaches a not negligible steady condition, leading to almost perfectly overlapping magnetization masks. In Fig. \ref{fig:fake} (b) this new steady condition is analyzed in terms of density matrix spectra, collected corresponding to the sites pair $(\lfloor N/2 \rfloor, \lfloor N/2 \rfloor+1)$. Since the two masks are no more independent, it is not possible to obtain the spectrum of $\varrho_{\lfloor N/2 \rfloor}$ by simply combining those of the density matrix projected into each label subspace. The evaporated eigenvalues converge to two different groups, rather than just one as observed in Fig. \ref{fig:evl}. The separation between these two groups is higher in the full density matrix with respect to the projected ones because of the aforementioned coherence.

\section{Volume to sub-volume law entanglement transition}\label{secA3}

\begin{figure}
    \centering
    \subfigure[]{\includegraphics[width=\linewidth]{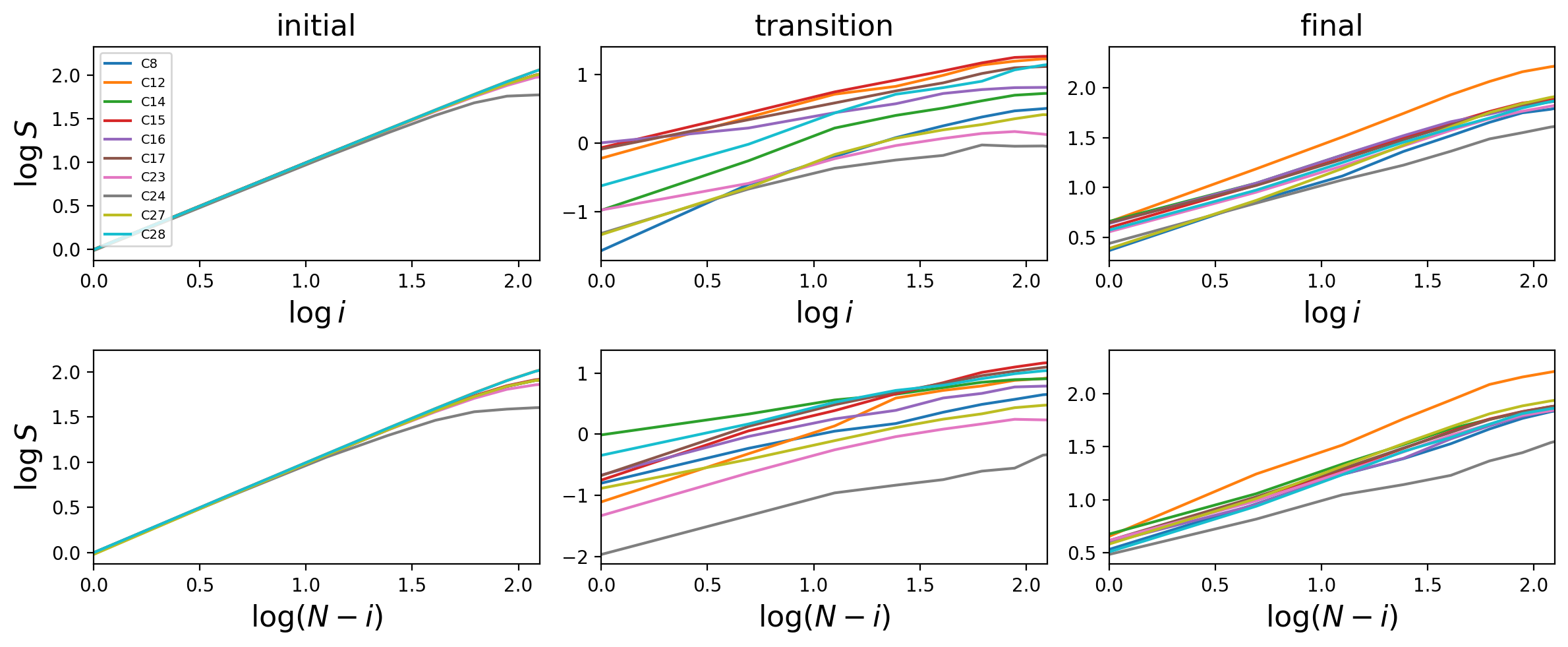}}
    \subfigure[]{\includegraphics[width=\linewidth]{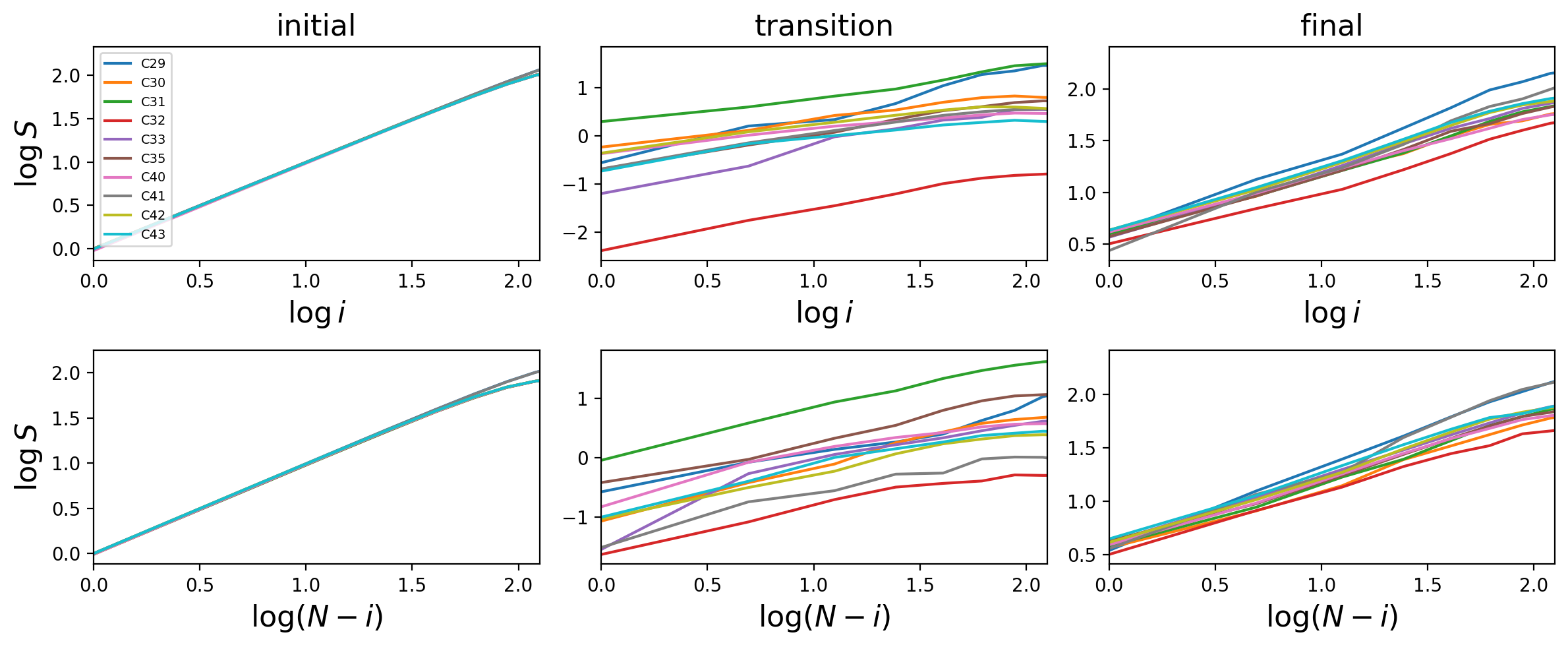}}
    \caption{Representation of entanglement entropy profiles with respect to the distance from the left and right boundaries of the one-dimensional lattice. In panel (a) we report communities from C8 to C28, ordered by number, while in panel (b) the remaining ones from C29 to C43. In both cases the top row is referred to the left boundary, while the bottom one to right boundaries. Each column contains data of the initial, transition or final sweep from left to right. We take into account 8 lattice sites from both boundaries in order to avoid the saturating entropy imposed by the finite bond dimension for the initial random MPS.}
    \label{fig:fit}
\end{figure}

The data discussed in Section \ref{subsec4a} regarding entanglement entropy scaling are shown in Fig. \ref{fig:fit}. In Table \ref{tab:left} and \ref{tab:right} we report both the slopes and coefficients of determination for linear fits, required in order to monitor the dependence of entanglement entropy on the distance from boundaries of the considered one-dimensional lattice on left and right, respectively. We choose to apply the linear fit in logarithmic scale for both axis by taking into account the eight lattice sites labeled by integer $i=1,\dots,8$ for the left boundary, as well as $N-i$ for the right one. This choice is motivated by the use of a bond dimension $\chi=400$ for almost all communities, with exceptions for C12, C28, C29 and C41 requiring $\chi=1000$, C23 with $\chi=300$ and C24 with $\chi=100$. As discussed in Section \ref{subsec4b}, the bond dimension influences the saturation value of entanglement entropy in random MPS with volume law: in order to establish a common number of sites for fitting every community, 8 sites results a proper choice to focus on the linear growth. Plots at left of each panel in Fig. \ref{fig:fit} confirm this assumption, since an almost indistinguishable behavior is verified for each community corresponding the random initialization.

The final behavior of entropy scaling corresponding to the preset number of sweeps shows an interesting trend for any community, since they collapse in an almost indistinguishable straight line in the logarithmic plane, as observed in plots at the right column of each panel in Fig. \ref{fig:fit}.

At the grokking transition the behavior is more heterogeneous as expected, because the feature extraction is highly dependent on both the community information content and on the particular permutation of gene expressions ordering. The observation of a gain in independent sets classification implies that an entanglement transition occurs, but the viceversa does not hold true: this statement based on our numerical experiments could justify the absence of any explicit relation of entropy scaling at transition with the measured evaluation metrics.

\section{Communities GSEA baseline without correlations thresholds}\label{secA4}

\begin{figure}
\centering
\captionof{table}[foo]{Baseline enriched gene subsets in C16, C28, C29, C40 and C41. \label{tab:old1}}
\includegraphics[width=\linewidth]{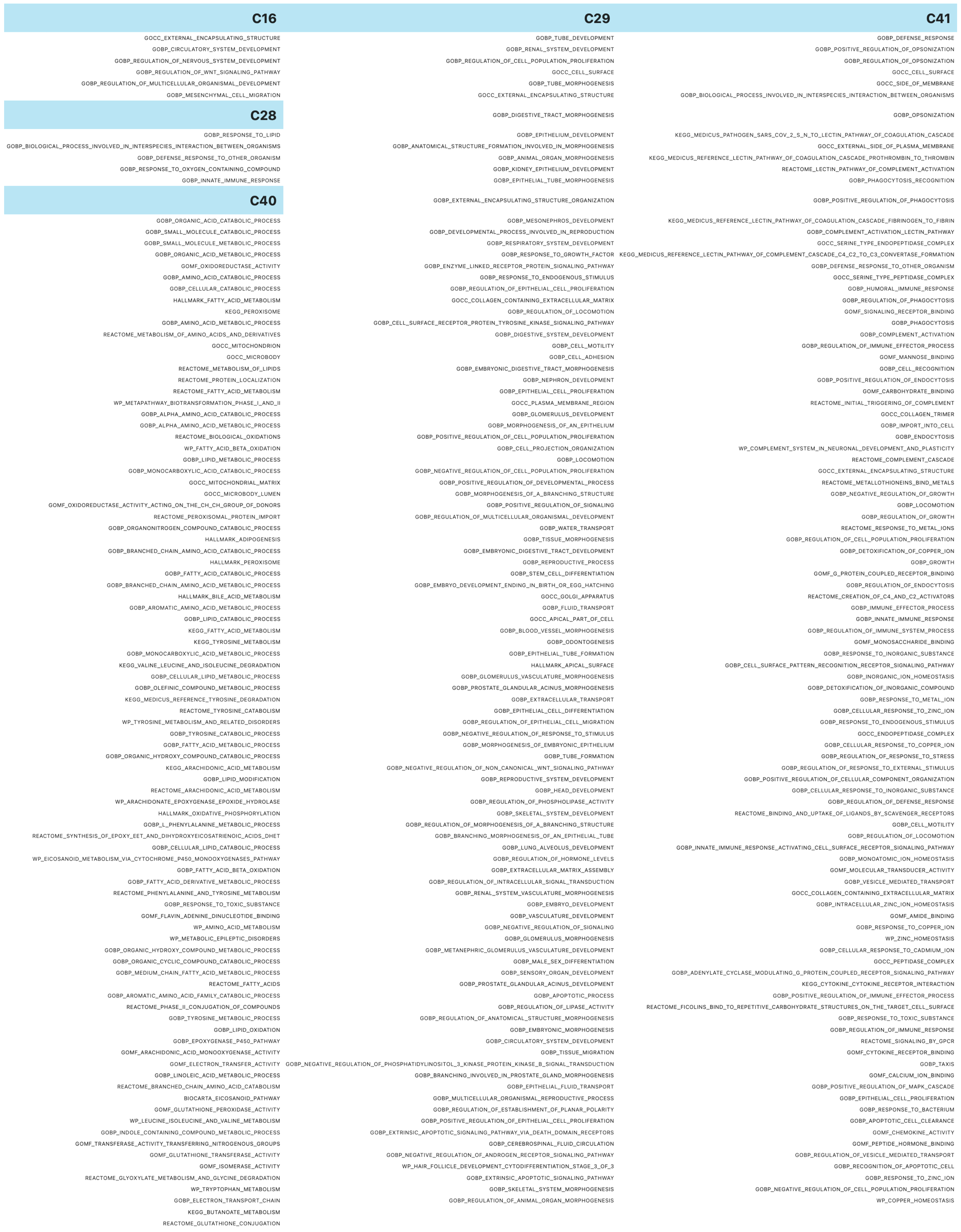}
\end{figure}

\begin{figure}
\centering
\captionof{table}[foo]{Baseline enriched gene subsets in C23, C30, C31, C33, C35 and C42. \label{tab:old2}}
\includegraphics[width=\linewidth]{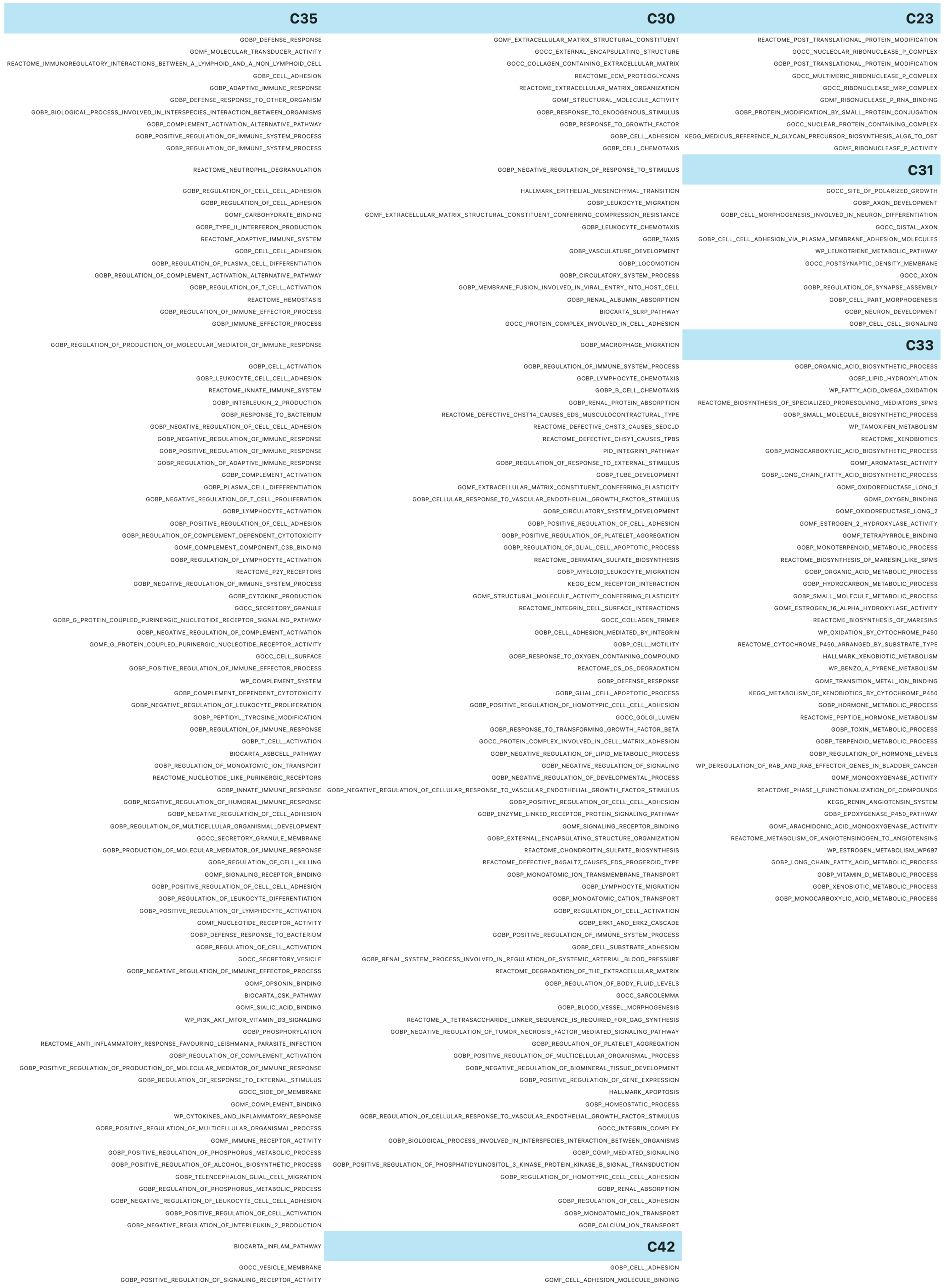}
\end{figure}

The considered gene expression communities determined in \cite{genescomm} are reduced to the common features subsets with the independent sets. This reduction of community sizes allows us to deduce new GSEA enrichments, representing the baseline with respect to thresholded sub-communities discussed in Section \ref{subsec4b}. These enriched gene subsets are listed comprehensively in Table \ref{tab:old1}, for enriched gene subsets in C16, C28, C29, C40 and C41,  and Table \ref{tab:old2}, for those in C23, C30, C31, C33, C35 and C42, where in C33 \texttt{GOMF\_OXIDOREDUCTASE\_LONG\_1} stands for \texttt{GOMF OXIDOREDUCTASE ACTIVITY ACTING ON PAIRED DONORS WITH INCORPORATION OR REDUCTION OF MOLECULAR OXYGEN NAD P H AS ONE DONOR AND INCORPORATION OF ONE ATOM OF OXYGEN} and \texttt{GOMF\_OXIDOREDUCTASE\_LONG\_2} for \texttt{GOMF OXIDOREDUCTASE ACTIVITY ACTING ON PAIRED DONORS WITH INCORPORATION OR REDUCTION OF MOLECULAR OXYGEN REDUCED FLAVIN OR FLAVOPROTEIN AS ONE DONOR AND INCORPORATION OF ONE ATOM OF OXYGEN}. We can check that C24 is the unique community achieving an accuracy higher than 0.65 without any enrichment for both the baseline and thresholded sub-communities.

We have to highlight that these enrichments contain some new information with respect to \cite{genescomm}, since we are considering the intersection of gene expressions in the training set GSE102079 and those of the independent set GSE54236. The immune microenvironment is confirmed in C29 and C41, as well as in C30, C35 and partially in C28, enriched in a more limited manner \cite{immune_cancer, immune1, immune2, immune3}. Another emerging confirmation regards metabolic abnormalities (obesity or diabetes), e.g. in C33 and C40, and alcohol consumption \cite{SHIN2023152134}. The vascular endothelial growth factor is enriched in C30, thus signaling a possible involvement of alpha-fetoprotein biomarker adopted in clinical practice \cite{AFP_endothelial, AFP}. The characterization of liver as a hormone-sensitive organ emerges in C33 with the enrichment of estrogen pathways, whose controversial action presents evidence suggesting a role as both a carcinogen and protective effect in liver \cite{BALDISSERA201667}.  The enrichment of Wnt pathways \cite{wnt} is observed in C16 and C29, even if this behavior is enforced in sub-communities presented in Section \ref{subsec4b}.



\end{appendices}

\bibliography{sn-bibliography}

\end{document}